\documentclass[11pt]{article}
\usepackage[T1]{fontenc}
\usepackage[utf8]{inputenc}
\usepackage{lmodern}
\usepackage{upgreek}
\usepackage{jheppub}
\usepackage{subcaption}
\usepackage{ragged2e}
 
\usepackage{amsmath,amssymb,amsfonts,dsfont,mathrsfs,amsthm,mathtools,bbm}
\definecolor{rossocorsa}{rgb}{0.83, 0.0, 0.0}
\definecolor{bleudefrance}{rgb}{0.19, 0.55, 0.91}

\usepackage{hyperref}
\hypersetup{
    colorlinks,
    citecolor=bleudefrance,
    filecolor=bleudefrance,
    linkcolor=rossocorsa,
    urlcolor=bleudefrance
}

\newcommand{\cT}{C_T}
\newcommand{\Ls}{L_{\star}}

\newcommand{\diff}{\text{d}}
\newcommand{\suniv}{s_{\text{univ}}}
\newcommand{\area}{\text{area}}
\newcommand{\eg}{\emph{e.g.,}}
\newcommand{\ie}{\emph{i.e.,}}
\newcommand{\qt}[2]{\left.#1\right|_{\text{#2}}}

\newcommand{\qti}[2]{\left.#1\right|_{#2}}

\newcommand{\cTFb}[1]{\left.\frac{C_{T}}{F_0}\right|_{#1}}
\newcommand{\cTb}[1]{\left.C_{T}\right|_{#1}}
\newcommand{\Fb}[1]{\left.F_0\right|_{#1}}

\newcommand{\Nf}{N_{\text{f}}}

\newcommand{\iu}{\text{i}}
\newcommand{\expp}[1]{\text{e}^{#1}}

\newcommand{\be}{\begin{equation}}
	\newcommand{\ee}{\end{equation}}
\newcommand{\bea}{\begin{eqnarray}}
	\newcommand{\eea}{\end{eqnarray}}

\begin{document}
\noindent\flushright
{YITP-26-30} 
\noindent\flushleft\justifying
\vspace{-2cm}
\title{\huge Entanglement entropy and conformal bounds for $d=5$ CFTs}

\author[a]{Pablo Bueno,}
\author[a]{Adam Fernández García,}
\author[b]{Francesco Gentile,}
\author[c]{Oscar Lasso Andino,}
\author[d,a]{Javier Moreno}

\affiliation[a]{Departament de F\'isica Qu\'antica i Astrof\'isica, Institut de Ci\'encies del Cosmos,\\
Universitat de Barcelona, Mart\'i i Franqu\'es 1, E-08028 Barcelona, Spain \vspace{0.1cm}}
\affiliation[b]{SISSA and INFN Sezione di Trieste, via Bonomea 265, 34136, Trieste, Italy
\vspace{0.1cm}}
\affiliation[c]{Escuela de Ciencias Físicas y Matemáticas, Universidad de Las Américas,\\ Redondel del ciclista, Antigua vía a Nayón, C.P. 170504, Quito, Ecuador
\vspace{0.1cm}}
\affiliation[d]{Center for Gravitational Physics and Quantum Information,\\
Yukawa Institute for Theoretical Physics, Kyoto University,\\
Kitashirakawa Oiwakecho, Sakyo-ku, Kyoto 606-8502, Japan}

\vspace{0.3cm}
\emailAdd{pablobueno@ub.edu}
\emailAdd{fgentile@sissa.it}
\emailAdd{oscar.lasso@udla.edu.ec}
\emailAdd{javier.moreno@yukawa.kyoto-u.ac.jp}

\abstract{
 The entanglement entropy of spacetime regions $A$ in odd-dimensional conformal field theories (CFTs) contains a universal constant term, $(-1)^{\frac{d-1}{2}}F(A)$. This quantity can be robustly defined by considering the mutual information of pairs of slightly deformed versions of $A$. In the case of general three-dimensional CFTs, $F(A)$ is positive definite and bounded below by the round disk result, $F(A)\geq F_0\equiv F(\partial A=\mathbb{S}^1)$. Additionally, strong evidence has been provided that for every region $A$, $F(A)/F_0$ is maximized, within the space of CFT$_3$'s, by the free scalar field result. In this paper we show that while $F(A)$ remains a local minimum around $F_0\equiv F(\partial A=\mathbb{S}^3)$ for small deformations of the spherical entangling surface, it can take values of arbitrarily large magnitude with either sign for more general regions, and hence it is neither upper- nor lower-bounded in general CFT$_5$'s. We argue that an analogous conjecture regarding the extremization of  $F(A)/F_0$ for general regions within the space of theories fails in $d=5$. We instead analyze the viability of the weaker bound,
 $F_{\epsilon}/F_0\leq  \qt{F_{\epsilon}/F_0}{free scalar}$, $\forall$CFT$_5$ for general small geometric deformations of the spherical entangling surface. This is equivalent to a general constraint involving the stress-tensor two-point function $C_T$ and the Euclidean partition function on the sphere, namely, $C_T/F_0\leq  \qt{C_T/F_0}{free scalar}\approx 0.314$, which we show to hold for all known CFT$_5$'s. We also comment on possible extensions of this result to higher dimensions.
}

\maketitle
\section{Introduction}

In the algebraic approach to quantum field theory (QFT)~\cite{Haag:1963dh}, the entanglement entropy (EE) of spatial subregions plays a canonical role as a statistical measure of vacuum fluctuations in the algebras associated to such regions~\cite{Casini:2022rlv}. Still, somewhat paradoxically, the EE is divergent in any state in a QFT. In the presence of a UV regulator $\delta$, the vacuum EE of a  general smooth entangling region $A$ admits an expansion of the form~\cite{Liu:2012eea}
\begin{equation}\label{eq: EE expression}
S_\delta(A)=c_{d-2} \frac{H^{d-2}}{\delta^{d-2}}+\cdots 
+\begin{cases}c_1 \frac{H}{\delta} +(-1)^{\frac{d-1}{2}}  F(A)\, , & \text {for odd } d\,, \\ c_2 \frac{H^2}{\delta^2}+(-1)^{\frac{d-2}{2}} \suniv(A) \log\frac{H}{\delta}+c_0\, , &\text {for even } d\,. \end{cases}
\end{equation}
In the above formula, $H$ is some characteristic length of $A$ and the $c_{i}$ coefficients are non-universal and have a local nature---with the exception of $c_0$, which may also capture long-range information. In even dimensions, the universal logarithmic coefficient $s^{\text{univ}}$ is local and, in the case of a conformal field theory (CFT) it involves a linear combination of theory-independent integrals of intrinsic and extrinsic curvatures of the entangling surface $\partial A$ weighted by the trace-anomaly coefficients of the corresponding theory~\cite{Calabrese:2004eu,Solodukhin:2008dh,Safdi:2012sn,Miao:2015iba}. On the other hand, $F(A)$ is an inherently non-local and state-dependent universal coefficient (when properly defined). In the prototypical case of round balls---or, equivalently, of spherical entangling surfaces---$F(\mathbb{B}^{d-1})\equiv F_0$ in the vacuum state equals the Euclidean partition function of the corresponding CFT on the round sphere~\cite{Dowker:2010bu,Casini:2011kv}---see~\eqref{eq:CHMfe}.

If instead of a single entangling region we consider two, $A,B$, we can easily define an intrinsically finite and universal quantity as a simple linear combination of EEs. This is the mutual information (MI), 
\begin{equation}
    I(A,B)= \lim_{\delta\rightarrow 0}\left[ S_{\delta}(A)+S_{\delta}(B)-S_{\delta}(AB) \right]\,,
\end{equation}
where we denoted $A\cup B \equiv AB$. Thus defined, the MI does not depend on $\delta$ and, in fact, it can be alternatively defined directly in the continuum in terms of the Araki relative entropy ---see \eg~\cite{Witten:2018zxz}. Interestingly, the MI allows one to extract CFT data from two complementary approaches. On the one hand, the long-distance expansion of MI for spherical regions gives rise to a---not fully developed yet---systematic procedure for identifying the scaling dimensions, spins and OPE coefficients of the underlying theory~\cite{Long:2016vkg,Chen:2016mya,Chen:2017hbk,Cardy:2013nua,Agon:2015ftl,Agon:2021zvp,Casini:2021raa,Agon:2021lus,Agon:2022efa,Agon:2024zae,Agon:2024xvs}. On the other hand, a convenient short-distance expansion of the MI provides a universal geometric regulator for the EE. This is achieved by considering the two regions as slightly reduced ($A^-$) and increased ($\overline{A^+}$) versions of some reference region $A$ and performing a short-distance expansion on their mutual separation, $\varepsilon$---see Figure~\ref{fig:MI} below. The resulting regulated entropy,
\begin{equation}
S_\text{reg}(A,\varepsilon)\equiv \frac{1}{2}I(A^+,A^-)= \frac{1}{2}\lim_{\delta\rightarrow 0} \left[S_{\delta}(\overline{A^+})+S_{\delta}(A^-)-S_{\delta}(\overline{A^+ A^-})\right]\, ,
\end{equation}
is a well-defined quantity in the continuum and it provides a robust definition of $F(A)$~\cite{Casini:2007dk,Casini:2008wt,Casini:2014yca,Casini:2015woa}.

Thus defined, $F(A)$ satisfies a number of interesting general properties in $d=3$. Firstly, it has been proved that $F_0$ globally minimizes $F(A)$ for arbitrary regions and for general CFT$_3$'s~\cite{Bueno:2021fxb}. Additionally, it was conjectured in~\cite{Bueno:2023gey} that for general theories and entangling regions, the ratio $F(A)/F_0$ is bounded above and below, respectively, by the free scalar and Maxwell theory results,
\begin{equation}\label{eq:bounds3d0}
\left.\frac{F(A)}{F_0}\right|_{\text {Maxwell }} \leq \frac{F(A)}{F_0} \leq\left.\frac{F(A)}{F_0}\right|_{\text {free scalar }}\,,\quad \forall A\,,\forall\text{CFT}_3\, .
\end{equation}
While the lower bound can be rigorously proved, the more interesting upper bound rests on a series of complementary arguments and proofs that, taken together, provide a compelling case for its validity---see Section~\ref{sec:bound3} for a recapitulation. Exploiting Mezei's formula for $F(A)$ in the case of small arbitrary geometric deformations of a spherical entangling surface in general dimensions~\cite{Mezei:2014zla,Faulkner:2015csl}, the conjecture becomes a constraint on the possible values of the ratio $C_T/F_0$, where $C_T$ is the stress-tensor two-point function coefficient~\cite{Bueno:2023gey}---see~\eqref{eq:CT} for its definition. This reads 
\begin{equation}\label{eq:bounds3CT0}
0 \leq \frac{C_T}{F_0} \leq\left.\frac{C_T}{F_0}\right|_{\text {free scalar}}=\frac{3}{4 \pi^2 \log 2-6 \zeta(3)} \approx 0.149\,, \quad \forall\text{CFT}_3\,.
\end{equation}
As summarized in Figure~\ref{fig:CTF0} below, all theories for which this ratio is known conform to the conjecture. 

It is the goal of this paper to determine the possible validity of analogous general bounds in the case of five-dimensional CFTs. As it turns out, explicit computations of EE in $d=5$ are so scarce that they can essentially be summarized in a few lines. On the one hand, for spherical regions and their general geometric perturbations, the Casini-Huerta-Myers~\cite{Dowker:2010bu,Casini:2011kv} and Mezei results~\cite{Mezei:2014zla,Faulkner:2015csl} are valid for general CFT$_d$'s, and therefore also apply in the particular case of $d=5$. General results for strip regions in the case of free scalars and fermions are also available~\cite{Casini:2009sr}. A greater variety of results exists in the case of five-dimensional holographic theories dual to Einstein gravity~\cite{Ryu:2006bv,Ryu:2006ef,Nishioka:2009un,Hung:2011xb,Myers:2012vs}. In particular, it was argued in~\cite{Anastasiou:2024rxe} that $F(A)$ could be written in that case as a functional on a duplicated version of the Ryu-Takayanagi surface embedded in $\mathbb{R}^5$, namely, $\qt{F(A)}{holo}\propto {\bf W}_5(2\Sigma_A)$. Here, ${\bf W}_5$ is a higher-dimensional version of the so-called Willmore energy~\cite{willmore1965note,willmore1996riemannian,marques2014min,Toda2017Willmore,Babich:1992mc,alexakis2008renormalized,Astaneh:2014uba,guven2005conformally,Graham:2017bew,Zhang:2017lcd}, which analogously captures $F(A)$ for three-dimensional holographic theories~\cite{Fonda:2014cca,Fonda:2015nma,Seminara:2018pmr,Anastasiou:2020smm,Anastasiou:2021swo}. In contradistinction to its three-dimensional cousin, ${\bf W}_5$ lacks a definite sign~\cite{Graham:2017bew,Martino:2023oog,lan2025analysis}, suggesting that $F(A)$ is bounded neither above nor below for general CFT$_5$'s~\cite{Anastasiou:2024rxe}. We confirm this explicitly below. 

The fact that $F_0$ is not a global minimum of $F(A)$ and that $F(A)$ can have either sign spoils the extrapolation to $d=5$ of several of the arguments which sustain the three-dimensional conjectural bounds (\ref{eq:bounds3d0}). In the absence of a clear replacement of the three-dimensional Maxwell theory as a candidate for the lower bound, we focus on the upper bound, hypothetically provided by the free scalar theory. By considering strip entangling regions, for which the sign of $F(A)$ flips with respect to $F_0$ in $d=5$, we argue that the five-dimensional version of the bound does not hold, namely,
\begin{equation}\label{eq:bounds5d0}
\frac{F(A)}{F_0} \not\leq \qt{\frac{F(A)}{F_0}}{free scalar}\,,
\quad \forall \text{CFT}_5\,,\ \forall A\,.
\end{equation}
On the other hand, $F(A)$ does behave very closely to its three-dimensional counterpart in the case of small geometric deformations of spherical regions, namely, $F_0$ locally provides a universal lower bound for $F(A)$, which suggests a milder version of the conjecture. Exploiting Mezei's formula, this is most elegantly expressed in terms of the coefficient which universally controls the leading correction to $F(A)$ arising from an arbitrary geometric deformation of the sphere $\mathbb{S}^3$, namely, $C_T$. In terms of this quantity, the conjecture reads
\begin{equation}\label{ctf050}
\frac{C_T}{F_0} \leq\qt{\frac{C_T}{F_0}}{free scalar}=\frac{45}{\pi ^4 \log 4+2 \pi ^2 \zeta (3)-15 \zeta (5)}\approx 0.314\, ,\quad \forall \text{CFT}_5 \,.
\end{equation}
We have collected the values of this ratio for all families of five-dimensional CFTs for which (to the best of our knowledge) it has been computed. As summarized in Fig.\,\ref{fig:CTF0}, they all satisfy the above constraint.

Moving to even higher odd dimensions---where the existence of interacting CFTs is dubious, but one can at least compare free and holographic models---we argue that $F(A)$ does not have a sign either in $d=7$, so the corresponding versions of (\ref{eq:bounds5d0}) most likely fail as well for $d\geq 7$. On the other hand, it seems that general-dimension versions of (\ref{ctf050}) are likely true. 

The remainder of the paper is organized as follows. In Section~\ref{sec:5dEEMI} we present a rigorous definition of the finite contribution to the EE in five-dimensional CFTs, $F(A)$, using MI as a geometric regulator in the short-distance expansion. Using this formula, we evaluate the corresponding universal quantities for several entangling surfaces, including the sphere, strip-like, and cylindrical geometries, within the Extensive Mutual Information model. The details of these computations are provided in Appendix~\ref{sect:EMIappendix}. Based on these results and several additional general observations, we argue that this finite contribution does not possess a definite sign. In Section~\ref{sec:bounds} we investigate the existence of bounds on the ratio between $F(A)$ and its value for a spherical entangling surface, the sphere free energy $F_0$, in five-dimensional CFTs. We begin by reviewing known results in three and four dimensions---for which we also present a new result in the case of strip regions, and then show that, in contrast to these lower-dimensional cases, such bounds cannot hold for general entangling regions in five dimensions. 
Then, we show that the perturbative version of these bounds in the vicinity of spherical entangling surfaces does hold for all known theories, and comment on its higher-dimensional extensions. 
A comprehensive list of universal coefficients relevant for the conformal bounds across dimensions is presented in Appendix~\ref{sec:appuq}. Some technical comments about the holographic entanglement entropy of a conical region in $d=5$ appear in Appendix~\ref{holoCorner}. 
 We conclude with some remarks in Section~\ref{conclu}.

\section{Five-dimensional entanglement entropy from mutual information}\label{sec:5dEEMI}
Mutual information (MI) can be used to provide a robust and universal\footnote{Both in the sense that the prescription is valid for general theories, and in the sense that all coefficients involved in the corresponding expansion capture information about the continuum theory.} definition of the EE universal coefficient $F(A)$ in general odd-dimensional theories. This idea was pioneered in~\cite{Casini:2007dk,Casini:2008wt,Casini:2014yca,Casini:2015woa} and has been exploited in previous works~\cite{Bueno:2021fxb,Huerta:2022cqw,Bueno:2023gey}, mostly in the case of three-dimensional theories. The motivation in all cases was to deal with potential ambiguities associated to the definition of constant universal terms in the EE and, in particular, to provide a procedure which: leads to convergent results in the lattice for general regions~\cite{Casini:2015woa,Bueno:2021fxb}; allows one to exploit certain general results regarding the interplay between mutual information, symmetries and superselection sectors~\cite{Huerta:2022cqw,Casini:2019kex,Casini:2019nmu}. We will extend this approach here to the case of five-dimensional CFTs. 

As opposed to the EE, the MI can be rigorously defined in the continuum using Araki's relative entropy, so it is free from ambiguities by construction. If we insist on thinking in terms of entanglement entropies, the MI of two subregions $A,B$ is defined as  
\begin{equation}
    I(A,B)= \lim_{\delta\rightarrow 0}\left[ S_{\delta}(A)+S_{\delta}(B)-S_{\delta}(AB) \right]\,,
\end{equation}
where we denoted $A\cup B \equiv AB$.  In the presence of any UV regulator $\delta$, all divergences appearing in the above combination cancel one another as $\delta \rightarrow 0$, rendering $I(A,B)$ finite for any pair of disjoint regions.\footnote{For general comments on the MI of touching regions, see~\cite{Bueno:2019mex}.} 

\begin{figure}
    \centering
    \includegraphics[width=0.64\linewidth]{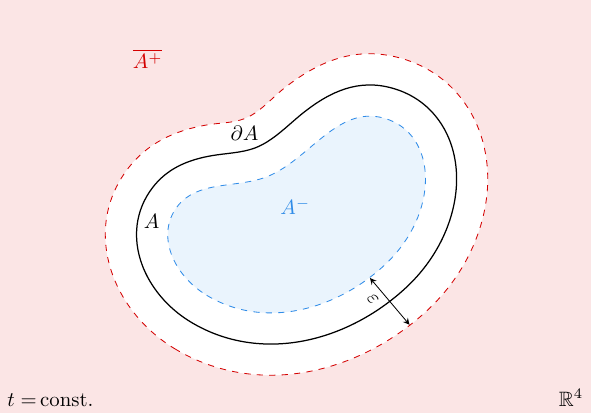}
    \caption{A smooth spatial region $A$ in $\mathbb R^4$, and two concentric regions $A^-$ and $\overline{A^+}$, obtained by displacing the entangling surface $\partial A$ along its normal direction at each point by distances $-\varepsilon/2$ and $\varepsilon/2$, respectively.}
    \label{fig:MI}
\end{figure}

Now, consider the ground state of a CFT in $d$-dimensional Minkowski spacetime $\mathbb R^{1,d}$ and let $A$ be a smooth spatial region on a fixed constant-time hypersurface.  At every point of the entangling surface $\partial A$, consider the two points separated a distance $\varepsilon/2$ outwards and inwards, respectively, along the normal direction. This defines the boundaries of two concentric regions, $A^-$ and $\overline{A^+}$ with a shape similar to $A$ and which would reduce to it as $\varepsilon \rightarrow 0$: $A^-$ is the region bounded by the points resulting from moving inwards, whereas $\overline{A^+}$ is the one bounded by the points resulting from moving outwards from $\partial A$---the setup is sketched in Figure~\ref{fig:MI}. In this situation, consider the MI
\begin{equation}\label{miee}
    I_{\varepsilon}(A)\equiv I(A^+,A^-)=\lim_{\delta\rightarrow 0} \left[S_{\delta}(A^+)+S_{\delta}(A^-)-S_{\delta}(A^+ A^-)\right]\,,
\end{equation} 
which for pure states is equivalent to
\begin{equation}
     I_{\varepsilon}(A)=\lim_{\delta\rightarrow 0} \left[S_{\delta}(\overline{A^+})+S_{\delta}(A^-)-S_{\delta}(\overline{A^+ A^-})\right]\,.
\end{equation}
In this formula, the last term is the entropy of the (hyper-)annulus region of width $\varepsilon$, so the three contributions correspond to bounded regions provided $A$ is bounded. It is also clear from the construction that we can identify
\begin{equation}
    S_\text{reg}(A,\varepsilon)\equiv  \frac{1}{2}I_{\varepsilon}(A)\, ,
\end{equation} 
whenever $\varepsilon$ is much smaller than any characteristic length scales of $A$, $\varepsilon\ll L_A$, namely, we can use $I_{\varepsilon}(A)/2 $ to define a new regulated version of the entropy.\footnote{If we were computing the various pieces appearing in~\eqref{miee} using some regulated version of the CFT with a UV regulator $\delta$---\eg\ in the lattice---we would also demand that $\varepsilon \gg \delta$. If the mutual information was computed directly in the continuum theory, no UV regulator would play any role in the construction.} In this expression, $\varepsilon$ is a physical distance and $S_\text{reg}(A,\varepsilon)$ is a well-defined quantity in the continuum theory.

Now, consider an expansion of this quantity for $\varepsilon\ll L_A$ in the case of a general five-dimensional CFT. The structure of terms is analogous to the UV-regulated EE one, namely,
\begin{equation}\label{eq:MI5}
I_{\varepsilon}(A)
= k\,\frac{\area(\partial A)}{\varepsilon^3}
- \frac{1}{\varepsilon} \int_{\partial A} \left[ \alpha_1\,\mathcal{R} + \beta_1K^2 + \beta_2\,K_{ab}K^{ab} \right] + 2F(A)+ \mathcal{O}(\varepsilon)\,,
\end{equation}
but, again, $\varepsilon$ is an actual physical distance, and the coefficients $k,\alpha_1,\beta_1,\beta_2,F(A)$ are all universal and characterize the CFT. In particular, $k$ is the universal coefficient appearing in the EE of thin strip regions in general dimensions \cite{Casini:2005zv,Casini:2007dk,Casini:2015woa}.\footnote{Although in higher dimensions the geometry is technically a hyperslab, we will continue referring to it as a ``strip'' for simplicity.} The coefficients $\alpha_1, \beta_1, \beta_2$ are not all independent. This is because the Gauss-Codazzi relation in a Minkowski ambient space, $\mathcal{R} = K^2 - K_{ab}K^{ab}$, implies that one of the coefficients can be expressed in terms of the other two. Taking this into account, we may rewrite~\eqref{eq:MI5} as\footnote{A definition of the universal coefficient for smooth entangling regions employing EE and an explicit UV regulator $\delta$ was provided in~\cite{Liu:2012eea}. Following that approach we have found a closed polynomial expression for the (universal) finite part $F(A)$ as well as for the coefficient of the logarithmic divergence $\suniv(A)$, namely,
\begin{alignat}{3}
    F(A)&=\frac{(-1)^{(d-1)/2}}{(d-1)!!}\sum_{i=0}^{(d-1)/2}2^{(d-1)/2-i}\left|s\left(\frac{d+1}{2},i+1\right)\right|(\delta\partial_\delta)^iS(A)\,,\quad &\text{if odd }d\,, \\
    \suniv(A)&=\frac{(-1)^{d/2}}{(d-2)!!}\sum_{i=1}^{d/2}2^{d/2-i}\left|s\left(\frac{d}{2},i\right)\right|(\delta\partial_\delta)^iS(A)\,,\quad & \text{if even }d\,,
\end{alignat}
where $s\left(x,m\right)$ are the Stirling numbers of the first kind.}
\begin{equation}\label{eq:MI5_rewritten}
F(A)=\frac{1}{2}I_{\varepsilon}(A)
-k\,\frac{\area(\partial A)}{2\varepsilon^3}
+ \frac{1}{2\varepsilon} \int_{\partial A} \Big[ \alpha\,\mathcal{R} + \beta \, (K^2 - 3 K_{ab}K^{ab}) \Big]+\mathcal{O}(\varepsilon)\,,
\end{equation}
where $\alpha$ and $\beta$ are some new coefficients. We have chosen the relative coefficient between $K^2$ and $K_{ab}K^{ab}$ so that the second term vanishes  for a spherical entangling surface $\partial A = \mathbb{S}_R^3$, isolating the contribution proportional to $\alpha$. Similarly, $\beta$ can be isolated by a cylinder region, $\partial A=\mathbb{S}_R^1\times\mathbb{R}_L^2$. More precisely, we have
\begin{align}\label{IS3}
   I_{\varepsilon}(\partial A=\mathbb{S}^3)&=k\cdot \left[\frac{2\pi^2R^3}{\varepsilon^3}\right]-  12 \pi^2 \alpha\cdot \left[\frac{R}{\varepsilon}\right]\, +2F_0\, , \\ \label{IS1R2}
   I_{\varepsilon}(\partial A=\mathbb{S}^1\times \mathbb{R}^2)&=k\cdot  \left[\frac{2\pi R L^2}{\varepsilon^3}\right]+4 \pi\beta \cdot  \left[\frac{ L^2}{R \varepsilon } \right]+2F(\partial A=\mathbb{S}^1\times \mathbb{R}^2)\, ,
\end{align}
where we use the notation $F_0\equiv F(\partial A=\mathbb{S}^3)$.
Hence, in order to compute $F(A)$ for a given theory using the above prescription, we may proceed as follows. First, consider a spherical entangling surface and construct the auxiliary spheres with interiors $A^-$ and $\overline{A^+}$. Evaluate $I_{\varepsilon}(A)$ as a function of $\varepsilon \ll R$ and fit the result to an expansion of the form~\eqref{eq:MI5}. Extract $k$ from the leading term and, using~\eqref{IS3}, identify $\alpha$. Next, consider a cylinder region with entangling surface $\partial A=\mathbb{S}^1\times\mathbb{R}^2$ and proceed analogously. The leading term must fit the one appearing in~\eqref{IS1R2} with the previously determined coefficient $k$. On the other hand, use the subleading term to determine $\beta$. Once $k$, $\alpha$ and $\beta$ have been characterized in this way---once and for all---for a given theory, we may use~\eqref{eq:MI5_rewritten} to evaluate $F(A)$ for any entangling region. In order to do so, we would proceed with the evaluation of $I_{\varepsilon}(A)$ in that case, and then we would subtract the pieces proportional to $\varepsilon^{-3}$ and $\varepsilon^{-1}$ respectively,  as in~\eqref{eq:MI5_rewritten}. The procedure must be consistent, in the sense that the resulting $F(A)$ should converge to a constant value as we make $\varepsilon$ decrease with respect to any characteristic scales of $A$. 

\subsection{Explicit example: the EMI model}\label{sec:exEMI} 

Remarkably, when one considers the set of axioms satisfied by MI in QFT and imposes the additional constraint that the tripartite information vanishes for all possible regions---which is equivalent to enforcing that the MI is an additive (or ``extensive'') function of its arguments, it is possible to find a simple explicit formula for the MI. This defines the so-called ``Extensive Mutual Information'' (EMI) model~\cite{Casini:2008wt}.  In particular, under the additional assumption of conformal symmetry, the MI for two regions $A_1,A_2$ in a time slice of $d$-dimensional Minkowski space is given in the EMI model by
\begin{equation}\label{eq:EMId_Info}
    \qt{I(A_1,A_2)}{EMI}=2\kappa\int_{\partial A_1}\diff^{d-2}\mathbf r_1\int_{\partial A_2}\diff^{d-2}\mathbf r_2\frac{\mathbf n_1\cdot\mathbf n_2}{\left|\mathbf r_1-\mathbf r_2\right|^{2(d-2)}}\,,
\end{equation}
where $\kappa$ is a dimensionless quantity which characterizes the model, and $\mathbf n_{1,2}$ are normal vectors to $\partial A_1$ and $\partial A_2$, respectively.

The EMI model has been often considered in the past as a computationally explicit toy model which produces physically reasonable results---\eg\, to probe geometric properties of MI and EE~\cite{Casini:2015woa,Bueno:2015rda,Witczak-Krempa:2016jhc,Bueno:2019mex,Berthiere:2019lks,Estienne:2021hbx}. In the cases in which the comparison is possible, the results always turn out to be very similar to the ones of a $d$-dimensional free fermion. In fact, the EMI model coincides with a free fermion in $d=2$~\cite{casini2005entanglement}. On the other hand, as argued in~\cite{Agon:2021zvp}, the EMI model does not correspond to any actual CFT in $d\geq 3$, which raises doubts on the completeness of the set of axioms known to be satisfied by MI in QFT.\footnote{It was shown in~\cite{Agon:2021zvp} that for two arbitrarily boosted spheres in general dimensions, the EMI result exactly matches the contribution to the MI produced by the free fermion current conformal block (whereas all the rest of contributions present in the free fermion theory are absent in the EMI model).} In the present context, the EMI model will allow us to illustrate the procedure outlined in the previous subsection.
In Appendix~\ref{sect:EMIappendix} we present numerous new computations of $F(A)$ in the EMI model for different regions.
Here we simply mention some of the relevant results as we need them. 

As explained before, the first step entails evaluating the coefficient $k$. We can do this \eg\, by considering the MI of two straight parallel blocks separated a small distance $\varepsilon$, or directly by considering two concentric spherical entangling regions, which also allows us to determine $\alpha$ at once. The former calculation yields
\begin{equation}\label{stripi}
    \qt{I_{\varepsilon}(\text{strip}) }{EMI}=  \frac{  \pi^2 \kappa}{2} 
   \bigg[\frac{\area_{\parallel}}{\varepsilon^3}   -\frac{\area_{\parallel}}{\ell_\perp^3} \bigg] \, ,
\end{equation}
where $\area_{\parallel}=\ell_\parallel^{3}$ is the (hyper)area of the parallel sides separated by a distance $\ell_\perp$.
On the other hand, the result for the spherical region is given by
\begin{align}\label{eq:EMI5Sph}
    \qt{I_{\varepsilon}({\partial A=\mathbb{S}^3})}{EMI} &= \frac{\pi^2\kappa}{2}\left[  \frac{2\pi^2 R^3}{\varepsilon^3} \right]
      - \frac{9 \pi^4  \kappa}{4} \left[\frac{R}{\varepsilon}\right]
      + 2\pi^4 \kappa\, .
\end{align}
Comparing with the general expression~\eqref{IS3}, we find
\begin{equation}\label{emik}
    \qt{k}{EMI}=\frac{\pi^2 }{2}\kappa\, ,\quad \qt{\alpha}{EMI}=\frac{3\pi^2}{16}\kappa\, , \quad \qt{F_0}{EMI}=\pi^4\kappa\, , \quad \qt{F(\text{strip})}{EMI}=-\frac{\pi^2 \area_{\parallel}}{4}\kappa\, .
\end{equation}
It is easy to check that the same $ \qt{k}{EMI}$ coefficient appears multiplying $[\area(\partial A)/\varepsilon^3]$ in the leading term for general regions.  In order to determine $\qt{\beta}{EMI}$, we may consider a cylindrical region with entangling surface $\partial A=\mathbb{S}_R^1\times\mathbb{R}_L^2$ (where $R$ is the disk radius and $L$ is an IR regulator for the cylinder length). The result reads
\begin{equation}\label{s1r2EMI}
    \qt{I_{\varepsilon}(\partial A=\mathbb{S}^1\times \mathbb{R}^2)}{EMI} =
   \qt{k}{EMI } \cdot \bigg[
    \frac{2\pi R L^2}{ \varepsilon ^3}\bigg] -\frac{\pi ^3  \kappa}{2  }\left[\frac{ L^2 }{R \varepsilon }\right]  \, ,
\end{equation}
which we may compare to the general formula~\eqref{IS1R2}. By doing this, we find
\begin{equation}
    \qt{\beta}{EMI}=  -\frac{\pi^2}{8}\kappa\, , \qquad \qt{F(\partial A=\mathbb{S}^1\times\mathbb{R}^2)}{EMI}=0\, .
\end{equation}
Once we have identified $\qt{k}{EMI},\qt{\alpha}{EMI},\qt{\beta}{EMI}$, we may define  $\qt{F(A)}{EMI}$ for general regions in the EMI model using~\eqref{eq:MI5_rewritten}. For example, consider a different cylindrical entangling region, $\partial A=\mathbb{S}_R^2\times \mathbb{R}_L^1$. Using the general MI formula for the EMI model~\eqref{eq:EMId_Info}, we find
\begin{equation}
\label{eq:MI_EMI_CylS2R}
    \qt{I_{\varepsilon}(\partial A=\mathbb{S}^2\times \mathbb{R}^1)}{EMI} =
    \kappa \bigg[ \frac{2 L \pi^3 R^2}{\varepsilon^3} 
- \frac{5 L \pi^3}{2 \varepsilon} 
+ \frac{5 L \pi^3 }{4 R}  \bigg]\, .
\end{equation}
Plugging this into~\eqref{eq:MI5_rewritten} and evaluating the relevant integrals over $\partial A$, we observe that all terms involving $\varepsilon$ cancel out when the previously identified values of $\qt{k}{EMI},\qt{\alpha}{EMI},\qt{\beta}{EMI}$ are inserted in the formula, and we are left with the result
\begin{equation}\label{s2r1}
    \qt{F(\partial A=\mathbb{S}^2\times\mathbb{R}^1)}{EMI}=\frac{5 L \pi^3 }{8 R} \kappa\, .
\end{equation}
The procedure can be extended in an obvious way to arbitrary entangling regions.  Similarly, as long as we are able to evaluate the corresponding MIs, the procedure allows one to evaluate $F(A)$ for general regions in any five-dimensional CFT. While this is extremely challenging in general, one may try to make some general comments about $F(A)$ without relying on explicit theory-dependent calculations.

\subsection{On the sign of $F(A)$}
From the results obtained in the previous subsection it is clear that $\qt{F(A)}{EMI}$ may be positive, negative or zero, depending on the entangling region. In fact, it may take both arbitrarily positive  and arbitrarily negative values, as is clear \eg\, from~\eqref{s2r1} and~\eqref{emik}. 

The fact that $F(A)$ may take both positive and negative values in $d=5$ CFTs is in sharp contrast with the $d=3$ case, where it is always positive, since~\cite{Bueno:2021fxb}
\begin{equation}\label{genva}
    F(A)\geq F_0\,,\quad \forall A\, , \forall \text{CFT}_3\,,
\end{equation}
and~\cite{Dowker:2010bu,Casini:2011kv}
\begin{equation}
    F_0> 0\,, \quad \forall \text{CFT}_3\, .
\end{equation}

Now, what is known about the sign of $F(A)$ in $d=5$ on general grounds? On the one hand 
\begin{equation}
F_0> 0\,, \quad \forall \text{CFT}_5\, ,
\end{equation} 
namely, it is always positive for spherical entangling surfaces, which follows from its relation to the Euclidean free energy on the sphere~\cite{Dowker:2010bu,Casini:2011kv}. Similarly, small geometric deformations of the spherical region always increase $F_0$, namely, 
\begin{equation}
 F_0\,\,     \text{is a local minimum of} \ F(A)\,,\ \forall\text{CFT}_5\, .
\end{equation}
Indeed, consider a perturbed spherical entangling surface $\partial A=\mathbb S_\epsilon^3$, whose radius is given, to linear order in the perturbation parameter $\epsilon$, by
\begin{equation}\label{eq:pertS30}
r(\mathbf{\Omega})=R\left[1+\epsilon\Phi(\mathbf{\Omega})\right]\,,\qquad 
\Phi(\mathbf{\Omega})\equiv\sum_{\ell,\mathbf m}a_{\ell,\mathbf m}Y_{\ell,\mathbf m}(\mathbf{\Omega})\,.
\end{equation}
where $Y_{\ell,\mathbf m}=Y_{\ell,\mathbf m}(\mathbf\Omega)$ denotes the scalar spherical harmonic on $\mathbb S^{3}$ with indices $\ell$ and $\mathbf{m}={m_1,m_2}$, satisfying $\Delta Y_{\ell,\mathbf m}=-\ell(\ell+2)Y_{\ell,\mathbf m}$, where $\Delta$ is the Laplace–Beltrami operator on $\mathbb S^{3}$. The coefficients $a_{\ell,\mathbf m}$ parameterize the deformation. For such an entangling surface, one finds~\cite{Mezei:2014zla,Faulkner:2015csl}
\begin{equation}\label{mezei}
    F_\epsilon=F_0+ \epsilon^2 C_T \cdot  \left[\frac{\pi^4}{2160} \sum_{\ell,\mathbf m} a_{\ell,\mathbf m }^2 
\prod_{k =1, \dots, 5} ( \ell +k -2)\right]\, , 
\end{equation}
where $C_T$ is the stress-tensor two-point function coefficient of the corresponding theory, which is positive definite, 
and the numerical coefficient inside the brackets is a positive-definite theory-independent function of the deformation parameters and $\{\ell,{\bf m}\}$.

On the other hand, strip-like regions consistently give rise to arbitrarily negative values of $F(A)$  for holographic theories~\cite{Ryu:2006ef}, free fields~\cite{Casini:2009sr} as well as for the EMI model studied above---see~\eqref{emik}, so it seems safe to claim that\footnote{\label{futi}We emphasize that we are choosing a convention such that the strip coefficient $k$ is positive in general dimensions and theories, in agreement with the previous literature. On the other hand, with the definition of $F(A)$ taken in~\eqref{eq: EE expression}---which is convenient because it makes $F(\partial A=\mathbb{S}^{d-2})\equiv F_0>0$ in general $d$---$F(\text{strip})$ is positive for $d=3,7,11,\dots$ and negative for $d=5,9,13,\dots$ Later in the paper we shall make use of the definition
\begin{equation}\label{fstripi}
    F_\text{strip}\equiv F(\text{strip}) \frac{\ell_\perp^{d-2}}{\area_{\parallel}}=\begin{cases} +k \quad & \text{for }d=3,7,11,\dots \\
    -k\quad & \text{for }d=5,9,13,\dots\end{cases}
\end{equation}}
\begin{equation}
    F(\text{strip})=-k\cdot \frac{\area_{\parallel}}{\ell_\perp^3}<0\,, \quad \forall \text{CFT}_5\, .
\end{equation}
Since $F({\text{strip}})$ is proportional to $[\area_{\parallel}/\ell_\perp^3]$, the above quantity tends to $-\infty$ as the strip is made increasingly thinner. Similarly, we expect that certain families of bounded regions with one dimension much thinner than the others will give rise to arbitrarily negative $F(A)$'s for general theories. For comparison, $F(\text{strip})=+ k\cdot \text{length}_{\parallel}/\ell $ becomes increasingly more positive as the strip becomes thinner in three-dimensional CFTs.

Recently, in~\cite{Anastasiou:2024rxe},  it has been argued that for $d=5$ holographic CFTs dual to Einstein gravity it is possible to achieve both arbitrarily negative as well as arbitrarily positive values of $F(A)$, namely, that no general (upper or lower) bounds on $F(A)$ analogous to~\eqref{genva} exists. In that case, however, the procedure required identifying the extremal surfaces  homologous to the three-dimensional entangling surfaces at the AdS$_6$ boundary~\cite{Ryu:2006bv} and then evaluating with them a higher-dimensional Willmore energy functional. Since the first step was computationally difficult for the relevant families of entangling regions, the authors of~\cite{Anastasiou:2024rxe} fed the functional with simpler families of non-extremal trial regions to test the boundedness of $F(A)$, and they found that neither bound exists. While it remains a logical possibility that restricting the analysis to extremal regions would restore the boundedness of  $F(A)$, it seems extremely unlikely. On the other hand,  the result found here for the EMI model in the case of the $\partial A=\mathbb{S}^2\times \mathbb{R}^1$ cylinder in~\eqref{s2r1} shows that also for this model it is possible to make $F(A)$ as large (and positive) as desired. An analogous behavior is expected in other theories.

\begin{figure}
    \centering
\includegraphics[width=0.7\textwidth]{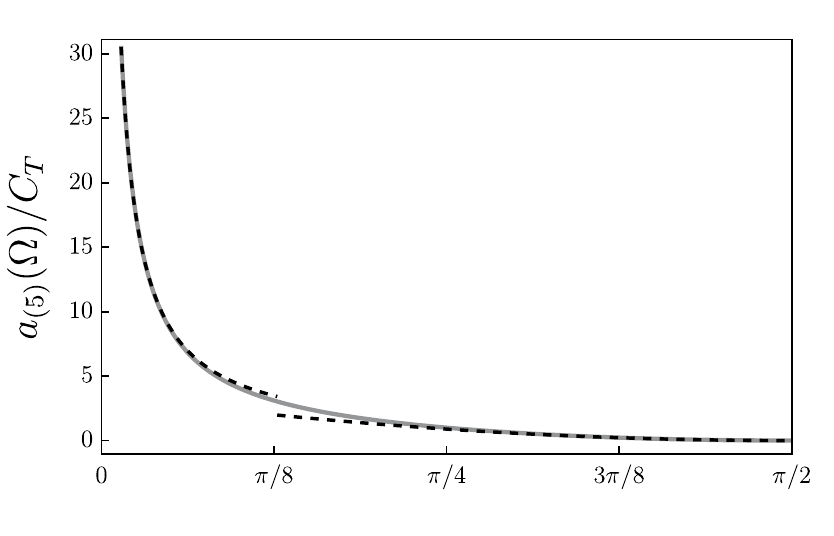}
    \caption{We plot the coefficient of the universal contribution to the EE generated by a conical entangling region as a function of the opening angle $\Omega$ for a $d=5$ holographic CFT dual to Einstein gravity. The result is normalized by the stress-tensor two-point function coefficient, $C_T$, which universally governs the leading behavior in the $\Omega \rightarrow \pi/2$ regime. The dashed lines correspond, respectively: to the exact result in that limit, $ a_{(5)}(\Omega)=  4\pi^4C_T(\Omega-\pi/2)^2/270+\dots$
\cite{Bueno:2015lza}; and to a fit of the form $a_{(5)}(\Omega)/C_T \approx 1.36/\Omega$. As is evident from the plot, $a_{(5)}(\Omega)$ remains positive for all values of the opening angle and approaches the exact result predicted in~\cite{Bueno:2015lza} as $\Omega \rightarrow \pi/2$. This is in fact the first explicit verification of the general results presented in that paper beyond $d=3$ theories. More details can be found in Appendix~\ref{holoCorner}.
}\label{fig:hee2}
\end{figure}

Another interesting setup which allows us to probe the sign of $F(A)$ entails considering regions with geometric singularities. The paradigmatic example takes place for $d=3$ CFTs and involves a sharp corner resulting from the intersection of two straight lines with an opening angle $\Omega$. In that case, the EE receives an additional logarithmic contribution, which can be thought of as arising from $F(A)$ as the geometry of $A$ is deformed to include the corner---see \eg~\cite{Bueno:2015lza}. In that case, we have
\begin{equation}
    F(A) = a_{(3)}(\Omega) \log\frac{L}{\varepsilon}+\dots
\end{equation}
where $ a_{(3)}(\Omega)$ is a positive-definite decreasing function of the opening angle, which behaves as~\cite{Fradkin:2006mb,Casini:2006hu,Hirata:2006jx,Casini:2008as,Bueno:2015rda}
\begin{equation}
    a_{(3)}(\Omega)=\begin{cases}k/ \Omega+\dots \quad &(\Omega\rightarrow 0)\, , \\ \frac{\pi^2}{24}C_T\cdot\left(\Omega-\pi \right)^2+\dots \quad &(\Omega \rightarrow \pi)\, , \end{cases}
\end{equation}
where $k$ is precisely the strip coefficient, and $C_T$ is the stress-tensor two-point function coefficient of the theory. This function is particularly interesting in the present context because it continuously connects, as $\Omega$ is varied, the strip result---which follows from the existence of a conformal mapping which relates the corner geometry to a strip on a cylindrical background $\mathbb{S}^2\times \mathbb{R}$~\cite{Myers:2012vs,Bueno:2015xda}---to the result corresponding to small geometric deformations of the disk region in~\eqref{mezei}---which in turn follows from adapting  Mezei's formula~\cite{Mezei:2014zla} to the case of singular deformations and using a conformal mapping which relates the sphere to the half plane~\cite{Faulkner:2015csl}. Now, what happens in the analogous situation of a conical entangling surface with spherical cross-section $\mathbb{S}^2$ in $d=5$? The entangling surface can be defined in cylindrical coordinates as $\{t_\text{E}=0,\rho=[0,+\infty),\theta= \Omega \}$, where the ambient metric reads\footnote{Here, $\diff \Omega_{2}^2$ is the metric of the round sphere, not to be confused with the cone opening angle $\Omega$.}
\begin{equation}
    \diff s_{\mathbb{R}^5}^2=\diff t_\text{E}^2+\diff\rho^2+\rho^2\left[\diff \theta^2+\sin^2\theta\diff \Omega_{2}^2 \right]  \, .
\end{equation}
In that case we have, analogously to the three-dimensional case,\footnote{Note that our conventions for $a_{(5)}(\Omega)$ differ by a sign from those used in~\cite{Myers:2012vs}. Hence, $a_{(5)}(\Omega)\propto-q_5(\Omega)$ as defined in that paper.}
\begin{equation}
    F(A) = a_{(5)}(\Omega) \log\frac{L}{\varepsilon}+\dots
\end{equation}
for general CFT$_5$'s. As shown in~\cite{Bueno:2015lza}, the regime of very open angles is again universally controlled by $C_T$. On the other hand, the holographic results of~\cite{Myers:2012vs} suggest that the very sharp cone regime is also controlled by a leading $\sim 1/\Omega$ term. Thus, we have
\begin{equation}\label{a5C}
    a_{(5)}(\Omega)=\begin{cases} \nu/ \Omega+\dots \quad &(\Omega\rightarrow 0)\, , \\ \frac{4\pi^4}{270}C_T\cdot\left(\Omega-\pi/2 \right)^2+\dots \quad &(\Omega \rightarrow \pi/2)\, . \end{cases}
\end{equation}
It is tempting to speculate that $\nu$ might also be connected to the strip coefficient $k$ in $d=5$. However, this is not the case. Indeed, as argued in~\cite{Myers:2012vs}, there exists a conformal mapping, analogous to the three-dimensional one, which transforms the cone into a cylinder $\mathbb{S}^2\times\mathbb{R}$ on a $\mathbb{S}^4\times \mathbb{R}$ background. In particular, the entangling region becomes a cylinder defined by $\{Y\in(-\infty,+\infty),\xi=\pi/2,\theta= \Omega \}$, embedded in an ambient space with metric
\begin{equation}
    \diff s_{\mathbb{S}^4\times\mathbb{R}}^2=\diff Y^2+L^2\left[\diff \xi^2+\sin^2\xi \left(\diff \theta^2+\sin^2\theta\,\diff \Omega_{2}^2 \right)\right]  \, .
\end{equation}
The induced metric on the cylinder reads then
\begin{equation}
    \diff s_{\mathbb{S}^2\times \mathbb{R}}^2=\diff Y^2+L^2\sin^2 \Omega\, \diff \Omega_{2}^2\, .
\end{equation}
It is then clear that as $\Omega\rightarrow 0$, the radius of the cylinder is much smaller than the radius of the ambient $\mathbb{S}^4\times \mathbb{R} $, and therefore the coefficient $\nu$ appearing in~\eqref{a5C} must be related to the one obtained for the same type of cylinder embedded in $\mathbb{R}^5$. As we saw in the previous section in~\eqref{s2r1}, this is actually positive, in contradistinction to the strip coefficient, which is negative. More generally, we expect $a_{(5)}(\Omega)$ to be a positive-definite and monotonically decreasing function of the opening angle. This is the case for holographic theories dual to Einstein gravity, as shown in Figure~\ref{fig:hee2}. Hence, conical regions provide us with instances of entangling regions for which $F(A)$ diverges to $+\infty$.

It is also clear that it should be possible to find families of entangling regions for which $F(A)$ flips sign as we continuously vary certain parameters. A simple example is given by regions formed as the union of pairs of identical, disjoint regions. Let $A$ and $A_{\ell}$ be each of the copies, the second of which is separated some distance $\ell$ from the first. By definition, we know that 
\begin{equation}
    I(A,A_{\ell})=2S(A)-S(A\cup A_{\ell})\, .
\end{equation}
Now, using our prescription for the EE universal piece, we can rewrite the above expression in $d=5$ as
\begin{equation}\label{IAAl}
    I(A,A_{\ell})=2F(A)-F(A\cup A_{\ell}) \quad \Rightarrow \quad  \frac{F(A\cup A_{\ell})}{F(A)}=2 - \frac{I(A,A_{\ell})}{F(A)}\, ,
\end{equation}
where all divergent terms cancel one another upon replacing $S(A)\rightarrow\frac{1}{2}I_\varepsilon(A^+,A^-)$. Now, since the MI is positive-semidefinite by definition, it follows that 
\begin{equation}\label{funio}
    F(A\cup A_{\ell})\leq 2 F(A)\, , \quad \forall A\, .
\end{equation}
Namely, for general CFT$_5$'s the universal term for the union of two disconnected identical regions separated an arbitrary distance $\ell$ is always smaller than twice the result corresponding to the original region. This is an immediate consequence of the strong subadditivity property of EE. Additionally, the MI decreases as the separation between the regions gets increased and it grows arbitrarily as the closest boundaries of $A$ and $A_{\ell}$ approach each other. Hence, we have
\begin{equation}\label{FAAL}
    \frac{ F(A\cup A_{\ell})}{F(A)}\rightarrow \begin{cases}+2 \quad &\text{for }\quad \ell/R_A \rightarrow \infty\, , \\
    -\infty  \quad &\text{for }\quad \ell/R_A \rightarrow 0\, , \end{cases}
\end{equation}
where $R_A$ is the largest characteristic scale of $A$. In particular, when the closest boundaries of $A$ and $A_{\ell}$ are parallel to each other, $\ell/R_A \rightarrow 0$, the dominant contribution is controlled by the strip coefficient 
\begin{equation}
    F(A\cup A_{\ell}) \sim -k \frac{\area_{\parallel}}{\ell^3}+\dots
\end{equation}

\begin{figure}
    \centering 
\includegraphics[width=0.7\textwidth]{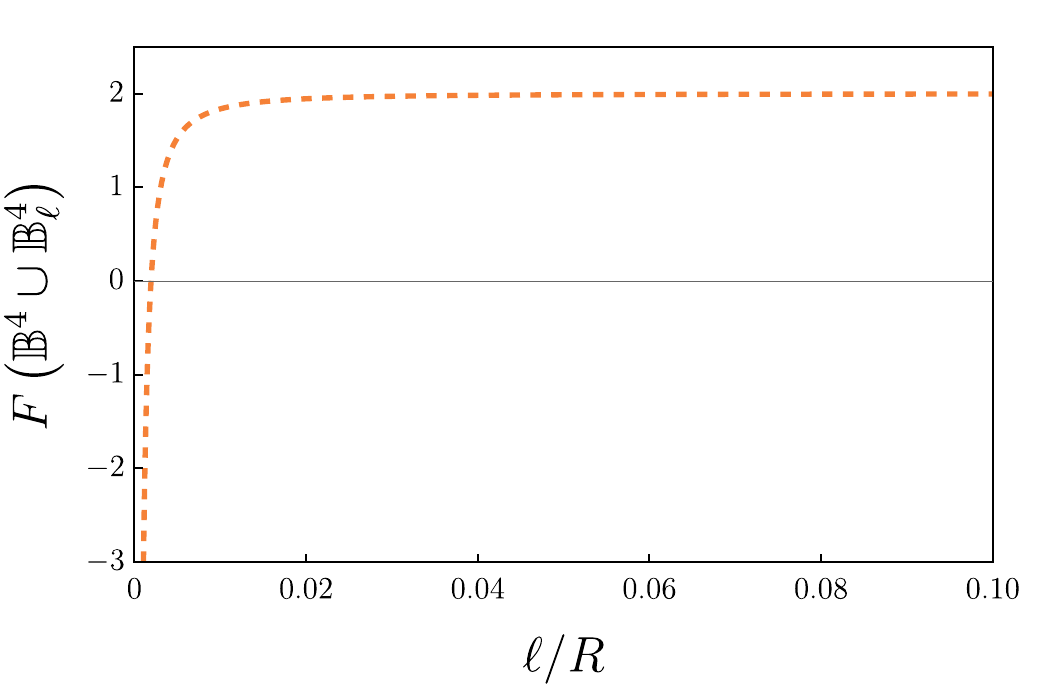}
    \caption{We plot the universal term $F(A_1 \cup A_2)$  in the EMI model for the union of two identical round balls, $A_1 = A_2 = \mathbb{B}^4$, each of radius $R$ and separated by a distance $\ell$, as a function of the relative separation $\ell/R$. The result is normalized by the single-ball value. At large separations, $F(A)/F_0$ is positive and approaches $+2$. It remains positive for most separations as the spheres move closer together, but it eventually vanishes for $\ell/R \simeq 0.00194$. As the separation decreases further and the spheres approach a point of contact, $F(A)/F_0$ diverges to $-\infty$.
  }\label{fig:heeF}
\end{figure}

If $F(A)<0$,   $F(A\cup A_{\ell})$ is negative for all separations, as is obvious from~\eqref{funio}. On the other hand, whenever $F(A)>0$, like in the spherical region case, $F(A\cup A_{\ell})$ is also positive when the two regions are sufficiently far apart from each other, it vanishes at some intermediate separation, beyond which it becomes negative, and it takes arbitrarily negative values as the two boundaries approach each other. 
In the case of the EMI model, the general formula for $I(A,A_{\ell})$ in the case of (arbitrarily boosted) spherical entangling surfaces is known in general dimensions and for arbitrary separations~\cite{Agon:2021zvp}, so we can easily compute $F(A\cup A_{\ell})$ from~\eqref{IAAl}. For coplanar spheres, the result reads
\begin{equation}
  \qt{\frac{F(\mathbb{B}^4\cup\mathbb{B}^4_{\ell})}{F_0}}{EMI}=
2-\frac{512\,( \ell^2 + 4\ell+2 )}
{315\,\pi^4\, \ell^{5/2}
(\ell+2)^5 (\ell+4)^{5/2}}\, \, {}_2F_1\!\left[1,\frac{5}{2},3,-\frac{4}{\ell(\ell+2)^2(\ell+4)}\right]\, ,
\end{equation}
where $\ell$ is the relative separation between the innermost boundaries of the spheres, which touch at a point for $\ell=0$. In Figure~\ref{fig:heeF}, this result is plotted as a function of the relative distance, clearly illustrating the general behavior discussed above.

In sum, $F(A)$ is unbounded both above and below for general CFT$_5$'s, with appropriate regions yielding arbitrarily large positive or negative values. On the other hand, as a consequence of Mezei's formula~\cite{Mezei:2014zla,Faulkner:2015csl}, regions with round spherical boundaries do provide a local minimum under general geometric perturbations of $A$ and for general CFT$_5$'s. 
\section{Conformal bounds from entanglement entropy}\label{sec:bounds}

\subsection{Three dimensions}\label{sec:bound3}
In~\cite{Bueno:2023gey}, it was conjectured that, for all CFTs in three dimensions, the universal part of EE, $F(A)$, normalized by the result for the disk $F_0$---that is, $F(A)/F_0$---is bounded from below by the (free) Maxwell theory and from above by the free scalar, for arbitrary regions $A$, this is
\begin{equation}\label{eq:bounds3d}
\qt{\frac{F(A)}{F_0}}{Maxwell} \leq \frac{F(A)}{F_0} \leq\qt{\frac{F(A)}{F_0}}{free scalar}\,,\quad \forall A\,,\forall \text{CFT}_3\, .
\end{equation}
In the same reference, the conjecture was supported by the following observations:
\begin{enumerate}
\item[i)] In $d=4$, an analogous pair of conjectural bounds for the universal part of EE---which includes the stress-tensor trace anomaly coefficients $a$ and $c$---turns out to be exactly equivalent to one of the so-called Hofman--Maldacena bounds~\cite{Hofman:2008ar,Hofman:2016awc}---see~\eqref{eq:HM bounds} in Section~\ref{sec:bounds4d} below.
\item[ii)] Quotienting by symmetry groups tends to reduce $F(A)/F_0$ in $d=3$. This follows from the relation between $F(A)$ and the MI which appears in its definition---the $d=3$ version of~\eqref{eq:MI5_rewritten}---and the fact that  the MI acquires a contribution proportional to the number of connected boundaries $n$ times the logarithm of the dimension of the symmetry group~\cite{Casini:2019kex}. This implies that \emph{complete} theories~\cite{Casini:2021zgr}---such as the free scalar---always have a greater $F(A)/F_0$ than any of their orbifolds in three dimensions. 
\item[iii)] The lower bound, saturated by Maxwell theory, can actually be proved: it follows from the fact that the three-dimensional Maxwell theory is the orbifold of the free scalar under the infinite group $\mathbb{R}$ implementing $\phi \to \phi+\lambda$~\cite{El-Showk:2011xbs}, so the logarithmic term dominates~\cite{Casini:2014aia} and the quotient reaches its minimal value $F(A)/F_0=n$.
\item[iv)] The upper bound from the scalar field can also be proved in the case of regions made from the union of subregions separated by long distances. In this regime, MI decays with an exponent fixed by the smallest scaling dimension $\Delta$ which, by virtue of general unitarity bounds, is minimized by the free scalar---see, \eg~\cite{Minwalla:1997ka,Bueno:2023gey}. As a result, the free scalar maximizes the MI and hence the quotient $F(A)/F_0$.
\item[v)] In the case of entangling regions with disconnected components such that one of them involves a thin deformation of its boundary on a null cone, the so-called \emph{pinching property} implies that $F(A)/F_0$ is smaller for any interacting CFT$_3$ than for any free theory, independently of the separation between the components~\cite{Schlieder:1972qr,Wall:2011hj,Benedetti:2022aiw,Bousso:2014uxa,Casini:2021raa}. In order to discriminate between free theories, one can resort to numerical calculations, which again show that the free scalar provides the absolute maximum~\cite{Agon:2022efa}.
\item[vi)] For small perturbations around the disk entangling region,~\eqref{eq:bounds3d} implies a derived conjecture involving the two-point stress energy-tensor charge, namely~\cite{Bueno:2023gey}
\begin{equation}\label{eq:bounds3CT}
0 \leq \frac{C_T}{F_0} \leq\left.\frac{C_T}{F_0}\right|_{\text {free scalar}}=\frac{3}{4 \pi^2 \log 2-6 \zeta(3)} \approx 0.149\,, \quad \forall\text{CFT}_3\,.
\end{equation}
Both quantities, $C_T$ and $F_0$, have been computed for many families of three-dimensional CFTs, and one can verify that the conjectural bounds~\eqref{eq:bounds3CT} hold for all them---see Appendix~\ref{sec:appuq}. This includes: free fields, holographic theories, the EMI model, $O(N)$ models, Gross--Neveu (GN) models, $\mathcal N=2$ Wess--Zumino (WZ) models, supersymmetric quantum electrodynamics (SQED) and Aharony--Bergman--Jafferis--Maldacena (ABJM) theories. As explained in the same appendix, an additional class of theories, corresponding to the so-called Atiyah--Drinfeld--Hitchin--Manin (ADHM) models~\cite{Atiyah:1978ri},\footnote{We thank Nikolay Bobev for suggesting this family of theories to us.} are considered here for the first time here, and they also fulfill the bounds. Interestingly, for CFTs that can be interdimensionally related between $d=3$ and $d=4$---such as free scalars and fermions, Maxwell fields, the EMI model and holographic theories, the quotients $C_T/F_0$ and $c/a$ follow a very similar hierarchy of values, as shown in Figure~\ref{fig:CTF0}. 

\item[vii)]  For elliptic and corner regions the conjecture is also satisfied for general values of the relevant geometric parameters---the eccentricity and the opening angle, respectively---for all CFT$_3$'s for which the relevant results are known. In both cases, the results interpolate between two regimes controlled, respectively by $C_T$---the almost round limit of the ellipse and the almost-smooth limit of the corner---and the strip coefficient $k$---the very thin limit of the ellipse and the very-sharp corner limit. The latter imply the general hierarchy~\cite{Bueno:2023gey}
\begin{equation}\label{eq:boundsk3}
    0\leq\frac{F_\text{strip}}{F_0}\leq\qt{\frac{F_\text{strip}}{F_0}}{free scalar}\approx0.622\,, \quad \forall\text{CFT}_3\,,
\end{equation}
where $F_\text{strip}$ was defined in~\eqref{fstripi}. Similarly to~\eqref{eq:bounds3CT}, this conjecture is satisfied for all theories for which $k$ has been computed.
\end{enumerate}

\subsection{Four dimensions}\label{sec:bounds4d}
As mentioned in the previous subsection, a conjecture analogous to~\eqref{eq:bounds3d} in the $d=4$ case is identical to the Hofman--Maldacena bounds~\cite{Hofman:2008ar,Hofman:2016awc} involving the trace-anomaly coefficients $a$, $c$---see~\eqref{eq:traceanom4} for their definition. Indeed, in that case the EE contains a logarithmic universal term of the form~\cite{Solodukhin:2008dh,Perlmutter:2015vma}
\begin{equation}\label{log4d}
 S_\text{reg}(A,\varepsilon)\supset -  \suniv(A)\log\frac{H}{\varepsilon}\, , \quad \text{where} \quad \frac{\suniv(A)}{a}=\frac{1}{\pi}\left[ \mathcal{W}_{\partial A}+\left(\frac{c}{a}-1 \right)\frac{\mathcal{K}_{\partial A}}{2} \right]\, ,
\end{equation}
where $\mathcal{W}_{\partial A}$ is a positive-semidefinite functional---sometimes referred to as the ``Willmore energy''~\cite{willmore1965note,willmore1996riemannian}---of the entangling surface $\partial A$, and $\mathcal{K}_{\partial A}$ is a positive-semidefinite integral involving a quadratic linear combination of extrinsic curvatures of $\partial A$. Observe that both $\mathcal{W}_{\partial A}$ and $\mathcal{K}_{\partial A}$ are theory-independent quantities, so all information about the theory under consideration is encoded in $a$ and $c$. These quantities are isolated, respectively, by spherical and cylindrical regions, and they are the $d=4$ versions of $F_0$ and $C_T$, respectively.\footnote{In fact, $a$ and $c$ are also related to the coefficients appearing in the stress-tensor three-point function in $d=4$. Extrapolating to three dimensions, where there are no trace anomalies, the ``analogous version'' of $c$ is clearly $C_T$, which exists in any dimension and is proportional to $c$ whenever the trace-anomaly is present. On the other hand, one could say that the ``analogous version'' of $a$---or some combination of $c$ and $a$---is either $F_0$ or the stress-tensor three-point function charge, $t_4$~\cite{Osborn:1989td,Hofman:2008ar}. While $F_0$ can be easily accessed via EE, there is no known way of capturing $t_4$ in that way---see~\cite{Bueno:2015ofa,Anastasiou:2022pzm}. On the other hand, energy positivity arguments can easily access $t_4$ but not $F_0$.  }
Now, from the  positivity properties of $\mathcal{W}_{\partial A}$ and $\mathcal{K}_{\partial A}$ it follows straightforwardly that a conjecture of the form
\begin{equation}\label{4HM}
    \qt{\frac{\suniv(A)}{a}}{Maxwell}\leq\frac{\suniv(A)}{a}\leq\qt{\frac{\suniv(A)}{a}}{free scalar}\,,\quad\forall A\,,\ \forall \text{CFT}_4\,,
\end{equation}
is exactly equivalent to 
\begin{equation}\label{eq:HM bounds}
\frac{18}{31} = \qt{\frac{c}{a}}{Maxwell} \leq \frac{c}{a} \leq\qt{\frac{c}{a}}{free scalar} = 3\, ,
\end{equation}
which are the aforementioned Hofman--Maldacena bounds.

The above observations were made in~\cite{Bueno:2023gey}. Here we would like to make an additional one. As a matter of fact, it is possible for  $\mathcal{W}_{\partial A}$ and $\mathcal{K}_{\partial A}$ to vanish simultaneously. This is precisely what happens for a strip region. In that case, the universal contribution is constant and it takes the form
\begin{equation}
   S_\text{reg}(A,\varepsilon)\supset-k\cdot \frac{\area_{\parallel}}{\ell^2}\,, \quad \forall \text{CFT}_4\, ,
\end{equation}
analogously to the $d=3$ and $d=5$ cases studied in the previous section. Then, the conjecture would read in this case
\begin{equation}\label{fsta}
    \qt{\frac{F_\text{strip}}{a}}{Maxwell}\leq\frac{F_\text{strip}}{a}\leq\qt{\frac{F_\text{strip}}{a}}{free scalar}\,,\quad\forall \text{CFT}_4\,,
\end{equation}
where we introduced the notation $F_\text{strip}\equiv k$ to facilitate the comparison with other dimensions---see footnote~\ref{futi}.
The above relations are clearly independent from~\eqref{eq:HM bounds}, since $k$ is not related in any simple way with $a$ and/or $c$ for general theories. In fact, $k$ has not been computed for many models, but at least there exist results for free scalars and fermions, the Maxwell field, the EMI model and holographic theories dual to Einstein gravity. As shown in Figure~\ref{fig:kF0}, the conjectural bounds in~\eqref{fsta} are satisfied in all known theories, and the hierarchy of theories is identical to the one encountered in the three-dimensional case.

\subsection{Five dimensions}\label{sec:bound5}

\subsubsection{(No) bounds for arbitrary regions}\label{sec:noboundA}

Let us now move on to the main case of interest in this paper, namely, five-dimensional CFTs.

\begin{figure}[t!]
    \centering
\includegraphics[width=0.9\textwidth]{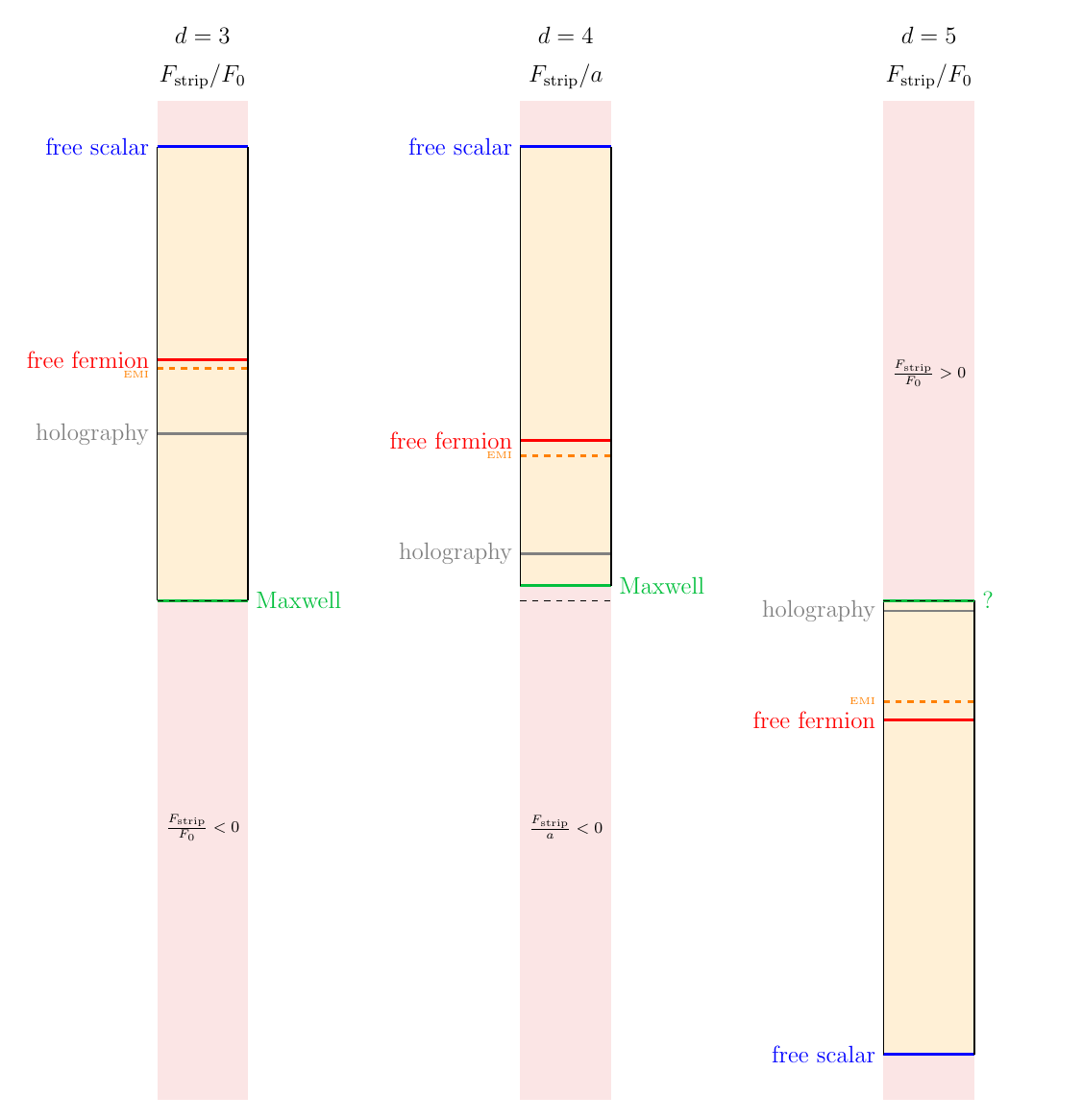}
    \caption{From left to right, we display the ratios $F_{\text{strip}}/F_0$ for a variety of CFTs in three, four, and five spacetime dimensions. In all cases, the values are normalized with respect to the free scalar result, which provides a convenient common reference across dimensions. In three and four dimensions, the free scalar sets the upper bound, whereas in five dimensions it appears to act instead as a lower bound. The lower bound is given by the Maxwell theory ratio in $d=3$ (trivially) and in $d=4$, where it provides a nontrivial constraint. Each panel includes the ratios corresponding to a free fermion, the EMI model, and holographic theories dual to Einstein gravity. We observe the same hierarchy in three and four dimensions, while in five dimensions this hierarchy is reversed. Regions corresponding to forbidden values of the ratio are shaded in light red.}
    \label{fig:kF0}
\end{figure}

A first noticeable difference with respect to three-dimensional theories is that  $F_0$ does not provide a ``canonical'' normalization for $F(A)$ in $d=5$, as it does not correspond to a global minimum of $F(A)$. Nevertheless, this fact does not necessarily rule out the possible validity of an analogue of the three-dimensional bounds~\eqref{eq:bounds3d} in five dimensions. A second difference with respect to $d=3$ is that the putative lower bound in $d=5$ cannot be related to the Maxwell theory for the simple reason that such a theory is not conformal in five dimensions.\footnote{However, as pointed out in~\cite{El-Showk:2011xbs}, conformal invariance in higher-dimensional Maxwell theory can be restored by embedding it into a non-unitary extension. This would further motivate a more systematic investigation into the existence of analogous conformal bounds in non-unitary CFTs. Although the literature on such theories remains relatively limited, interest in them has grown in recent years~\cite{Beccaria:2015uta,Osborn:2016bev,Giombi:2024zrt}. In particular, within holographic frameworks~\cite{Strominger:2001pn,Witten:2001kn,Maldacena:2002vr}, it has been argued that pseudoentropy  and timelike entanglement entropy for ball-shaped regions encode information about the (generally complex) central charge $a^\star$~\cite{Nakata:2020luh,Mollabashi:2020yie,Mollabashi:2021xsd,Doi:2022iyj,Narayan:2022afv,Doi:2023zaf,Caputa:2024gve,Fujiki:2025rtx} and, more recently, about the coefficient $C_T$ under perturbations in non-unitary CFTs~\cite{Anastasiou:2025rvz,Anastasiou:2026bbf}.} As a consequence, we focus on a conjectural general upper bound of the form
\begin{equation}\label{eq:bounds5d}
\frac{F(A)}{F_0} \stackrel{?}\leq \qt{\frac{F(A)}{F_0}}{free scalar}\,,
\quad \forall \text{CFT}_5\,,\ \forall A\,.
\end{equation}
However, it is easy to see that this fails for general regions. Indeed, let $A$ be a strip region. Then, the above conjecture would become\footnote{The negative relative sign between $F_\text{strip}$ and $F_0$ in $d=5$ was overlooked in the discussion section of~\cite{Bueno:2023gey}, where it was incorrectly claimed that the bound involving a strip region holds for all known five-dimensional free theories.}
\begin{equation}\label{eq:bounds5dk}
\frac{F_\text{strip}}{F_0} \stackrel{?}\leq \qt{\frac{F_\text{strip}}{F_0}}{free scalar} \quad \Longleftrightarrow \quad\frac{k}{F_0} \stackrel{?}\geq \qt{\frac{k}{F_0}}{free scalar}
\quad \forall \text{CFT}_5\,,\ \forall A\,,
\end{equation}
where we used~\eqref{fstripi}. Now, in the case of a free fermion, we have
\begin{equation}\label{eq:viold5}
0.0602\approx\qt{\frac{k}{F_0}}{free fermion}\ngeq
\qt{\frac{k}{ F_0 }}{free scalar}\approx0.2282
\, ,
\end{equation}
which therefore violates the conjecture. In Figure~\ref{fig:kF0} we have plotted all known values of $F_\text{strip}/F_0$ in $d=3,4,5$---the explicit values corresponding to all theories are collected in Appendix~\ref{sec:appuq}. While for $d=3,4$ the expected hierarchy is respected, for $d=5$ we observe an inverted hierarchy, namely, the free scalar yields the \emph{most negative} value for the quotient among all known theories.

Note also that the argument iv) in Section~\ref{sec:bound3} only works in $d=5$ for the union of identical regions such that the individual values of $F(A)$ are negative. Indeed, let $A^\star$ be some entangling region such that $F(A^\star)<0$. Then, from eq.\,(\ref{IAAl}) it follows that
\begin{equation}
   \frac{ F(A^\star\cup A^\star_\ell)}{F(A^\star)}=2+\frac{I(A^\star\cup A^\star_\ell)}{|F(A^\star)|}\overset{\ell \gg R_A}{\leq} 2+\qt{\frac{I(A^\star\cup A^\star_\ell)}{|F(A^\star)|}}{free scalar}=  \qt{\frac{ F(A^\star\cup A^\star_\ell)}{F(A^\star)}}{free scalar}\, ,
\end{equation}
where in the intermediate inequality we used the fact that the free scalar mutual information has the smallest possible inverse scaling with $\ell$ among all CFT$_5$'s for sufficiently large separations. Hence, (\ref{eq:bounds5d}) holds for general CFT$_5$'s in such cases. The argument fails however for entangling regions such that the individual subregions have positive $F(A)$---like the union of two round balls.

The above observations suggest a way of trying to salvage the general conjecture, namely, 
\begin{equation}\label{eq:bounds5da}
\frac{|F(A)|}{F_0} \stackrel{?}\leq \qt{\frac{|F(A)|}{F_0}}{free scalar}\,,
\quad \forall \text{CFT}_5\,,\ \forall A\,.
\end{equation}
However, there exist highly compelling reasons to believe that this modified general conjecture does not hold either. Indeed, as we argued earlier, given a particular CFT, $F(A)$ will generically vanish for multiple choices of the region $A$---\eg \, for $A$ being the union of two disconnected subregions separated by a certain theory-dependent distance, as in~\eqref{FAAL}. However, there is a priori no reason to believe that it will vanish for the same set of regions for different theories (in fact, the expectation is the opposite). Consider then the (infinite) set  $\{A_o\}$ of regions for which $\qt{F(A_o)}{free scalar}=0$. Generically, $F(A_o)\neq 0$ for any other CFT$_5$, and therefore 
\begin{equation}
    \frac{|F(A_o)|}{F_0}>    \qt{\frac{|F(A_o)|}{F_0}}{free scalar}=0\, ,
\end{equation}
thus violating the conjecture.

\subsubsection{Bounds from small deformations around the ball}
Not all hope is lost though. As we saw earlier, the round ball does provide a \emph{local} minimum of $F(A)$ for arbitrary deformations of $A$ and for general theories. Hence, it is natural to inquire whether a milder version of~\eqref{eq:bounds5d} may hold when restricted to small deformations of the spherical entangling surface, namely,
\begin{equation}
    \frac{F_{\epsilon}}{F_0} \stackrel{?}\leq \qt{\frac{F_{\epsilon}}{F_0}}{free scalar}\,,
\quad \forall \text{CFT}_5\,,\ \forall\partial A=\mathbb{S}_\epsilon^3\,,
\end{equation}
where recall that $F_{\epsilon}$ is given by~\eqref{mezei} for general five-dimensional CFTs~\cite{Mezei:2014zla,Faulkner:2015csl}. While this conjecture makes reference to an infinite set of possible entangling regions, it actually collapses to a single relation due to the fact that the geometric dependence on the deformation details is completely disentangled from the theory-dependent coefficient which controls the leading correction to $F_0$, namely, the stress-tensor correlator coefficient $C_T$---this is analogous to the four-dimensional relation between~\eqref{4HM} and~\eqref{eq:HM bounds}. Hence, the above conjecture can be immediately rewritten as 
\begin{equation}\label{ctf05}
   \frac{C_T}{F_0} \leq    \qt{\frac{C_T}{F_0}}{free scalar}=\frac{45}{\pi ^4 \log 4+2 \pi ^2 \zeta (3)-15 \zeta (5)}\approx 0.314\, ,\quad \forall \text{CFT}_5 \,.  
\end{equation}
Luckily, there exists a decent amount of theories for which both quantities have been evaluated in the literature. We have collected the results in appendix~\ref{sec:appuq}, but let us summarize them here. In the case of free fermions~\cite{Osborn:1993cr,Klebanov:2011gs}, the EMI model~\cite{Casini:2008wt,Agon:2021zvp} and holographic theories dual to Einstein gravity~\cite{Ryu:2006bv,Ryu:2006ef} we have, respectively,
\begin{alignat}{5}
&\qt{\frac{C_T}{F_0}}{free fermion}&\ =\ &   \frac{90}{\pi ^4 \log 64+10 \pi ^2 \zeta (3)+15 \zeta (5)}&\ \approx\ &\ 0.167 \, , \\
&\qt{\frac{\cT}{F_0}}{EMI}&\ =\ &   \frac{144}{\pi ^6} &\ \approx\ &\ 0.150\, , \\
&\qt{\frac{\cT}{F_0}}{holo}&\ =\ &   \frac{90}{\pi ^6}&\ \approx\ &\ 0.0936\,.
\end{alignat}

The results for infinite families of theories corresponding to $O(N)$ and GN five-dimensional fixed points are also available~\cite{Giombi:2014xxa,Diab:2016spb}. Using them, we observe
\begin{alignat}{3}
  & \qti{\frac{C_T}{F_0}}{O(N)}&\overset{\text{large-}N}{\approx}&\qt{\frac{C_T}{F_0}}{free scalar}\left(1-\frac{0.369}{N}\right)+\mathcal O(1/N^2)\, , \quad &0.285\lesssim\qti{\frac{C_T}{F_0}}{O(N)}\leq0.314\,,\ \forall N\, , \\ 
  & \qt{\frac{C_T}{F_0}}{GN}&\overset{\text{large-}N}{\approx}&\qt{\frac{C_T}{F_0}}{free fermion} \left(1+ \frac{0.970}{N} \right)+\mathcal{O}(1/N^2)
\, , \quad & 0.167 \leq \qt{\frac{C_T}{F_0}}{GN} \lesssim 0.251\,,\ \forall N\,,
\end{alignat}
where in each case the first result corresponds to the leading large-$N$ correction, and the second to the full range of values spanned for arbitrary values of $N\geq 1$. Notice, in particular, the crucially negative sign that appears in the large-$N$ subleading correction to the free scalar result obtained for the $O(N)$ model. Had it been a plus instead, conjecture~\eqref{ctf05} would have failed.\footnote{A similarly crucial minus sign appears in the analogous three-dimensional version of the conjecture---see~\eqref{eq:CTF0ON3}.}

Additional results are available for five-dimensional supersymmetric CFTs. In particular, for the Seiberg theories with exceptional $E_n$, ($n=1,2,\dots,8$) flavor symmetry groups~\cite{Seiberg:1996bd} and the Morrison--Seiberg $\tilde E_1=U(1)$ theory~\cite{Morrison:1996xf}. From the results obtained in~\cite{Chang:2017cdx} one finds
\begin{equation}
 0.0923 \lesssim  \left. \frac{C_T}{F_0}\right|_{E_n}\lesssim 1.115\, ,\, \quad \text{for }n=1,\ldots,8\,,\quad \text{and} \quad \left. \frac{C_T}{F_0}\right|_{\tilde E_1}\approx0.0944\, .
\end{equation}

\begin{figure}[t!]
    \centering
  \centering
\includegraphics[width=\textwidth]{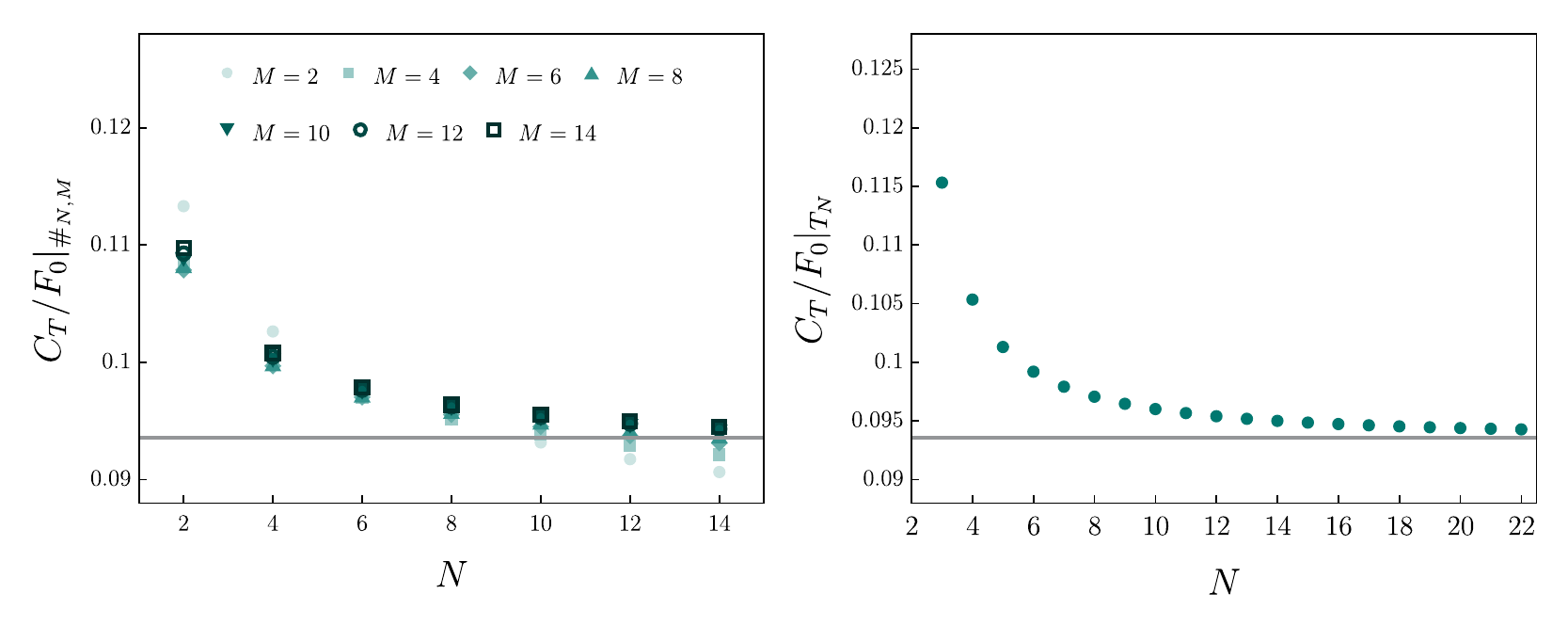}
    \caption{Quotient $C_T/F_0$ for two families of $d=5$ SCFTs corresponding, respectively, to $\#_{N,M}$~\cite{Aharony:1997bh} and $T_N$~\cite{Benini:2009gi,Mitev:2014isa} models, respectively. All data points for $C_T$ and $F_0$---as well as the approximate interpolation shown for the $T_N$ model---are extracted from~\cite{Fluder:2018chf}. }
    \label{fig:TNalmo}
\end{figure}

There also exist results for infinite families of $d=5$ superconformal field theories (SCFTs) engineered by $(p,q)$ 5-brane webs in Type IIB string theory. In particular, the $T_N$ theories are realized by a junction of $N$ D5-branes, $N$ NS5-branes, and $N$ $(1,1)$ 5-branes forming a balanced brane web~\cite{Benini:2009gi,Mitev:2014isa}. Using the numerical results from~\cite{Fluder:2018chf}, we have:
\begin{equation}
0.0936\leq  \left. \frac{C_T}{F_0}\right|_{T_N}\lesssim  0.115\,, \quad \forall N\,.
\end{equation}
The results are plotted in Figure~\ref{fig:TNalmo}, where it is apparent that the corresponding ratio approaches the holographic result in the large-$N$ limit. Similarly, for the SCFTs engineered by $(p,q)$ 5-brane webs with $N$ D5-branes and $M$ NS5-branes (the so-called $\#_{N,M}$ theories), we employ the results from~\cite{Fluder:2018chf} to find 
\begin{equation}
0.0906 \lesssim  \left. \frac{C_T}{F_0}\right|_{\#_{N,M}}\lesssim 0.113\,,\quad \forall N\,,\,M\,,
\end{equation}
where in all cases the results approach the holographic one when the number of branes becomes large. We show various series of values corresponding to different $M$ in Figure~\ref{fig:TNalmo}. In this case, some of the theories yield values of the quotient smaller than the leading holographic result, and are the only examples we have found with this behavior.

\begin{figure}[t!]
    \centering
\includegraphics[width=1\textwidth]{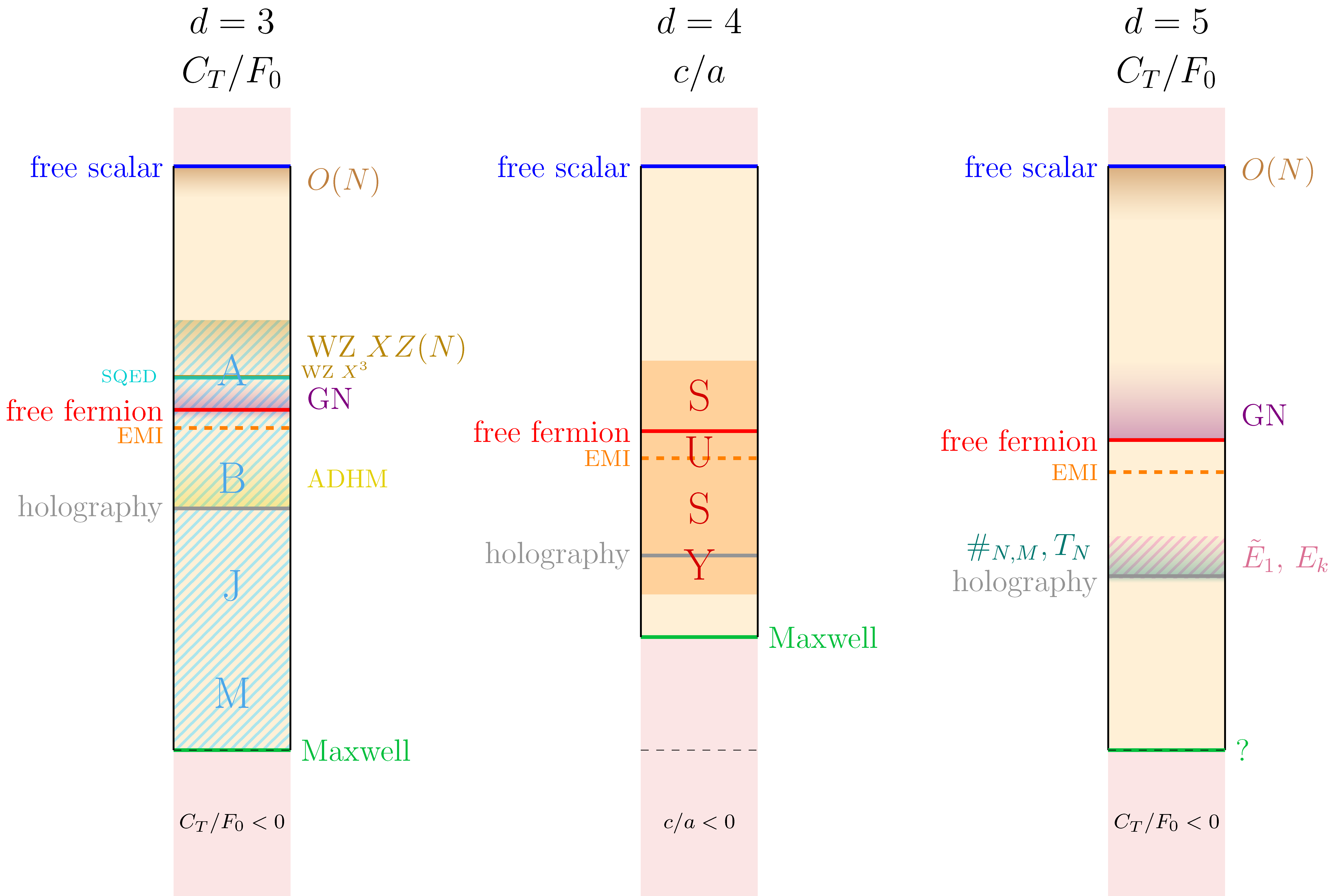}
    \caption{We display the ratios $C_T/F_0$ in three and five dimensions, as well as $c/a$ in four dimensions, for several theories. In all cases, the values are normalized by the free scalar result, which provides an upper bound in every dimension. In three and four dimensions the lower bound is set by Maxwell theory, whereas in five dimensions the putative lower bound remains unknown, which is indicated with a question mark. In each diagram we also include the values of these ratios for the free fermion, the EMI model, and holographic theories, observing the same hierarchy across dimensions. In the three-dimensional diagram we additionally show: i) the discrete values of $C_T/F_0$ obtained in the WZ $X^3$ and SQED models; ii) a shaded region corresponding to the range of ratios realized by the ABJM model; iii) fading bands representing large-$N$ approximations for the $O(N)$, GN, and ADHM models—the fading indicates that the results become progressively less reliable. In four dimensions, we highlight with a solid-colored region the more stringent Hofman--Maldacena bounds that apply to supersymmetric (SUSY) models. Finally, in five dimensions we include: i) a shaded region delimiting the ratios realized by the superconformal rank-one Seiberg exceptional theories $E_n$ $(n=1,\ldots,8)$ together with the Morrison--Seiberg superconformal theory $\tilde E_1$; ii) fading regions corresponding to large-$N$ approximations in the $O(N)$, GN, $\#_{N,M}$, and $T_N$ models—the latter two are combined into a single region due to their very similar values. Regions corresponding to forbidden values of the ratios are highlighted in light red.
    }\label{fig:CTF0}
\end{figure}

All the results are collected in Figure~\ref{fig:CTF0}, where we also present the analogous $d=3,4$ ratios. The similarity between the plots in different dimensions is striking. In particular, the free scalar ratio always provides the greatest value for all known theories in all dimensions. Secondly, the free fermion  value is always greater than the EMI model and holographic results, and the Maxwell one serves as the lower limit both in $d=3$ and $d=4$, but does not seem to play a role in $d=5$ (where recall that the theory is not a CFT). Also, supersymmetric theories always seem to fill some subset of values which extends above the free fermion result and below the holographic one. Finally, GN theories provide greater values than the free fermion in $d=3,5$, which are nevertheless comfortably lower than the corresponding $O(N)$ model results. The latter always saturate the free scalar upper bounds in the large-$N$ limit, but they never exceed them. In the $d=5$ case it is unclear how close to zero the bound can be. It seems plausible that additional families of supersymmetric CFTs may involve models for which $C_T/F_0$ vanishes in certain limit, analogously to three-dimensional ABJM models in the large Chern--Simons level limit---see~\eqref{eq:CTF0ABJMN1}.

\subsection{Higher dimensions}
In view of the above results, it is only natural to conjecture that 
\begin{equation}\label{ctf0d}
   \frac{C_T}{F_0} \leq    \qt{\frac{C_T}{F_0}}{free scalar}\, ,\quad \forall \text{CFT}_d \,, 
\end{equation}
namely, that the free scalar result is an upper bound for this ratio across all CFTs in general dimensions. 

While the existence of unitary interacting CFTs is dubious beyond $d=6$~\cite{Parisi:1975im,Bekaert:2011cu,Fitzpatrick:2013sya,Cordova:2018eba,Fei:2014yja}, the question can at least still be posed for free theories and (formal?) holographic duals to Einstein gravity.
In those cases it is easy to verify that such a conjecture holds for general $d$, and that the relative separation between the free scalar results and the others grows with $d$. In particular, in the large-$d$ limit one finds
\begin{align}
    &\qt{\frac{C_T}{F_0}}{free scalar}\overset{d\rightarrow\infty}{\approx}\sqrt{\frac{\pi^3d}{8}}\left[\frac{d}{\pi\text e}\right]^d\, ,\\ &\qt{\frac{C_T}{F_0}}{free fermion}\overset{d\rightarrow\infty}{\approx}\sqrt{\frac{\pi^3d}{32}}\left[\frac{d}{\pi\text e}\right]^d\,, \\
   & \qt{\frac{C_T}{F_0}}{holo}\overset{d\rightarrow\infty}{\approx}\sqrt{\frac{d}{2\pi}}\left[\frac{d}{\pi\text e}\right]^d\,,
\end{align}
and therefore
\begin{align}
  \lim_{d\rightarrow \infty} \frac{\left[\qt{C_T/F_0}{free scalar}\right]}{\left[\qt{C_T/F_0}{free fermion}\right]}= 2\, ,\qquad
    \lim_{d\rightarrow \infty} \frac{\left[\qt{C_T/F_0}{free scalar}\right]}{ \left[\qt{C_T/F_0}{holo}\right]}=\frac{\pi^2}{2}\, .
\end{align}
Observe also that in the $d=7,11,\dots$ cases, the strip coefficient takes again the same sign as $F_0$---see~\ref{fstripi}, so it may seem a priori possible that many of the arguments which hold in the $d=3$ case are valid again in those dimensions. However, we do not expect $F(A)$ to have a definite sign beyond three dimensions for general regions. We have explicitly verified this for the $d=7$ EMI model, where different types of cylindrical entangling surfaces give rise to arbitrarily positive, arbitrarily negative or vanishing values of $F(A)$---see eqs.\,(\ref{s2r3}), (\ref{s4r1}), (\ref{s3r2}) and (\ref{s1r4})---similarly to the five-dimensional case. Hence, it seems that (\ref{ctf0d}) remains the only plausible candidate bound for odd-dimensional theories beyond $d=3$.

The remaining case of special interest corresponds to $d=6$ CFTs. For those, the entanglement entropy universal term is again logarithmic. The universal coefficient takes a form similar to the four-dimensional Solodukhin formula (\ref{log4d}), namely, it involves a linear combination of theory-independent functionals of intrinsic and extrinsic curvatures on the entangling surface weighted by the four trace-anomaly coefficients that exist in $d=6$~\cite{Safdi:2012sn,Miao:2015iba}. The six-dimensional version of (\ref{4HM}) would involve a constraint on some combination of the charges and the geometric functionals. However, it is far from obvious that the latter will play as trivial a role as they do in the four-dimensional case.  Consequently, it remains unclear whether there exists a relation between this putative constraint and the $d=6$ conformal collider bounds, analogous to the one found in $d=4$. 

One may also wonder about the hierarchy of theories regarding the $F_\text{strip}/F_0$ quotient beyond $d=5$. This can be more easily expressed in terms of the coefficient $k$, which will appear with a minus sign in $F_\text{strip}$ in $d=5,9,\dots$---therefore reversing the hierarchy in such cases---and with a plus in all the rest of cases. We have explicitly verified that $k/F_0$ is indeed such that the hierarchy
\begin{equation}
\qt{\frac{k}{F_0}}{free scalar}>\qt{\frac{k}{F_0}}{free fermion} >\qt{\frac{k}{F_0}}{EMI} > \qt{\frac{k}{F_0}}{holo}>  \qt{\frac{k}{F_0}}{Maxwell}\, ,
\end{equation}
holds $\forall d$ (where the Maxwell case is meant to apply only in $d=3,4$). In particular, in the large-$d$ limit, one finds\footnote{Note that the free fermion result deviates more and more from the EMI model one as $d$ grows~\cite{Agon:2021zvp}.}
\begin{equation}
    \lim_{d\rightarrow \infty} \frac{\qt{k/F_0}{free fermion}}{\qt{k/F_0}{free scalar}}=  \frac{1}{d}\, , \quad   \lim_{d\rightarrow \infty}  \frac{\qt{k/F_0}{EMI}}{\qt{k/F_0}{free scalar}}=\frac{32}{\pi d^2}\, , \quad \lim_{d\rightarrow \infty} \frac{\qt{k/F_0}{holo}}{\qt{k/F_0}{free scalar}} = \frac{8\text e}{\pi^2d}\left[\frac{\pi}{d}\right]^d\, .
\end{equation}

\section{Concluding remarks}\label{conclu}
Charting the space of CFTs compatible with general physical principles is a fundamental and outstanding problem. Several kinds of constraints and bounds in the space of theories have been unveiled based on different criteria such as unitarity, energy-positivity, causality, or crossing-symmetry---see   \eg\, 
\cite{Minwalla:1997ka,Zamolodchikov:1986gt,Hofman:2008ar,Buchel:2009sk,Casini:2012ei,Hartman:2015lfa,Hofman:2016awc,El-Showk:2012cjh,Rattazzi:2008pe,Poland:2018epd,Casini:2021raa,Casini:2017vbe,Komargodski:2011vj,Rychkov:2016iqz} and references therein. In contrast to these more or less thoroughly studied and well-established principles, a somewhat alien set of conjectural bounds was presented in~\cite{Bueno:2023gey}, based on the general structural properties of EE and MI in three-dimensional CFTs. If true, (\ref{eq:bounds3d}) would in fact represent an infinite set of constraints---namely, one for each choice of entangling region $A$---for every CFT$_3$.  In this paper we have explored the possibility that an analogous set of bounds may hold for general CFT$_5$'s. We have seen that, as opposed to the three-dimensional case~\cite{Bueno:2021fxb}, $F(A)$ does not have a sign in $d=5$, which spoils several of the putative arguments that suggest the general validity of the three-dimensional bounds and, indeed,  we have explicitly shown that  (\ref{eq:bounds5d}) does not hold for general regions. On the other hand, the five-dimensional version of the arguably most relevant particular bound following from (\ref{eq:bounds3d}), namely, the one for the ratio $C_T/F_0$, is satisfied for all CFT$_5$'s for which these quantities (to the best of our knowledge) have been evaluated in the literature---see (\ref{ctf05}).  While the theoretical basis for the validity of this five-dimensional conjecture remains admittedly weak---relying as it does on a heuristic extrapolation from $d=3,4$, and the observation that $F_0$ constitutes a local minimum for $F(A)$ in $d=5$---the cumulative evidence presented in this work suggests a robust universal underlying principle (which likely extends to general dimensions).

In the absence of a clear route to unveiling such a principle, it would be interesting to find new five-dimensional models for which the ratio $C_T/F_0$ can be computed, in order to further test our conjecture. An obvious subclass of candidates are SCFTs. For those, the stress-tensor charge can be obtained from the leading (quadratic) correction to the Euclidean free energy on a squashed sphere around the round-sphere minimum~\cite{Chang:2017cdx,Imamura:2011wg}. Hence, in those cases, the bound on $C_T/F_0$ would translate into a bound on how much the squashed-sphere partition function can deviate from the round-sphere result at the lowest non-trivial order in the deformation.

The three-dimensional lower bound for $C_T/F_0$ is saturated by the Maxwell theory, which can be thought of as a shift-orbifold of the free scalar. There is no analogous theory which may serve as a candidate for saturating the lower bound in $d=5$. On the other hand, there is at least another way of approaching  $C_T/F_0=0$ in $d=3$, namely, considering the large  Chern--Simons level limit, $k\rightarrow \infty$, of the $U(N)_k\times U(N)_{-k}$ ABJM model~\cite{Bueno:2023gey}. Hence, it is tempting to speculate that certain class of five-dimensional SCFTs may behave analogously. However, the absence of a tunable parameter analogous to the Chern--Simons level, as well as the intrinsically strongly-coupled nature of known five-dimensional fixed points, makes it unclear whether such a limit can be realized. Nevertheless, it remains conceivable that an alternative mechanism could lead to a similar suppression of $C_T/F_0$ in $d=5$.

\section*{Acknowledgments}

We thank César A. Agón, Nikolay Bobev, Horacio Casini, Rajeev Erramilli, Yu Nakayama, Tadashi Takayanagi, Guido van der Velde and Nicolò Zenoni for useful discussions. P.B. was supported by a Proyecto de Consolidación Investigadora (CNS 2023-143822) from Spain's Ministry of Science, Innovation and Universities, and by the grant PID2022-136224NB-C22, funded by MCIN/AEI/ 10.13039/501100011033/FEDER, UE. The work of J.M. is supported by the Beatriu de Pinós fellowship BP 2024 00033 of the Agència de Gestió d'Ajuts Universitaris i de Recerca, Generalitat de Catalunya.

\appendix

\section{Computations in the EMI model}
\label{sect:EMIappendix}

\subsection{Entangling surfaces in general dimensions}

\subsubsection{Sphere $\partial A=\mathbb S^{d-2}$}
Let us first compute the regulated EE in the EMI model for a ball entangling region $A=\mathbb B^{d-1}$ of radius $R$, whose entangling surface is the $\partial A=\mathbb S^{d-2}$ sphere. We parametrize the $(d-1)$-dimensional ball as
\begin{equation}\label{eq:Bddef}
\rho = \sqrt{x_1^2 + \dots + x_{d-1}^2},
\qquad
\mathbb B^{d-1} = \{ (x_1,\dots,x_{d-1}) \mid \rho < R \}\,.
\end{equation}
Using the canonical angular coordinates $\theta_1,\ldots,\theta_{d-3}\in[0,\pi]$ and $\theta_{d-2}\in[0,2\pi)$, we can write the components of the position vectors $\mathbf r_{i}=R_i\mathbf \Omega_{i}=R_i\cdot\left(\Omega_{i,1},\ldots,\Omega_{i,d-1}\right)$ describing two balls $A_1=A^-$ and $A_2=\overline{A^+}$, respectively, where
\begin{equation}\label{eq:rij}
\Omega_{i,j}=\cos\theta_{i,j}\prod_{k=1}^{j-1}\sin\theta_{i,k}\,,\quad j=1,\ldots,d-2\,,\quad \text{and} \quad \Omega_{i,d-1}=\prod_{k=1}^{d-2}\sin\theta_{i,k}\,,
\end{equation}
and their radii are given by
\begin{equation}\label{eq:R12eps}
 R_1 = R- \varepsilon/2\,,\quad  R_2 = R+ \varepsilon/2\,.
\end{equation}

\begin{figure*}
\centering
\begin{subfigure}{.5\textwidth}
  \centering
\includegraphics[height=0.95\linewidth]{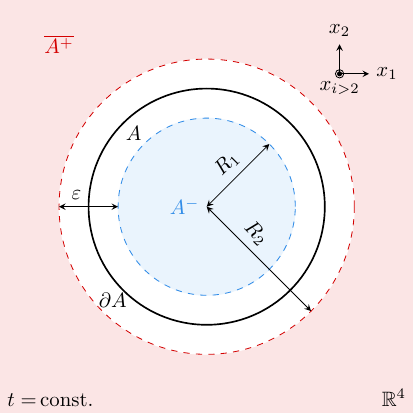}
\end{subfigure}%
\begin{subfigure}{.5\textwidth}
  \centering
  \includegraphics[height=0.95\linewidth]{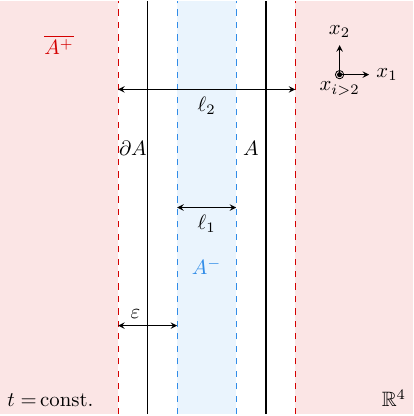}
\end{subfigure}
\caption{We sketch specific setups used to evaluate the short-distance limit of the MI, from which the finite contribution $F(A)$ arises. (Left) We consider a ball region $A^-$ of radius $R_1$ and a slightly larger one, $A^+$, of radius $R_2 = R_1 + \varepsilon$, with $\varepsilon \ll R_1$. (Right) We consider a strip-like region $A^-$ that is infinitely extended along the $x_2$, $x_3$, and $x_4$ directions and has width $\ell_1$, together with a slightly larger region $A^+$ of width $\ell_2 = \ell_1 + \varepsilon$, with $\varepsilon \ll \ell_1$. In both setups, the complementary region $\overline{A^+}$ is used in the MI to extract $F(A)$.}
\label{fig:MIstripcircle}
\end{figure*}

The setup is described in Figure~\ref{fig:MIstripcircle} (Left), considering $\mathbb B^{d-1}$ instead of $\mathbb B^{3}$. The corresponding normal vectors to $\mathbf r_i$ are given by $\mathbf n_i=\mathbf r_i/R_i$. Given the symmetry of the setup, we can fix $\mathbf r_2=(R_2,0,\ldots,0)$ and in turn $\mathbf n_2=\left(1,0,\ldots,0\right)$ while $\mathbf r_1$ is given by setting $i=1$ in~\eqref{eq:rij}. This simplifies greatly the expressions of $\mathbf n_1  \cdot \mathbf n_2 = \cos \theta_{1,1}$ and $|\mathbf r_1 - \mathbf r_2|^2 =R_1^2+R_2^2-2R_1R_2\cos\theta_{1,1}$. Taking this into account, together with the surface elements corresponding to the ones of the unit $(d-2)$-sphere,\footnote{The surface element in the $(d-2)$-unit sphere and its surface integral are, respectively, given by \begin{equation}\label{eq:Sd}
\diff\Omega_{d-2}=\diff\theta_{d-2}\prod_{j=1}^{d-3}\sin^{d-2-j}\theta_{j}\,\diff\theta_{j}\,,\quad \Omega_{d-2}\equiv\int_{\mathbb S^{d-2}}\diff\Omega_{d-2}=\frac{2\pi^{(d-1)/2}}{\Gamma\left[(d-1)/2\right]}\,.
\end{equation}} $\diff^{d-2}\mathbf r_i=R_i^{d-2}\diff\Omega_{i,d-2}$, the integral in~\eqref{eq:EMId_Info} reads
\begin{align}
  \qt{I_\varepsilon(\partial A=\mathbb S^{d-2})}{EMI}&=2R_1^{d-2}R_2^{d-2}\kappa \int_{\mathbb S^{d-2}}\int_{\mathbb S^{d-2}}\frac{\cos\theta_{1,1}\diff\Omega_{1,d-2}\diff\Omega_{2,d-2}}{(R_1^2+R_2^2- 2 R_1 R_2 \cos \theta_1)^{d-2}}\\
  &=2\Omega_{d-2}\Omega_{d-3}R_1^{d-2} R_2^{d-2} \kappa 
  \int_0^\pi \frac{(\sin \theta_1)^{d-3} \cos \theta_1\diff \theta_1 }{(R_1^2+R_2^2- 2 R_1 R_2 \cos \theta_1)^{d-2} }\,,
\end{align}
where in the second line we have integrated and used the splitting property of the surface element $\diff\Omega_{i,d-2}=\sin^{d-3}\theta_{i,1}\diff\theta_{i,1}\,\diff\Omega_{i,d-3}$---see~\eqref{eq:Sd}---as well as dropped the first label in the coordinate $\theta_{1,1}$ for simplicity. After integrating $\theta_1$, we obtain
\begin{equation}\label{eq:Idsphere}
\qt{I_\varepsilon (\partial A=\mathbb S^{d-2})}{EMI}  =  \frac{4 \pi ^{(d-2)/2} \Gamma \left(d/2\right) \Omega_{d-2}\kappa}{(d-1) \Gamma (d-2)}u^{d-1}{}_2F_1\left(\frac{d-1}{2},\frac{d}{2};\frac{d+1}{2};u^2\right)\,,
\end{equation}
where $u=2R_1R_2/(R_1^2+R_2^2)$. While a similar expression to~\eqref{eq:Idsphere} was derived in greater detail in~\cite{Agon:2021zvp}, that reference does not explicitly present the universal coefficients appearing in the short-distance expansion beyond the strip coefficient $\qt{k}{EMI}$ and the central charge $\qt{a^\star}{EMI}$. To extract the remaining coefficients, we express~\eqref{eq:Idsphere} in terms of $\varepsilon$ according to~\eqref{eq:R12eps}, and expand in the regime $\varepsilon \ll R$, obtaining, up to order $\mathcal O(\varepsilon)$,
\begin{align}\label{eq:MIEMIdSph}
  \qt{I_\varepsilon(\partial A=\mathbb S^{d-2})}{EMI} =& \qt{k}{EMI}\left[\frac{\Omega_{d-2}R^{d-2}}{\varepsilon^{d-2}} \right] +\sum_{i=2}^{\lfloor d/2\rfloor}\qt{\gamma^{(i)}}{EMI}\left[\frac{R}{\varepsilon}\right]^{d-2i}\notag\\ 
  &+ 
    2\times\begin{cases}(-1)^{(d-1)/2} \qt{2\pi a^\star}{EMI}\quad &\text{for }d\text{ odd},\\
    (-1)^{(d-2)/2}\qt{4a^\star}{EMI}\log\frac{2R}{\varepsilon} &\text{for }d\text{ even},
    \end{cases}
\end{align}
where the coefficients in the expansion read
\begin{alignat}{2}
    \qt{k}{EMI} &= \frac{2 \pi^{\frac{d-2}{2}} \Gamma[(d-2)/2]\kappa}{\Gamma(d-2) }\,,\label{eq:stripcoefd}\\
    \qt{\gamma^{(i)}}{EMI}&=\frac{(d-2)\Gamma(d-1)\pi^{d-1}}{\Gamma[(d-1)/2]^2}\frac{(-1)^{i-1}2^{7-d-2i}\kappa}{\Gamma(d-i)\Gamma(i)(d-2i)}\,,\quad \text{for }i=2,\ldots,\Bigl\lfloor\frac{d}{2}\Bigl\rfloor\,.\\
    \qt{a^\star}{EMI}&=\frac{\pi^{d-2}\kappa}{\Gamma(d-2)}\,.    \label{eq:astar}
\end{alignat}
Unlike the EE expansion, the coefficients $\qt{k}{EMI}$---\ie~the $d$-dimensional strip coefficient---and $\qt{\gamma^{(i)}}{EMI}$ appearing in the MI expansion~\eqref{eq:MIEMIdSph} are universal, as $\varepsilon$ is a physical regulator and not a UV cutoff.\footnote{This also applies to the finite term appearing in even-dimensional CFTs. Whereas in EE it can be polluted by the choice of UV regulator, in the MI expansion it is also (unambiguously) controlled by the central charge $a^\star$, as indicated by the factor of 2 in the logarithmic term.} From this expression we can particularize the dimensions so that we can compare our results with those presented in~\cite{Casini:2015woa}, and contextualize the novel five-dimensional one in~\eqref{eq:EMI5Sph}, this is
\begin{alignat}{3}
     \qt{I_\varepsilon(\partial A=\mathbb S^{d-2})}{EMI}  &= 2\pi\kappa\cdot\left[\frac{2\pi R}{\varepsilon}\right]-4\pi\cdot\left(\pi\right)\kappa+\mathcal O(\varepsilon)\,,&&\text{if }d=3\,, \\ 
     \qt{I_\varepsilon(\partial A=\mathbb S^{d-2})}{EMI} &= 2\pi\kappa\cdot\left[\frac{4\pi R^2}{\varepsilon^2}\right] 
      -8 \pi^2 \kappa \,  \log\frac{2R}{\varepsilon} +O(\varepsilon)\,,&&\text{if } d=4\,, \\
    \qt{I_\varepsilon(\partial A=\mathbb S^{d-2})}{EMI}&= \frac{\pi^2\kappa}{2} \cdot \left[\frac{2\pi^2R^3}{\varepsilon^3}\right] 
      - \frac{9 \pi^4 \kappa}{4} \frac{R}{\varepsilon}
      + 4\pi\left( \frac{\pi^3 \kappa}{2} \right)
      +O(\varepsilon)\,,\quad&&\text{if }d=5\,.
      \label{eq:IEMI_sphere_d5}
\end{alignat}
Here, the coefficient $\qt{\gamma^{(2)}}{EMI}= -9\pi^4/4\kappa$ appearing in~\eqref{eq:IEMI_sphere_d5} is related to $\qt{\alpha}{EMI}$ in~\eqref{eq:EMI5Sph} through $\qt{\gamma^{(2)}}{EMI} = 12\pi^2 \qt{\alpha}{EMI}$, where the normalization of the latter is adapted to its five-dimensional origin, as discussed in Section~\ref{sec:exEMI}.

\subsubsection{Strip}
Let us next compute the MI of two very close strips using~\eqref{eq:EMId_Info}. We choose a set of coordinates in $d$ dimension whose spatial components are $\mathbf r_{i}=(x_{i,1}, \dots x_{i,d-1})$. The regions $A_1$ and $A_2$ in~\eqref{eq:EMId_Info} are strips defined as
\begin{equation}
A_i= \left\{ x_{i,1}  \in [-\ell_i/2, \ell_i/2] , \, x_{i,k} \in[0,\ell_\parallel]\right\}\,,\quad \text{for }i=1,2\,,\, k=2,\ldots,d-1,
\end{equation}
 and we consider $\ell_\perp$ such that $\ell_1<\ell_\perp<\ell_2$. We may choose $\mathbf{r}_2 = ( \ell_2 /2, 0, \dots , 0)$ and, consequently, $\mathbf{n}_2 = ( 1,0,\dots,0)$ as well as $\mathbf{r}_1^\pm  = ( \pm \ell_1/2, x_{1,2}, \ldots, x_{1,d-1} ) $, and thus $\mathbf{n}_1^\pm  = (\pm 1, 0, \dots ,0)$. With this parametrization it is easy to check that $\mathbf{n}_1^\pm \cdot \mathbf{n}_2  = \pm 1$ and therefore $|\mathbf{r}_2 -\mathbf{r}_1^\pm |^2 = 
\left[(\ell_2\mp \ell_1)^2/4+ x_{1,2}^2+ \ldots + x_{1,d-1}^2\right]$, and thus the MI is computed from $\qt{I_\varepsilon}{EMI}(\text{strip})=2\left[\qt{I_\varepsilon^+(\text{strip})}{EMI}+\qt{I_\varepsilon^-(\text{strip})}{EMI}\right]$, where
\begin{equation} 
\label{eq:Istrips_pm}
\qt{I_\varepsilon^\pm(\text{strip})}{EMI}= 
2 \kappa \int_{\partial A_1}\diff^{d-2}\mathbf r_1\int_{\partial A_2}\diff^{d-2}\mathbf r_2\frac{\pm 1}{(v_\mp^2+ x_{1,2}^2+ \ldots +  x_{1,d-1}^2 )^{d-2}}\,,
\end{equation}
where $v_\mp=(\ell_2\mp \ell_1)/2$. Now, we need to regularize the infinite volume spanned by the coordinates $(x_{i,2},\dots,x_{i,d-1})$. 
For this purpose, we reparametrize the integral~\eqref{eq:Istrips_pm} in terms of the radius $\rho$, obtaining $\diff^{d-2}\mathbf r_i=\diff\Omega_{i,d-3}\rho_i^{d-3}\diff\rho_i$. After this, the integral in~\eqref{eq:Istrips_pm} becomes
\begin{equation}\label{eq:strippreexp}
\qt{I_\varepsilon^\pm(\text{strip})}{EMI}=\pm2\,\area_\parallel\Omega_{d-3}\kappa\int_0^\infty\frac{\rho^{d-3}\diff \rho  }{[v_\mp^2 + \rho^2 ]^{d-2}}=\pm\frac{\Gamma[(d-2)/2)]^2}{\Gamma(d-2)}\Omega_{d-3}\kappa\frac{\area_\parallel}{v_\mp^{d-2}}\,,
\end{equation}
where area$_\parallel=\ell_\parallel^{d-2}$ is the infrared-regularized area of the strip. Finally, if the strips are at a distance $\varepsilon$ from each other, \ie~$\ell_1=\ell_\perp-\varepsilon$, $\ell_2=\ell_\perp+\varepsilon$ we obtain\footnote{ Note that the factor of $2$ in front of the $k\cdot [\text{ area}_{\parallel}/\varepsilon^{d-2}]$ piece is a result of the fact that our regulated strip actually possesses two boundaries, as opposed \eg\, to the regulated sphere---compare both plots in Fig.\,\ref{fig:MIstripcircle}. On the other hand, the same factor 2 in front of the $-k\cdot[ \text{ area}_{\parallel}/\ell_{\perp}^{d-2}]$ term is the expected one.}~\cite{Agon:2021zvp}
\begin{equation}
\label{eq:IEMI_strip}
\qt{I_\varepsilon(\text{strip})}{EMI}= 
 2k \left[ 
\frac{\area_\parallel}{\varepsilon^{d-2}} - 
\frac{\area_\parallel}{\ell_\perp^{d-2}}\right]\,,
\end{equation}
where we have written the result in terms of the universal coefficient $k$ is given in~\eqref{eq:stripcoefd}. Expression~\eqref{eq:IEMI_strip} particularized for the five-dimensional case is given in~\eqref{stripi}.

\subsection{Five dimensions}
Let us now particularize the spacetime dimension to $d=5$ and study a few additional cases of interest.
\subsubsection{Cylinder  $\partial A=\mathbb S^{1} \times \mathbb{R}^2$}

We consider now a cylindrical entangling surface with geometry $\partial A=\mathbb S^{1} \times \mathbb{R}^2$. 
We choose the set of coordinates $(r,\phi,  z,w)$ in four dimensions, related to the usual Cartesian coordinates as
$x_1 = r \cos \phi$, $x_2 = r \sin  \phi$, $x_3 =z$, $x_4 = w$. The regions $A_1$ and $A_2$ are parametrized as 
\begin{equation}
A_i= \left\{   r_i \in [0,R_i]\,, \phi_i\in[0,2\pi)\,,  z_i\,, w_i \in \left[-L/2,L/2\right]  \right\}\,,\qquad \text{for }i=1,2\,,
\end{equation} 
where $R_1$ and $R_2$ are given in~\eqref{eq:R12eps} and $L \gg R_1,R_2$.
Based on the symmetries of the setup we may choose $\mathbf{r}_2 = ( R_2 , 0, 0 , 0)$ and, consequently $\mathbf{n}_2 = ( 1,0,\dots,0)$ as well as $\mathbf{r}_1  = ( R_1 \cos \phi_1 ,R_1 \sin \phi_1, z_1, w_1 )$ and thus $\mathbf{n}_1  =(  \cos \phi_1 , \sin \phi_1, 0, 0 ) $. With this parametrization it is easy to check that $
\mathbf{n}_1 \cdot \mathbf{n}_2 = \cos{\phi_1}$ and $|\mathbf{r}_1 - \mathbf{r}_2|^2  = R_1^2 +R_2^2 -2 R_1 R_2 \cos \phi_1 + z_1^2+w_1^2$,
and thus, using that $\diff^3\mathbf r_i=R_i\diff \phi_i\diff z_i\diff w_i$, the MI can be computed expressing the integral in~\eqref{eq:EMId_Info} as 
\begin{align}
\label{eq:IEMI_cyl2}
  \qt{I_{\varepsilon}(\partial A =\mathbb{S}^1\times \mathbb{R}^2)}{EMI} & 
  =
 \int_{-\frac{L}{2}}^{\frac{L}{2}}  
\int_{-\frac{L}{2}}^{\frac{L}{2}} 
\int_0^{2 \pi }
\frac{ 4 \pi L^2 R_1 R_2\kappa   \cos \phi\,\diff \phi\,\diff z\,\diff w}{(R_1^2 +R_2^2 -2 R_1 R_2 \cos \phi + z^2+w^2)^{3}   }\,.
\end{align}
Proceeding as before, we renamed the dummy variable and dropped the subscript $i=1$ in the coordinates. Now, we integrate over $z$ and then over $w$, performing an expansion in the limit $L \to \infty$ at each step. We then integrate over $\phi$ and obtain 
\begin{equation}
\label{eq:IEMI_Qcyl3bis} 
  \qt{I_{\varepsilon}(\partial A =\mathbb{S}^1\times \mathbb{R}^2)}{EMI} = 
\frac{8 \pi ^3 \kappa \,  L^2 R_1^2 R_2^2}{(R_2^2-R_1^2 )^3}\,,
\end{equation}
We finally express~\eqref{eq:IEMI_Qcyl3bis} in terms of $\varepsilon$ according to~\eqref{eq:R12eps}
and obtain~\eqref{s1r2EMI}. 

\subsubsection{Cylinder  $\partial A=\mathbb S^{2} \times \mathbb{R}^1$}

Let us now compute the MI for two very close entangling regions whose shape is $\partial A=\mathbb S^{2} \times \mathbb{R}^1$. 
We choose the set of coordinates $(r,\theta, \phi, z)$ related to the usual Cartesian coordinates following $x_1 = r\cos\theta$, $x_2 = r\sin\theta\cos\phi$, $x_3 = r \sin\theta\sin\phi$, $x_4 = z$. The regions $A_1$ and $A_2$ are parameterized as 
\begin{equation}
A_i= \left\{   r_i \in [0,R_i]\,, \theta_i\in[0,\pi]\,,\phi_i\in[0,2\pi)\,,  z_i \in \left[-L/2,L/2\right]  \right\}\,, \quad \text{for }i=1,2\,,
\end{equation} 
where $R_1$ and $R_2$ are given in~\eqref{eq:R12eps} and $L \gg R_1,R_2$. Once again, based on the symmetries of the setup we may choose $\mathbf{r}_2 = (R_2,0, 0, 0)$ and, consequently $\mathbf{n}_2 = (1,0,0,0)$ as well as $\mathbf{r}_1  = ( R_1 \cos \theta_1, R_1 \sin \theta_1 \cos \phi_1, R_1 \sin\theta_1\sin\phi_1, z_1 ) $, and $\mathbf{n}_1  =(\cos \theta_1,\sin \theta_1 \cos \phi_1, \sin\theta_1\sin\phi_1, 0)$. With this parametrization it is easy to check that $\mathbf{n}_1 \cdot \mathbf{n}_2 = \cos{\theta_1}$, 
and $|\mathbf{r}_1 - \mathbf{r}_2|^2  = R_1^2 +R_2^2 -2 R_1 R_2 \cos \theta_1 + z_1^2$. Using that $\diff^3\mathbf r_i=R_i^2\sin\theta_i\,\diff\theta_i\,\diff\phi_i\,\diff z_i$ the MI can be computed expressing the integral in~\eqref{eq:EMId_Info} as
\begin{equation}
\label{eq:IEMI_cyl2}
  \qt{I_{\varepsilon}(\partial A =\mathbb{S}^2\times \mathbb{R}^1)}{EMI}=
\int_{-\frac{L}{2}}^{\frac{L}{2}}\int_0^\pi
\frac{16 \pi^2 L R_2^2 R_1^2 \kappa\cos \theta \sin \theta \,\diff \theta \,\diff  z}{(R_1^2 +R_2^2 -2 R_1 R_2 \cos \theta + z^2)^{3}}\,.
\end{equation}
First we integrate $z_1$, then expand in $L \to \infty $ and finally we integrate $\theta$, obtaining
\begin{equation}\label{eq:IEMI_cyl3bis}
 \qt{I_{\varepsilon}(\partial A =\mathbb{S}^2\times \mathbb{R}^1)}{EMI} = 
\frac{4\pi^3 LR_1^3  (5R_2^2-R_1^2)\kappa}{(R_2^2-R_1^2)^3}\,,   
\end{equation}
We finally express~\eqref{eq:IEMI_cyl3bis} in terms of $\varepsilon$ according to~\eqref{eq:R12eps}
and obtain~\eqref{eq:MI_EMI_CylS2R}. 

\subsubsection{Perturbed sphere $\partial A=\mathbb S^3_\epsilon$}

Consider a perturbed spherical entangling surface $\partial A=\mathbb S_\epsilon^3$, whose radius is given, to linear order in the perturbation parameter $\epsilon$, by
\begin{equation}\label{eq:pertS3}
r(\mathbf{\Omega})=R\left[1+\epsilon\Phi(\mathbf{\Omega})\right]\,,\qquad 
\Phi(\mathbf{\Omega})\equiv\sum_{\ell,\mathbf m}a_{\ell,\mathbf m}Y_{\ell,\mathbf m}(\mathbf{\Omega})\,
\end{equation}
where $Y_{\ell,\mathbf m}=Y_{\ell,\mathbf m}(\mathbf \Omega)$ denotes the scalar spherical harmonic on $\mathbb S^{3}$ with labels $\ell$ and $\mathbf{m}={m_1,m_2}$, satisfying $\Delta Y_{\ell,\mathbf m}=-\ell(\ell+2)Y_{\ell,\mathbf m}$,\footnote{The scalar harmonics on $\mathbb S^3$ can be written as
\begin{equation}
Y_{\ell,\mathbf m}=\mathcal N_{\ell,\mathbf m}\text{e}^{\iu m_2 \phi}\,
C_{\ell - m_1}^{(m_1 + 1)}(\cos\theta_1)\,
P_{m_1}^{m_2}(\cos\theta_2)\,
\sin^{m_1}\theta_1\,,
\end{equation}
where $C_n^{(\alpha)}(x)$ denotes the Gegenbauer (ultraspherical) polynomial of degree $n$, parameter $\alpha$ and $P_{\ell}^{m}(x)$ are the associated Legendre polynomials, and the normalization constant reads
\begin{equation}
\mathcal N_{\ell,\mathbf m}=\sqrt{\frac{2^{2 m_1 - 1} (\ell + 1) (\ell - m_1)! (2 m_1 + 1) (m_1!)^2 (m_1 - m_2)!}{\pi^2 \Gamma(\ell + m_1 + 2) (m_1 + m_2)!}}\,.
\end{equation}
} with $\Delta$ is the Laplace–Beltrami operator on $\mathbb S^{3}$. The coefficients $a_{\ell,\mathbf m}$ parameterize the amplitude of the deformation. In this setup, computing MI following~\eqref{eq:EMId_Info} is considerably more challenging that EE.\footnote{In general dimension $d$, the EE in the EMI model is given by
\begin{equation}\label{eq:EMId}
    \qt{S_\delta(A)}{EMI}=\kappa\int_{\partial A}\diff^{d-2}\mathbf r_1\int_{\partial A}\diff^{d-2}\mathbf r_2\frac{\mathbf n_1\cdot\mathbf n_2}{\left|\mathbf r_1-\mathbf r_2\right|^{2(d-2)}}\,,
\end{equation}
while the MI between two regions $A_1$ and $A_2$ which are $\varepsilon/2$ close to $A$, $I_\varepsilon(A_1,A_2)$, is given by particularizing this setup as in~\eqref{eq:EMId_Info}.}  Because of this, we consider the latter quantity and employ expression~\eqref{eq:EMId}. In doing so, we focus on extracting the universal contribution to the finite piece coming from the deformation.

The position vectors $\mathbf{r}_i$ are given by
\begin{equation}
    \mathbf{r}_i = r(\mathbf{\Omega}_i)\,\mathbf{\Omega}_i \,,
\end{equation}
where $\mathbf{\Omega}_i = (\cos\psi_i,\sin\psi_i \cos\theta_i ,\sin\psi_i \sin\theta_i \cos\phi_i,\sin\psi_i \sin\theta_i \sin\phi_i)$ as the unit vector on $\mathbb S^3$. The normal vector reads $\mathbf n_i=\left(r_i\,\mathbf\Omega_{i}-\nabla_ir_i\right)/\sqrt{r_i^2+(\nabla_ir_i)^2}$, where $\nabla_i$ is the gradient in spherical coordinates. Finally, the surface element is given by $\diff^3\mathbf r=r^{2}_i\sqrt{r_i^2+(\nabla_ir_i)^2}\diff\Omega_{i}$. With all these ingredients, we obtain
\begin{align}
    \mathbf n_1\cdot\mathbf n_2\,\diff^3\mathbf r_1\diff^3\mathbf r_2&=\left(r_1r_2\cos\omega+\nabla_1r_1\cdot \nabla_2r_2-r_1\mathbf\Omega_1\cdot\nabla_2r_2-r_2\mathbf\Omega_2\cdot\nabla_1r_1\right)\diff\Omega_1\diff\Omega_2\,,\\
    \left|\mathbf r_1-\mathbf r_2\right|^{6}&=\left(r_1^2+r_2^2-2r_1r_2\cos\omega\right)\,,
\end{align}
where $\omega$ is the geodesic angle between two points $\mathbf\Omega_1\cdot\mathbf\Omega_2=\cos\omega$. Now, particularizing~\eqref{eq:pertS3} for these two expressions we obtain, at quadratic order in the perturbation parameter,
\begin{align}
  \mathbf n_1\cdot\mathbf n_2\,\diff^3\mathbf r_1\diff^3\mathbf r_2&=R^2\left(\cos\omega+\epsilon\left[\cos\omega\left(\Phi_1+\Phi_2\right)-\mathbf\Omega_1\cdot\nabla_2\Phi_2-\mathbf\Omega_2\cdot\nabla_1r_1\right]\right.  \nonumber\\
  &\left.+\epsilon^2\left[\Phi_1\Phi_2\cos\omega+\nabla_1\Phi_1\cdot\nabla_2\Phi_2-\Phi_1\mathbf\Omega_1\cdot\nabla_2\Phi_2-\Phi_2\mathbf\Omega_2\cdot\nabla_1\Phi_1\right]\right)\,,\\
  \left|\mathbf r_1-\mathbf r_2\right|^{6}&=16R^6\left(4\sin^6(\omega/2)+12\epsilon\sin^6(\omega/2)\left(\Phi_1+\Phi_2\right)\right.\nonumber\\
  &\left.+3\epsilon^2\sin^4(\omega/2)\left[3\Phi_1^2+3\Phi_2^2-4\Phi_1\Phi_2+2\cos\omega\left(\Phi_1+\Phi_2\right)\right]\right)\,.
\end{align}
Proceeding with the EE calculation and expanding in $\epsilon$, we isolate---after some manipulations---the first non-vanishing perturbative contribution to the finite part of the EE. This contribution appears at quadratic order and reads
\begin{equation}
 \qt{S_\delta\left(\partial A=\mathbb S_\epsilon^3\right)}{EMI}\supset\qt{C_T}{EMI}\frac{\epsilon^2\pi^4}{2160} \sum_{\ell,\mathbf m} a_{\ell,\mathbf m }^2 
\prod_{k =1, \dots, 5} ( \ell +k -2)\,.
\end{equation}
In deriving this expression, we have been able to factor out the stress–tensor two-point function coefficient using the general CFT result of Mezei~\cite{Mezei:2014zla,Faulkner:2015csl}. In particular, we find
\begin{equation}\label{eq:cTEMI5}
\qt{C_T}{EMI}=\frac{144}{\pi^2}\kappa\,,\qquad \text{for }d=5\,.    
\end{equation}

\subsection{Seven dimensions}
Let us move now to $d=7$, which will allow us to make a few general comments about the sign of $F(A)$ in that case. We will consider four different cylindrical entangling surfaces.
\subsubsection{Cylinder $\partial A=\mathbb{S}^1\times \mathbb{R}^4$}
The first geometry is given by $\partial A=\mathbb{S}^1\times \mathbb{R}^4$. We introduce the coordinates $(r, \phi, z,w,v,u)$, so the concentric cylinders of radii $R_1$ and $R_2$ can be defined as
\begin{equation}
    A_i=\left\{r\in [0,R_i], \phi_i\in [0,2\pi), z_i,w_i,v_i,u_i\in \left[-L/2,L/2\right] \right\}\,, \quad \text{for }i=1,2\,.
\end{equation}
As before, $R_1$ and $R_2$ are defined in~\eqref{eq:R12eps} and we take  $\textbf{r}_2 = (R_2, 0,0,0,0,0)$ together with
$\textbf{r}_1 = (R_1 \cos\phi_1, R_1 \sin\phi_1,z_1, w_1, v_1, u_1)$. then the corresponding normal vectors are
$\mathbf{n}_2 = (1, 0,0,0,0,0)$ and
$\mathbf{n}_1= (\cos\phi_1,\sin\phi_1,0, 0, 0, 0). $
Therefore $\mathbf{n}_1 \cdot \mathbf{n}_2 = \cos \phi_1$ and $|\mathbf{r}_1 - \mathbf{r}_2|^2 = \textbf{y}_1^2+ R_1^2 + R_2^2 - 2 R_1 R_2 \cos\phi_1$, where $\textbf{y}_1^2=z_1^2+w_1^2+v_1^2+u_1^2$, then  with $d^5\textbf{r}_i = R_i \, \diff \phi_i \, \diff z_i \, \diff w_i\, \diff v_i \,\diff u_i$  we can calculate the MI using the EMI model, thus
\begin{equation}
  \qt{I_{\varepsilon}(\partial A=\mathbb{S}^1\times \mathbb{R}^4)}{EMI}=
  \int_{-\frac{L}{2}}^{\frac{L}{2}}   \int_{-\frac{L}{2}}^{\frac{L}{2}}   \int_{-\frac{L}{2}}^{\frac{L}{2}}   \int_{-\frac{L}{2}}^{\frac{L}{2}}   \int_{0}^{2 \pi } 
  \frac{4\pi R_2  R_1 \kappa L^4\cos\phi\,\diff\phi \,\diff z\, \diff w\, \diff v\, du}{\left( R_1^2 + R_2^2 - 2 R_1 R_2 \cos\phi + \textbf{y}^2 \right)^5} \, 
\end{equation}
where $\textbf{y}^2=z^2+w^2+v^2+u^2$. We integrate the variables $z,w,v,u$ and take the limit $L\to \infty$. Then we integrate the angular variables, obtaining
\begin{equation}
    \qt{I_{\varepsilon}(\partial A=\mathbb{S}^1\times \mathbb{R}^4)}{EMI} = \frac{2\pi^4 \kappa L^4 R_1^2 R_2^2 (R_1^2 + R_2^2)}{(R_2^2 - R_1^2)^5}\, .
\end{equation}
Replacing $R_1$ and $R_2$ with their expressions as a function of $\varepsilon$  and taking the limit $\varepsilon\to 0$  we get
\begin{equation}\label{s1r4}
 \qt{I_{\varepsilon}(\partial A=\mathbb{S}^1\times \mathbb{R}^4)}{EMI}= \pi^4 \kappa L^4 \left[ \frac{R}{8 \varepsilon^5} - \frac{1}{32 R \varepsilon^3} - \frac{1}{128 R^3 \varepsilon} \right]+\mathcal{O}(\varepsilon)\, .
\end{equation}

\subsubsection{Cylinder $\partial A=\mathbb{S}^2\times \mathbb{R}^3$} 
For our next cylinder we have $\partial A=\mathbb{S}^2\times \mathbb{R}^3$. Using  coordinates $(r, \theta, \phi,z,w,v)$, the two concentric cylinders of radii $R_1$ and $R_2$ are defined as
\begin{equation}
    A_i=\left\{r_i\in[0,R_i], \theta_i\in[0, \pi], \phi_i\in[0, 2\pi), z_i,w_i,v_i\in\left[-L/2,L/2\right]  \right\}\,, \quad \text{for }i=1,2\,,
\end{equation}
for $i=1,2$ and with $R_1$ and $R_2$ given in~\eqref{eq:R12eps}. The positions vectors can be written as $\textbf{r}_1 =(R_1 \cos \theta_1,R_1 \sin \theta_1 \cos \phi_1, R_1 \sin \theta_1 \sin \phi_1, z_1, w_1, v_1)$ and $\textbf{r}_2 =(R_2, 0, 0, 0,0,0)$. Thus, we have $\mathbf{n}_2= (1, 0, 0, 0,0,0)$ and $\mathbf{n}_1=(\cos \theta_1,\sin \theta_1 \cos \phi_1, \sin \theta_1 \sin \phi_1, 0, 0, 0)$ and $\mathbf{n}_1 \cdot \mathbf{n}_2 = \cos \theta_1$, $ \lvert \textbf{r}_1-\textbf{r}_2\lvert^2 =R_1^2 + R_2^2 - 2R_1 R_2 \cos \theta_1 + z_1^2 + w_1^2 + v_1^2$. Now we can use equation~\eqref{eq:EMId_Info} for the computation of MI. Employing $\diff ^5\mathbf{r}_i = R_i^2 \sin\theta_i\, \diff\theta_i \,\diff \phi_i \,\diff z_i\,\diff w_i\,\diff v_i$, we obtain
\begin{equation}\label{mutual7cy:1}
  \qt{I_\varepsilon(\partial A=\mathbb{S}^2\times \mathbb{R}^3)}{EMI} =\int_{-\frac{L}{2}}^{\frac{L}{2}}   \int_{-\frac{L}{2}}^{\frac{L}{2}}   \int_{-\frac{L}{2}}^{\frac{L}{2}}  \int_{0}^{ \pi }  
  \frac{16\pi^2 R_1^2R_2^2 L^3\kappa\sin\theta \cos\theta\,\diff\theta\, \diff z\,  \diff w\, \diff v}{\left( R_1^2 + R_2^2 - 2 R_1 R_2 \cos\theta+ z^2 + w^2 + v^2 \right)^5} \,.
\end{equation}
We now integrate and take the limit $L\to \infty$ in the integral and find
\begin{equation}
   \qt{I_\varepsilon(\partial A=\mathbb{S}^2\times \mathbb{R}^3)}{EMI} = \frac{5\pi^4R_1^2R_2^2 L^3\kappa }{4} \int_{0}^{\pi} \frac{\sin\theta \cos\theta\,\diff \theta}{\left( R_1^2 + R_2^2 - 2 R_1 R_2 \cos\theta \right)^{7/2}}\,,
\end{equation}
which reduces to
\begin{equation}
\qt{I_\varepsilon(\partial A=\mathbb{S}^2\times \mathbb{R}^3)}{EMI} = \frac{\pi^4  \kappa L^3 R_1^3 (35 R_2^4 + 14 R_1^2 R_2^2 - R_1^4)}{6 (R_2^2 - R_1^2)^5}\,.
\end{equation}
Finally, replacing $R_1$ and $R_2$ given in~\eqref{eq:R12eps} we get
\begin{equation}\label{s2r3}
 \qt{I_\varepsilon(\partial A=\mathbb{S}^2\times \mathbb{R}^3)}{EMI}=\frac{\pi^4 \kappa L^3}{12} \left( \frac{3R^2}{\varepsilon^5} - \frac{7}{4\varepsilon^3} + \frac{7}{32R^3}\right)+\mathcal O(\varepsilon)\,.
\end{equation}

\subsubsection{Cylinder $\partial A=\mathbb{S}^3\times \mathbb{R}^2$}
We move now to a entangling surface defined by $\partial A=\mathbb{S}^3\times \mathbb{R}^2$. We use the coordinates $(r, \psi, \theta,\phi,z,w)$. Then the concentric cylinders  of radii $R_1$ and $R_2$ are respectively defined as
\begin{equation}
    A_i=\left\{r_i\in[0,R_i],\psi_i,\theta_i\in [0,\pi],  \phi_i\in [0,2\pi), z_{i},w_{i}\in\left[-L/2,L/2\right]  \right\} \,, \quad \text{for }i=1,2\,,
\end{equation}
where $R_1$ and $R_2$ are presented in~\eqref{eq:R12eps}. Due to the symmetries involved, the vectors that localize the surfaces of the cylinders can be fixed to  $\textbf{r}_2 = (R_2, 0, 0, 0,0, 0)$  and
$\textbf{r}_1 = (R_1 \mathbf \Omega_1,z_1, w_1)$, where here, $\mathbf\Omega_1=(\cos\psi_1,\sin\psi_1 \cos\theta_1 ,\sin\psi_1 \sin\theta_1 \cos\phi_1,\sin\psi_1 \sin\theta_1 \sin\phi_1)$. The corresponding normal unitary vectors are $\mathbf{n}_2 = (1, 0, 0, 0,0,0)$, and $\mathbf{n}_1=(\mathbf \Omega_1,0,0)$, hence $\mathbf{n}_1 \cdot \mathbf{n}_2 = \cos \psi_1$ and $|\mathbf{r}_1 - \mathbf{r}_2|^2 = z_1^2 + w_1^2 + R_1^2 + R_2^2 - 2 R_1 R_2 \cos\psi_1$. With $
d^5\textbf{r}_i = R_i^3 \sin^2\psi_i \sin\theta_i \, \diff\psi_i \, \diff\theta_i \, \diff\phi_i \, \diff z_i \, \diff w_i$ the MI in the EMI model can be calculated, and one finds
\begin{equation}
   \qt{I_{\varepsilon}(\partial A=\mathbb{S}^3\times \mathbb{R}^2)}{EMI} = 
       \int_{-\frac{L}{2}}^{\frac{L}{2}}   \int_{-\frac{L}{2}}^{\frac{L}{2}}  \int_{0}^{ \pi } \int_{0}^{ \pi }  
   \frac{ 8\pi^3\kappa R_1^3 R_2^3 L^2 \cos\psi \sin^2\psi \sin\theta \, \diff\psi \, \diff\theta  \, \diff z \, \diff w}{\left( R_1^2 + R_2^2 - 2 R_1 R_2 \cos\psi + z^2 + w^2 \right)^5} 
\end{equation}
After integrating and taking the limit $L\to \infty$  we get
\begin{equation}\label{s3r2}
\qt{I_{\varepsilon}(\partial A=\mathbb{S}^3\times \mathbb{R}^2)}{EMI} = \frac{4\pi^5 \kappa L^2 R_1^4 R_2^4}{(R_2^2 - R_1^2)^5}.
\end{equation}
Writing $R_1$ and $R_2$ as in~\eqref{eq:R12eps} and in the limit $\varepsilon\to 0$ the previous expression can be written as
\begin{equation}
 I_{\varepsilon}(\partial A=\mathbb{S}^3\times \mathbb{R}^2)\rvert_{EMI}  =\pi^5 \kappa L^2 \left( \frac{R^3}{8 \varepsilon^5} - \frac{R}{8\varepsilon^3} + \frac{3}{64 R \varepsilon} \right)+\mathcal{O}(\varepsilon)\, .
\end{equation}

\subsubsection{Cylinder $\partial A=\mathbb{S}^4\times \mathbb{R}^1$}
Finally, we consider a cylinder with entangling surface $\partial A=\mathbb{S}^4\times \mathbb{R}^1$. We use the coordinates $(r, \xi, \psi,\theta,\phi,z)$, hence the two concentric cylinders of radius $R_1$ and $R_2$ are defined
\begin{equation}
 A_i=\left\{r_i\in[0,R_i],\xi_i, \psi_i,\theta_i\in [0,\pi],\phi_i\in [0,2\pi),z_i\in\left[-L/2,L/2\right]\right\}\,, \quad \text{for }i=1,2\,, 
\end{equation}
with again $R_1$ and $R_2$ given in~\eqref{eq:R12eps}. Due to the symmetries involved, the vectors that localize the surfaces of the cylinders can be fixed to  $\textbf{r}_2 = (R_2, 0, 0, 0, 0,0)$  and $\textbf{r}_1 = (R_1\mathbf\Omega_1,z)$ where $\mathbf\Omega_1=(\cos\xi_1, \sin\xi_1 \cos\psi_1, \sin\xi_1 \sin\psi_1 \cos\theta_1, \sin\xi_1 \sin\psi_1 \sin\theta_1 \cos\phi_1,
\sin\xi_1 \sin\psi_1 \sin\theta_1 \sin\phi_1)$. Thus, the unit normal vectors can be written $
\mathbf{n}_2 =(1, 0, 0, 0, 0,0)$ and $\mathbf{n}_1=(\mathbf \Omega_1,0)$. Therefore $\mathbf{n}_1 \cdot \mathbf{n}_2 = \cos\xi_1$ and $|\mathbf{r}_1 - \mathbf{r}_2|^2 = z_1^2 + R_1^2 + R_2^2 - 2 R_1 R_2 \cos\xi_1$. Using the integration measures  $
  d^5\mathbf{r}_i = R_i^4 \sin^3\xi_i \sin^2\psi_i \sin\theta_i \, \diff\xi_i \, \diff\psi_i \, \diff\theta_i \, \diff\phi_i \, \diff z_i$ now we can proceed to calculate the mutual information using equation~\eqref{eq:EMId_Info}, thus
 \begin{equation}
 \qt{I_{\varepsilon}(\partial A=\mathbb{S}^4\times \mathbb{R}^1)}{EMI} = 
        \int_{-\frac{L}{2}}^{\frac{L}{2}}       \int_{0}^{ \pi } \int_{0}^{ \pi }  \int_{0}^{ \pi } 
 \frac{32\pi^3  R_2^4 R_1^4 L\kappa\cos\xi \sin^3\xi \sin^2\psi \sin\theta\, \diff\xi \, \diff\psi \, \diff\theta \, dz}{3\left( R_1^2 + R_2^2 - 2 R_1 R_2 \cos\xi + z^2 \right)^5}\, .
\end{equation}
In the limit $L\to \infty$ we get, after integration, the following expression 
\begin{equation}
    \qt{I_{\varepsilon}(\partial A=\mathbb{S}^4\times \mathbb{R}^1)}{EMI}  = \frac{\pi^5 \kappa L R_1^5 (R_1^4 - 6 R_1^2 R_2^2 + 21 R_2^4)}{3 (R_2^2 - R_1^2)^5}
   \end{equation}
As usual, we replace $R_1$ and $R_2$ by the expressions given in~\eqref{eq:R12eps} and 
in the limit $\varepsilon\to 0$  we finally arrive to 
\begin{equation}\label{s4r1}
 \qt{I_{\varepsilon}(\partial A=\mathbb{S}^4\times \mathbb{R}^1)}{EMI}  = \frac{L \pi^5 \kappa}{2} \left[ \frac{R^4}{3 \varepsilon^5} - \frac{R^2}{2 \varepsilon^3} + \frac{7}{16 \varepsilon} - \frac{7}{32 R} \right] + \mathcal{O}(\varepsilon)\, .
\end{equation}

\section{Universal quantities across theories and dimensions}\label{sec:appuq}

In this appendix, we compile the values of the universal quantities that appear in the entropic conformal bounds for different theories and dimensions, as reported in the literature. These are:
\begin{itemize}
\item The coefficient $\cT$, which controls the two-point, stress-energy tensor correlator, which in arbitrary dimension reads
\begin{equation}\label{eq:CT}
\left\langle T_{a b}(x) T_{c d}(0)\right\rangle=\frac{C_T}{x^{2 d}} \frac{1}{2}\left[I_{a c}(x) I_{b d}(x)+I_{a d}(x) I_{b c}(x)\right]-\frac{1}{d} \eta_{a b} \eta_{c d}\, ,
\end{equation}
where $I_{a b}(x)\equiv\eta_{a b}-2 x_a x_b/x^2$~\cite{Osborn:1993cr}.
\item The free energy $F_0$ of the $d$-dimensional CFT when the theory is placed on a sphere $\mathbb S^d$, \ie~\cite{Dowker:2010bu,Casini:2011kv}
\begin{equation}\label{eq:CHMfe}
    F_0=-\log Z_{\mathbb S^d}\,.
\end{equation}
\item The strip---or more generally, the (hyper-)slab in $d>3$---coefficient $k$, appearing in the expansion for strip-like entangling regions~\cite{Casini:2005zv}. In terms of the regulated EE, one has
\begin{equation}\label{stripi}
    S_\text{reg}(\text{strip},\varepsilon)=k
   \left[\frac{\area_{\parallel}}{\varepsilon^{d-2}}   -\frac{\area_{\parallel}}{\ell_\perp^{d-2}} \right] \, ,
\end{equation}
where $\area_\parallel=\ell_\parallel^{d-2}$ is the infrared-regulated area of the extended directions, transverse to $\ell_\perp$.

\item The type A and B central charges in four-dimensional CFTs, $a$ and $c$, respectively, appearing in the stress-tensor trace anomaly~\cite{Deser:1993yx,Duff:1977ay}
\begin{equation}\label{eq:traceanom4}
    \langle T_a{}^a\rangle=-\frac{a}{16\pi^2}\mathcal X_4+\frac{c}{16\pi^2}W_{abcd}W^{abcd}\,,\quad \forall\text{CFT}_{4}\,,
\end{equation}
where $\mathcal X_4=R^2-4R_{ab}R^{ab}+R_{abcd}R^{abcd}$ is the Euler density in four dimensions---or Gauss-Bonnet density---and $W_{abcd}$ is the Weyl tensor. Both charges, $a$ and $c$, appear in the EE functional of general entangling regions.
\end{itemize}

\begin{table}
\centering
\begin{minipage}[t]{0.48\textwidth}
\centering
\begin{tabular}[t]{|l||c|c|}
\hline
        Theory& $C_T/F_0$ & $k/F_0$\\\hline\hline
        Free scalar &$0.149$&$0.622$\\
        $O(N)$ &$(0.141,0.149)$&\\
        $Z^4$&0.110& \\
        WZ $XZ^2(N)$&$(0.0950,0.110)$&\\
        SQED&$0.0950$&\\
        GN &$(0.0844,0.0940)$&\\
        Free fermion &$0.0868$&$0.330$\\
        EMI model &0.0821&0.318\\
        Holography &$0.0616$&$0.228$\\
        ABJM &$(0,0.110)$& \\
        ADHM &$(0.0616,0.0769)$& \\
        Maxwell &$0$& $0$\\\hline
    \end{tabular}
    \end{minipage}
\begin{minipage}[t]{0.40\textwidth}
\centering
    \begin{tabular}[t]{|l||c|c|}
    \hline
        Theory& $c/a$ & $k/a$\\\hline\hline
        Free scalar & 3.00 & $1.99$ \\
        Free fermion & 1.64 & 0.703 \\
        EMI model & 1.50 & 0.637\\
        Holography & 1.00 & 0.204 \\
        Maxwell & 0.581& 0.0645 \\\hline
    \end{tabular}
    \end{minipage}
    \caption{Ratios $C_T/F_0$ and $k/F_0$ for the corresponding three-dimensional CFTs, as well as $a/c$ and $k/a$ for four-dimensional theories. When the quotient takes values within a certain interval, we indicate this using parentheses.}
    \label{tab:CTF034}
\end{table}
\begin{table}
    \centering
    \begin{tabular}{|l||c|c|}
    \hline
        Theory & $C_T/F_0$ & $k/F_0$ \\\hline\hline
        Free scalar     & 0.314&0.228 \\
        Free fermion    &$0.167$ & 0.0601\\
        EMI model       & 0.150 &0.0507\\
        $E_n$ $(n=1,\ldots,8)$  &$(0.0923,0.115)$& \\
        $\tilde E_1$  & $0.0944$ & \\
        $T_N$  & $(0.0936,0.115)$ & \\
        $\#_{N,M}$  & $(0.0906,0.113)$  & \\
        Holography& $0.0936$ &$5.25\cdot10^{-3}$\\ \hline
    \end{tabular}
    \caption{Ratios $C_T/F_0$ and $k/F_0$ for five-dimensional CFTs. Similarly as in Table~\ref{tab:CTF034}, we use parathenteses notation for intervals.}
    \label{tab:CTF0}
\end{table}

\subsection{Free theories}
For free field theories, the coefficient of the two-point function of the stress-energy tensor has been computed in general dimension $d$. For a theory with $n_\phi$ real scalar fields and $n_\psi$ Dirac fermions, one finds~\cite{Osborn:1993cr}
\begin{equation}\label{eq:CTfreefields}
\qt{C_T}{free scalar} = n_\phi\frac{d}{d-1}\frac{1}{\Omega_{d-1}^{2}}\,,\qquad
\qt{C_T}{free fermion} = n_\psi\frac{d}{2}\frac{\text{tr}\,\mathbf 1}{\Omega_{d-1}^{2}}\,,
\end{equation}
where $\Omega_d$ is given in~\eqref{eq:Sd} and $\text{tr}\,\mathbf 1=2^{\lfloor d/2\rfloor}$ is the is the trace of the identity in the Dirac matrices space. We evaluate these expressions for a single field in each case, obtaining
\begin{alignat}{4}
    \qt{C_T}{free scalar}&=\frac{3}{32\pi^2}\,, \qquad &&\qt{C_T}{free fermion}=\frac{3}{16\pi^2}\, ,\qquad &&& \text{for }d=3\,,\\
    \qt{C_T}{free scalar}&=\frac{45}{256 \pi ^4}\,, \qquad && \qt{C_T}{free fermion}=\frac{45}{32 \pi ^4}\, ,\quad &&& \text{for }d=5\,.
\end{alignat}

On the other hand, the free energy of a conformally coupled free scalar field on the round sphere $\mathbb S^d$ in odd dimensions is given by
\begin{equation}
\label{eq:F0_freescalar}
    \qt{F_0}{free scalar}=\frac{1}{\Gamma(d+1)}\int_0^1\diff u\, u\sin(\pi u)\Gamma\left(\frac{d}{2}+u\right)\Gamma\left(\frac{d}{2}-u\right)\,.
\end{equation}
From this expression, we see that particularizing dimensions we obtain
 \begin{alignat}{3}
    \qt{F_0}{free scalar} &=\frac{1}{16}\left(2\log2-\frac{3\zeta(3)}{\pi^2}\right)\, ,\qquad && \text{for }d=3\, ,\label{eq:F0fscal3}\\
    \qt{F_0}{free scalar} &=\frac{1}{256}\left(2\log2+\frac{2\zeta(3)}{\pi^2}-\frac{15\zeta(5)}{\pi^4}\right)\, ,\qquad && \text{for }d=5\,.
\end{alignat}   
In the case of a conformally coupled free fermion field, the sphere free energy in odd dimensions is given by~\cite{Giombi:2014xxa}
\begin{equation}\label{eq:F0ferd}
\qt{F_0}{free fermion}=\frac{\text{tr}\,\mathbf 1}{\Gamma(d+1)}\int_0^1\diff u\, \cos\left(\frac{\pi u}{ 2}\right)\Gamma\left(\frac{d+1+u}{2}\right)\Gamma\left(\frac{d+1-u}{2}\right)\,.
\end{equation}
Again, for the dimensions of our interest this gives
 \begin{alignat}{3}
\qt{F_0}{free fermion} &=\frac{1}{8}\left(2\log2+\frac{3\zeta(3)}{\pi^2}\right)\,,\qquad &&\text{for } d=3\,,\\
\qt{F_0}{free fermion} &=\frac{1}{64}\left(6\log2+\frac{10\zeta(3)}{\pi^2}+\frac{15\zeta(5)}{\pi^4}\right)\,,\qquad &&\text{for } d=5\,.
\end{alignat}

Moving now to four dimensions, the values of the central charges $c$ and $a$ controlling the trace anomaly~\eqref{eq:traceanom4} for a free scalar field are~\cite{Dowker:1976zf,Duff:1977ay}
\begin{equation}
    \qt{c}{free scalar}=\frac{1}{120}\,, \qquad \qt{a}{free scalar}=\frac{1}{360}\,,
\end{equation}
respectively, whereas for a free fermion field, they are given by 
\begin{equation}
    \qt{c}{free fermion}=\frac{1}{20}\,, \qquad \qt{a}{free fermion}=\frac{11}{360}\,.
\end{equation}
Finally, in these theories, the (numerical) strip coefficient reads, in each dimension,~\cite{Casini:2009sr}
\begin{alignat}{6}
&\qt{k}{free scalar}\approx 0.0397\,, &\qquad & \qt{k}{free fermion}\approx 0.0722\,,&\qquad& \text{for }d=3\,,\\
&\qt{k}{free scalar}\approx  5.54\cdot10^{-3}\,, & \qquad & \qt{k}{free fermion}\approx 0.0215\,,&\qquad &\text{for }d=4\,,\\
&\qt{k}{free scalar}\approx1.31\cdot10^{-3}\,,&\qquad & \qt{k}{free fermion}\approx 5.20\cdot10^{-3}\,,&\qquad & \text{for }d=5\,.
\end{alignat}
Combining all these results, we find the ratios~\cite{Bueno:2023gey}
\begin{alignat}{8}
&\qt{\frac{C_T}{F_0}}{free scalar}&=&\frac{3}{ 4\pi^2\log 2- 6\zeta(3)}\approx 0.149\,,&\quad& \qt{\frac{k}{F_0}}{free scalar}\approx0.622\,,\quad &\text{for }d=3\,.\\
&\qt{\frac{C_T}{F_0}}{free fermion}&=&\frac{3}{ 4\pi^2\log 2+6\zeta(3)}\approx 0.0868\,,&\quad& \qt{\frac{k}{F_0}}{free fermion}\approx0.330\,,\quad &\text{for }d=3\,.
\end{alignat}
On the other hand, in four-dimensions we find~\cite{Hofman:2008ar}
\begin{alignat}{5}
&\qt{\frac{c}{a}}{free scalar}&&=3\, , \qquad &&\qt{\frac{k}{a}}{free scalar}&&\approx1.99\,,\quad &\text{for }d=4\,,\\
&\qt{\frac{c}{a}}{free fermion}&&=\frac{18}{11}\approx1.64\,,\qquad &&\qt{\frac{k}{a}}{free fermion}&&\approx0.703\,,\quad &\text{for }d=4\,,
\end{alignat}
Finally, in five-dimensional CFTs, the ratios take the value
\begin{alignat}{4}
&\qt{\frac{C_T}{F_0}}{free scalar}&=&\ \frac{45}{2 \pi ^2 \zeta (3)-15 \zeta (5)+\pi ^4 \log 4}&\approx 0.314\, , &\quad &\quad \text{for }d=5\,,\\
&\qt{\frac{C_T}{F_0}}{free fermion}&=&\ \frac{90}{10 \pi ^2 \zeta (3)+15 \zeta (5)+\pi ^4 \log 64}&\approx 0.167\,,&\quad &\quad \text{for }d=5\,,\\
&\qt{\frac{k}{F_0}}{free scalar}&\approx&\ 0.228\,,&&&\quad \text{for }d=5\,,\\
&\qt{\frac{k}{F_0}}{free fermion}&\approx&\ 0.0601\,,&&&\quad \text{for }d=5\,.
\end{alignat}

As an additional remark, notice that the expressions for $C_T$ in~\eqref{eq:CTfreefields} and for $F_0$ in~\eqref{eq:F0_freescalar} and~\eqref{eq:F0ferd} old for any odd $d$ and therefore can be considered in the limit $d \to \infty$: 
\begin{alignat}{5}
    &\qt{C_T}{free scalar}&\overset{d\rightarrow\infty}{\approx}&\frac{\pi}{d}\left[\frac{d}{2\pi\text e}\right]^d\,, \qquad  &\qt{C_T}{free fermion}&\overset{d\rightarrow\infty}{\approx}\frac{\pi}{2}\left[\frac{d}{\sqrt{2}\pi\text e}\right]^d\,,\\
    &\qt{F_0}{free scalar}&\overset{d\rightarrow\infty}{\approx}&\sqrt{\frac{8}{\pi d^3}}\frac{1}{2^d}\,, \qquad &\qt{F_0}{free fermion}&\overset{d\rightarrow\infty}{\approx}\sqrt{\frac{8}{\pi d}}\frac{1}{2^d}\,.
    \end{alignat}
The expressions for $k$ in the $d \to \infty$ limit are given in~\cite{Casini:2009sr}
\begin{alignat}{5}
    &\qt{k}{free scalar}&\overset{d\rightarrow\infty}{\approx}&
    \sqrt{\frac{\pi^3}{d^3}}\left[\frac{d}{8\pi\text e}\right]^{d/2}\, , \qquad &\quad\qt{k}{free fermion}&\overset{d\rightarrow\infty}{\approx}
    \sqrt{\frac{\pi^3}{d^3}}\left[\frac{d}{4\pi\text e}\right]^{d/2}\
    \end{alignat}
From these limits, it is straightforward to obtain the following asymptotic ratios
\begin{alignat}{5}
      &\qt{\frac{C_T}{F_0}}{free scalar}&\overset{d\rightarrow\infty}{\approx}&\sqrt{\frac{\pi^3d}{8}}\left[\frac{d}{\pi\text e}\right]^d\, , \qquad &\qt{\frac{C_T}{F_0}}{free fermion}&\overset{d\rightarrow\infty}{\approx}\sqrt{\frac{\pi^3d}{32}}\left[\frac{d}{\pi\text e}\right]^d\,, \label{eq:CTF0ratio_free} \\
   &\qt{\frac{k}{F_0}}{free scalar}&\overset{d\rightarrow\infty}{\approx}& \frac{\pi^2}{\sqrt{8}}\left[\frac{d}{2\pi \text e} \right]^{d/2}\, , \qquad &\qt{\frac{k}{F_0}}{free fermion}&\overset{d\rightarrow\infty}{\approx} \frac{\pi^2}{\sqrt{8}d}\left[\frac{d}{2\pi \text e} \right]^{d/2}\,, \label{eq:kF0ratio_free}
\end{alignat}
which imply
\begin{equation}
    \frac{\qt{C_T/F_0}{free fermion}}{\qt{C_T/F_0}{free scalar}}\overset{d\rightarrow\infty}{\approx} \frac{1}{2}\, ,\qquad \quad 
      \frac{\qt{k/F_0}{free fermion}}{\qt{k/F_0}{free scalar}}\overset{d\rightarrow\infty}{\approx}  \frac{1}{d}\, .
\end{equation}

\subsection{Extensive mutual information model}
For the EMI model, the coefficient of the stress–energy tensor two-point function is not known in arbitrary dimension $d$; it has only been computed in $d=3$~\cite{Bueno:2021fxb} and in $d=5$ in~\eqref{eq:cTEMI5},\footnote{$C_T$ for the EMI model is also known in $d=4$, as this quantity is related to the type B central charge $C_T=40/\pi^4c$ in general CFT---see~\cite{Osborn:1993cr} or, more recently,~\cite{Aros:2026gms}. The latter is given below in~\eqref{eq:EMI4c}.} this is, 
\begin{alignat}{2}
\qt{C_T}{EMI} &=\frac{16}{\pi^2}\kappa\,,\qquad& \text{for }d=3\,,\\
\qt{C_T}{EMI} &=\frac{144}{ \pi ^2} \kappa\,, \qquad& \text{for }d=5\,,
\end{alignat}
On the other hand, the free energy of the sphere in the EMI model is known for all odd dimension $d$, and, as indicated in~\eqref{eq:MIEMIdSph}, it is related to $a^\star$ through~\cite{Agon:2021zvp}
\begin{equation}
\label{eq:F0_EMI}
\qt{F_0}{EMI} =\qt{2\pi a^\star}{EMI}= 
\frac{2 \pi^{d-1}}{\Gamma(d-2)} \kappa \,,\quad \text{for odd }d\,.    
\end{equation}
This means that in our dimensions of interest, we have
\begin{alignat}{3}
    \qt{F_0}{EMI} &= 2 \pi^2 \kappa \,,\qquad & \text{for } d=3\,,\\
   \qt{F_0}{EMI} &= \pi^4 \, \kappa\,,\qquad & \text{for } d=5\,.
\end{alignat}

Now, moving to four dimensions, the trace-anomaly charges take the values~\cite{Agon:2021zvp}
\begin{equation}\label{eq:EMI4c}
\qt{c}{EMI}=\frac{3 \pi^2}{2} \kappa\,, \quad \qt{a}{EMI}=\pi^2\kappa\,.
\end{equation}

Finally, the value of the strip coefficient, which is known for arbitrary dimension, is given by~\cite{Agon:2021zvp}
\begin{equation}
\label{eq:k_EMI}
\qt{k}{EMI} = 
\frac{ 2 \pi^{\frac{d-2}{2}} \Gamma[(d-2)/2]  }{\Gamma(d-2) } \kappa\,.
\end{equation}
Hence, it is straightforward to see that
\begin{alignat}{3}
\qt{k}{EMI} &=  2 \pi \kappa\,,  \qquad & \text{for } d=3\,, \\
\qt{k}{EMI} &= 2 \pi \kappa\,, \qquad &\text{for } d=4\,, \\
\qt{k}{EMI} &= \frac{\pi^2}{2} \kappa\,, \qquad& \text{for } d=5\,.
\end{alignat}

Using all these values we find the following quotients for the EMI model. First, in three dimensions,  
\begin{equation}
    \qt{\frac{\cT}{F_0}}{EMI}=\frac{8}{\pi ^4}\approx\, 0.0821,\qquad \qt{\frac{k}{F_0}}{EMI}= \frac{1}{\pi}\approx\, 0.318,\qquad \text{for }d=3\,,
\end{equation}
while in four dimensions we obtain
\begin{equation}
    \qt{\frac{c}{a}}{EMI} = \frac{3}{2} \approx 1.50 , \qquad \qt{\frac{k}{a}}{EMI}= \frac{2}{\pi} \approx 0.637 \,,\qquad \text{for }d=4\,,
\end{equation}
and finally, for five-dimensional EMI model we obtain
\begin{equation}
    \qt{\frac{\cT}{F_0}}{EMI}=\frac{144}{\pi ^6}\approx 0.150 \,,\qquad \qt{\frac{k}{F_0}}{EMI}=\frac{1}{2 \pi^2 }\approx \,0.0507,\qquad \text{for }d=5\,.
\end{equation}

Similarly as in the free fields case, the expression for $ \qt{k}{EMI}$ in~\eqref{eq:k_EMI} and that for $\qt{F_0}{EMI}$ in~\eqref{eq:F0_EMI} hold for any odd $d$ and in the $d \to \infty $ limit read
\begin{equation}
    \qt{k}{EMI}\overset{d\rightarrow\infty}{\approx}\frac{\sqrt{32}d}{\pi}\left(\frac{\pi\text e}{2d}\right)^{d/2}\kappa\,,\qquad \qt{F_0}{EMI}\overset{d\rightarrow\infty}{\approx}\sqrt{\frac{2d^5}{\pi^3}}\left(\frac{\pi \text e}{d}\right)^d\kappa\,,
\end{equation}
leading to the following asymptotic ratio
\begin{equation}
    \qt{\frac{k}{F_0}}{EMI}\overset{d\rightarrow\infty}{\approx}4\sqrt{\frac{\pi}{d^3}}\left(\frac{d}{2\pi\text e}\right)^{d/2}\,,
\end{equation}
Such asymptotic ratio can be compared to those in~\eqref{eq:kF0ratio_free} as follows:
\begin{equation}
      \frac{\qt{k/F_0}{EMI}}{\qt{k/F_0}{free scalar}}\overset{d\rightarrow\infty}{\approx}  \frac{32}{\pi d^2}\, .
      \qquad \quad 
      \frac{\qt{k/F_0}{EMI}}{\qt{k/F_0}{free fermion}}\overset{d\rightarrow\infty}{\approx}  \frac{32}{\pi d}\, .
\end{equation}

\subsection{Theories with a holographic Einstein dual}
For CFTs dual to Einstein gravity in the bulk, the universal quantities of interest have been computed in arbitrary dimensions. First, the value of the coefficient $C_T$ we have~\cite{Liu:1998bu}
\begin{equation}
\label{eq:CTholo}
    \qt{\cT}{holo}=\frac{\Gamma(d+2)}{8\pi^{(d+2)/2}(d-1)\Gamma(d/2)}\frac{\Ls^{d-1}}{G}\,,
\end{equation}
where $G$ is the Newton constant and $L_\star$ is the AdS$_{d+1}$ radius.
Hence, the values in $d=3$ and $d=5$ read
 \begin{alignat}{3}
    \qt{\cT}{holo} &=\frac{3}{\pi^3}\frac{\Ls^2}{G}\,,&\qquad&\text{for }  d=3\,,\label{eq:CTholo3}\\
    \qt{\cT}{holo} &=\frac{30}{\pi^4}\frac{\Ls^4}{G}\,,&\qquad&\text{for } d=5\,,\label{eq:CTholo5}
\end{alignat}
On the other hand, the free energy in odd dimensions is given by~\cite{Ryu:2006bv,Ryu:2006ef,Casini:2011kv}
\begin{equation}
\label{eq:F0holo}
    \qt{F_0}{holo}=2\pi \qt{a^\star}{holo}=\frac{\pi^{d/2}}{4\Gamma(d/2)}\frac{\Ls^{d-1}}{G}\,,\quad\text{for odd }d\,,
\end{equation}
thus
\begin{alignat}{2}
\qt{F_0}{holo} &= \frac{\pi}{2}\frac{\Ls^2}{G}
    \,, &\qquad& \text{for } d=3\,, \label{eq:F0holo3}\\
\qt{F_0}{holo} &= \frac{\pi^2}{3}\frac{\Ls^4}{G}
\,,& \qquad& \text{for } d=5\,,\label{eq:F0holo5}
\end{alignat}
For four-dimensional CFTs with an Einstein gravity dual, the conformal anomaly coefficients are known to coincide~\cite{Henningson:1998gx}
\begin{equation}
    \qt{a}{holo}=\qt{c}{holo}=\frac{\pi}{8}\frac{\Ls^3}{G}\approx0.393\frac{\Ls^3}{G}\,,
\end{equation}
which follows from the general expression for the type-A anomaly in arbitrary dimensions, 
\begin{equation}
\qt{a^\star}{holo}=\frac{\pi^{(d-2)/2}\Ls^{d-1}}{8\Gamma(d/2)G}\, .    
\end{equation}
In the case of the strip, the universal coefficient for holographic CFTs reads, in arbitrary dimensions,~\cite{Ryu:2006bv,Ryu:2006ef}
\begin{equation}
\label{eq:kholo}
\qt{k}{holo}=\frac{2^{d-3}\pi^{(d-1)/2}\Gamma[d/(2(d-1))]^{d-1}}{(d-2)\Gamma(1/(2(d-1)))^{d-1}}\frac{\Ls^{d-1}}{G}\,. 
\end{equation}
Thus, we can particularize the dimension and immediately find
\begin{alignat}{3}
\qt{k}{holo} &= \frac{\pi \Gamma \left(3/4\right)^2}{\Gamma \left(1/4\right)^2}\frac{\Ls^2}{G}\,, \qquad & \text{for } d=3\,, \\
\qt{k}{holo} &= \frac{\pi^{3/2} \Gamma \left(2/3\right)^3}{\Gamma \left(1/6\right)^3}\frac{\Ls^3}{G} \qquad & \text{for } d=4\,, \\
\qt{k}{holo} &= \frac{4\pi^2 \Gamma \left(5/8\right)^4}{3\Gamma \left(1/8\right)^4}\frac{\Ls^4}{G}\,, \qquad &\text{for } d=5\,.
\end{alignat}

Using these values, the conformal ratios for holographic three-dimensional CFTs become
\begin{equation}\label{eq:CTFO_holo}
    \qt{\frac{\cT}{F_0}}{holo}=\frac{6}{\pi ^4}\approx0.0616\,,\qquad \qt{\frac{k}{F_0}}{holo}=\frac{2 \Gamma \left(3/4\right)^2}{\Gamma \left(1/4\right)^2}\approx0.228\,,\qquad \text{for }d=3\,,
\end{equation}
while for four-dimensional theories one finds
\begin{equation}
    \qt{\frac{c}{a}}{holo}=1\,,\qquad \qt{\frac{k}{a}}{holo}=\frac{8 \sqrt{\pi } \Gamma \left(2/3\right)^3}{\Gamma \
\left(1/6\right)^3}\approx0.204\,,\qquad \text{for }d=4\,,
\end{equation}
and finally, for five-dimensional CFTs,
\begin{equation}
    \qt{\frac{\cT}{F_0}}{holo}=\frac{90}{\pi ^6}\approx0.0936\,,\qquad \qt{\frac{k}{F_0}}{holo}=\frac{4 \Gamma \left(5/8\right)^4}{\Gamma \left(1/8\right)^4}\approx5.26\cdot10^{-3}\,,\qquad \text{for }d=5\,.
\end{equation}

In the large $d \to \infty$ limit, the expressions for $\qt{C_T}{holo}$, $\qt{k}{holo}$ and $\qt{F_0}{holo}$, respectively in~\eqref{eq:CTholo},~\eqref{eq:kholo} and~\eqref{eq:F0holo}, reduce to
\begin{align}
\qt{C_T}{holo}&\overset{d\rightarrow\infty}{\approx}\frac{d}{\sqrt{2}\pi}\left[\frac{2d}{\pi\text e}\right]^{d/2}\frac{\Ls^{d-1}}{8G}\,,\\ \qt{k}{holo}&\overset{d\rightarrow\infty}{\approx}\frac{\text e}{\pi}\left[\frac{\pi}{d}\right]^d\frac{\Ls^{d-1}}{8G}\,,\\
\qt{F_0}{holo}&\overset{d\rightarrow\infty}{\approx}\sqrt{\frac{d}{\pi }} \left[\frac{2 \pi  \text e}{d}\right]^{d/2} \frac{\Ls^{d-1}}{8 G}\,,
\end{align}
which lead to the following asymptotic ratios
\begin{equation}
\qt{\frac{C_T}{F_0}}{holo}\overset{d\rightarrow\infty}{\approx}\sqrt{\frac{d}{2\pi}}\left[\frac{d}{\pi\text e}\right]^d\,,\qquad\qt{\frac{k}{F_0}}{holo}\overset{d\rightarrow\infty}{\approx}\frac{\text e}{\sqrt{\pi  d}}\left[\frac{\pi }{2 d\text e }\right]^{d/2}\,.
\end{equation}
Such asymptotic ratios can be compared to those in~\eqref{eq:CTF0ratio_free} and~\eqref{eq:kF0ratio_free} as
\begin{align}
\frac{\qt{C_T/F_0}{holo}}{\qt{C_T/F_0}{free scalar}} &\overset{d\rightarrow\infty}{\approx} \frac{2}{\pi^2}\,,\quad  \qquad \qquad   
\frac{\qt{C_T/F_0}{holo}}{\qt{C_T/F_0}{free fermion}} \overset{d\rightarrow\infty}{\approx} \frac{4}{\pi^2}\,,\\
\frac{\qt{k/F_0}{holo}}{\qt{k/F_0}{free scalar}} &\overset{d\rightarrow\infty}{\approx}  \frac{8\text e}{\pi^2d}\left[\frac{\pi}{d}\right]^d\, ,\quad \qquad
\frac{\qt{k/F_0}{holo}}{\qt{k/F_0}{free scalar}}  \overset{d\rightarrow\infty}{\approx}  \frac{8\text e}{\pi^2}\left[\frac{\pi}{d}\right]^d\,.
\end{align}

\subsection{(Free) Maxwell theory}
As discussed in Section~\ref{sec:bound3}, in $d=3$ the Maxwell theory is an orbifold of the free massless scalar field under the continuous shift symmetry, $\phi \to \phi + \mathbb{R}$. This can be thought of as the theory of the derivatives of the scalar, $\partial_a\phi=\epsilon_{abc}F^{bc}$, where $\varepsilon_{abc}$ is the Levi-Civita symbol---see also~\cite{El-Showk:2011xbs}. 
By contrast, the bare Maxwell field strength $F_{ab}$ is scale invariant only in four dimensions, and this special three-dimensional duality has no straightforward extension to dimensions $d \ge 5$. In higher dimensions, restoring conformal symmetry requires a non-unitary extension of the theory. Based on these observations, we collect the available results in the literature. 

First, we point out that in~\cite{Casini:2014aia} it was shown that in $d=3$, the finite piece of EE diverges with the regularization scale $\delta$ as $\qt{F(A)}{Maxwell}\sim\qt{F(A)}{free scalar}+n/4\log(-\log\delta)$, where $n$ is the number of connected boundaries of the region. From this expression, it is immediate to see that~\cite{Bueno:2023gey}
\begin{equation}
\qt{\frac{F(A)}{F_0}}{Maxwell}=n\,,\qquad \forall A\,,\text{ for }d=3\,.
\end{equation}
This saturation happens for all orbifold theories whose parent theory has been quotiented by a infinite symmetry group~\cite{Casini:2019kex,Bueno:2021fxb}---see ABJM case below. Combining this with $\qt{C_T}{Maxwell}$, which, based on the duality discussed above coincides with the free scalar, we have~\cite{Bueno:2023gey}
\begin{equation}
\qt{\frac{C_T}{F_0}}{Maxwell}=0\,,\qquad \text{ for }d=3\,.
\end{equation}

On the other hand, in four dimensions, the central charges $c$ and $a$ are given by~\cite{Duff:1977ay}
\begin{equation}
\qt{c}{Maxwell}=\frac{1}{10}\,,\qquad \qt{a}{Maxwell}=\frac{31}{180}\,,
\end{equation}
which implies 
\begin{equation}
\qt{\frac{c}{a}}{Maxwell}=\frac{18}{31}\approx0.581\,.
\end{equation}
As mentioned in the main text, this value saturates the lower bound of the Hofman--Maldacena constraints~\cite{Hofman:2008ar,Hofman:2016awc}.

\subsection{$O(N)$ models}

The coefficient $C_T$ has been computed for the $d$-dimensional critical scalar theory $O(N)$---this is, in the Wilson-Fisher point---using large $N$ methods~\cite{Wilson:1971dc,tHooft:1973alw} and operator product expansion in order to determine up to the leading $1/N$ corrections for the quartic interaction theory. Here, we focus only on one-loop, corrections, namely~\cite{Petkou:1994ad}
\begin{equation}\label{eq:CTONd}
 \qti{C_T}{O(N)}=\qt{C_T}{free scalar}N+\qti{\delta C_T(d)}{O(N)}+\mathcal O(1/N)\,,
\end{equation}
where $\qt{C_T}{free scalar}$ is given in~\eqref{eq:CTfreefields} and the leading correction, $\delta C_T(d)$, reads
\begin{equation}
\qti{\delta C_T(d)}{O(N)}=-\frac{(d-4)\Gamma(d-1)\sin(\pi d/2)}{(d-1)(d+2)\pi^{d+1}}\left[\qti{\Psi(d)}{O(N)}+\frac{d^2+6d-8}{2d(d-2)}\right]\,,
\end{equation}
with $\qti{\Psi(d)}{O(N)}=\psi\left(3-d/2\right)+\psi(d-1)-\psi(1)-\psi\left(d/2\right)$ and $\psi(x)=\Gamma'(x)/\Gamma(x)$ being the digamma function.
If we particularize expression~\eqref{eq:CTONd} to the dimensions of our interest, we obtain
\begin{alignat}{2}\label{eq:CTOlargeN}
\qti{C_T}{O(N)}&= \qt{C_T}{free scalar} \left( N-\frac{40}{9\pi^2}\right) +\mathcal{O}(1/N)\,, \quad &&\text{for }d=3\,,\\
\qti{C_T}{O(N)}&= \qt{C_T}{free scalar} \left( N-\frac{1408}{1575\pi^2}\right) +\mathcal{O}(1/N)\,, \quad &&\text{for }d=5\,.
\end{alignat}
These results are also reproduced within the $d-\epsilon$ expansion framework~\cite{Wilson:1971dc}. In particular, one may consider weakly coupled quartic interactions in $4-\epsilon$ dimensions and cubic interactions in $6-\epsilon$ dimensions. While the results for the former are consistent with standard perturbative analyses~\cite{Jack:1983sk,Cappelli:1990yc}, in the latter case the Wilson-Fisher fixed point has been shown to exist only for sufficiently large $N$: specifically, for $N>1038$ at one loop, and for values as low as $N>64$ at three loops~\cite{Fei:2014yja,Fei:2014xta,Fei:2015kta}.

Away from the weakly coupled regime, we focus on the $N=1$ theory, whose Wilson-Fisher fixed point coincides with the Ising CFT. While in $d=3$, the value is known to a great degree of numerical precision employing bootstrap techniques~\cite{Rattazzi:2008pe,Simmons-Duffin:2016gjk,Poland:2018epd,Poland:2022qrs,Hartman:2022zik}, obtaining~\cite{El-Showk:2012cjh,Kos:2013tga,Giombi:2014xxa,El-Showk:2014dwa,Chang:2024whx}
\begin{equation}\label{eq:CTO1}
\qti{C_T}{O(1)}\approx0.947\qt{C_T}{free scalar}\,,\quad\text{for }d=3\,,
\end{equation}
this is not the case for $d=5$.

Now, we move to the free energy. Again, there are available results in the literature using $d-\epsilon$ techniques to calculate the free energy in the Wilson-Fisher fixed point of the $O(N)$ vector model in the large $N$ limit. In particular, we can write\footnote{The overall sign in~\eqref{eq:F0O(N)d} differs from~\cite{Giombi:2014xxa} due to a choice of convention. In~\cite{Giombi:2014xxa}, $F_0$ alternates sign with the dimension while $\delta F_0$ does not. In our convention, we take $F_0$ to be always positive and instead absorb the alternating sign into $\delta F_0$, preserving the same relative sign between the two terms.}~\cite{Giombi:2014xxa}
\begin{equation}\label{eq:F0O(N)d}
\qti{F_0}{O(N)}=N \qt{F_0}{free scalar}+\delta F_0(\Delta=d-2)+\mathcal O(1/N)\,, 
\end{equation}
where
\begin{equation}\label{eq:deltaF}
    \delta F_0(\Delta)=\frac{1}{\Gamma(d+1)}\int_0^{\Delta-d/2}\diff u\,u\sin(\pi u)\Gamma\left(\frac{d}{2}+u\right)\Gamma\left(\frac{d}{2}-u\right)\,,
\end{equation}
with $\Delta$ being the conformal scaling dimension. In~\eqref{eq:F0O(N)d}, we used that $O(N)$ model with quartic interaction has $\Delta=d-2$. Using this expression we can proceed similarly as with $C_T$ and obtain $F_0$, namely,
\begin{alignat}{3}
    \qti{F_0}{O(N)}&=\qt{F_0}{free scalar}\left(N-\frac{2\zeta(3)}{\pi^2\log4-3\zeta(3)}\right)+\mathcal O(1/N)\,,\quad &&\text{for }d=3\,,\\
    \qti{F_0}{O(N)}&=\qt{F_0}{free scalar}\left(N+\frac{8 \left[\pi ^2 \zeta (3)+3 \zeta (5)\right]}{\pi ^4
   \log 64+6 \pi ^2 \zeta (3)-45 \zeta (5)}\right)+\mathcal O(1/N)\,,\quad &&\text{for }d=5\,,
\end{alignat}
in the $N\gg1$ expansion. Finally, in the case of the Ising model, one has~\cite{Giombi:2014xxa}
\begin{equation}
\qt{F_0}{Ising}\approx0.957\qt{F_0}{free scalar}\,, \quad\text{for }d=3\,.
\end{equation}

With all the relevant values at hand, we find the quotients
\begin{alignat}{3}
\qti{\frac{C_T}{F_0}}{O(N)}\approx\qt{\frac{C_T}{F_0}}{free scalar}\left(1-\frac{0.212}{N}\right)+\mathcal O(1/N^2)\,,&   \quad &&\text{for }d=3\,,\label{eq:CTF0ON3}\\
\qti{\frac{C_T}{F_0}}{O(N)}\approx\qt{\frac{C_T}{F_0}}{free scalar}\left(1-\frac{0.369}{N}\right)+\mathcal O(1/N^2)\,,&   \quad &&\text{for }d=5\,,\label{eq:CTF0ON5}
\end{alignat}
while in the case of the Ising model we find
\begin{equation}
\qt{\frac{C_T}{F_0}}{Ising}\approx0.993\qt{\frac{C_T}{F_0}}{free scalar}\,,\quad \text{for }d=3\,.
\end{equation}
Evaluating~\eqref{eq:CTF0ON3} and~\eqref{eq:CTF0ON5} for sufficiently small $N$---in particular $N=4$, we obtain the bounds
\begin{alignat}{3}
&0.141\lesssim\qti{\frac{C_T}{F_0}}{O(N)}\leq0.149\,, \quad &\text{for } d=3\,,\\
&0.285\lesssim\qti{\frac{C_T}{F_0}}{O(N)}\leq0.314\,, \quad &\text{for } d=5\,,
\end{alignat}

\subsection{Gross--Neveu models}
There are also results in the literature for the interacting fixed point of the GN model, which is an $O(N)$-invariant fermionic theory. In particular, in the large-$N$ limit one finds~\cite{Klebanov:2011gs,Diab:2016spb}
\begin{equation}\label{eq:CTONd}
 \qt{C_T}{GN}=\qt{C_T}{free fermion}N+\qt{\delta C_T(d)}{GN}+\mathcal O(1/N)\,,
\end{equation}
where the one-loop correction is given by
\begin{equation}
\qt{\delta C_T(d)}{GN}=\frac{2^{\lfloor (d-2)/2\rfloor}(d-2)\Gamma(d-1)\sin(\pi d/2)}{(d+2)\pi^{d+1}}\left(\qt{\Psi(d)}{GN}+\frac{d-2}{d(d-1)}\right)\,,
\end{equation}
and he have introduced the short-hand notation $\qt{\Psi(d)}{GN}=\psi\left(2-d/2\right)+\psi(d-1)-\psi(1)-\psi\left(d/2\right)$. From~\eqref{eq:CTONd} we obtain
\begin{alignat}{3}\label{eq:CTGNepsilon3}
\qt{C_T}{GN}&= \qt{C_T}{free fermion}\left(N+\frac{8}{9\pi^2}\right) +\mathcal{O}(1/N)\, ,&&\quad \text{for }d=3\,,\\
\qt{C_T}{GN}&= \qt{C_T}{free fermion}\left(N+\frac{2528}{525 \pi ^2}\right) +\mathcal{O}(1/N)\, ,&&\quad \text{for }d=5\,.\label{eq:CTGNepsilon5}
\end{alignat}

Numerical estimates for $\qt{C_T}{GN}$ in $d=3$ at small and intermediate values of $N$ are also available from the $d-\epsilon$ expansion~\cite{Diab:2016spb}. In particular, one finds
\begin{align}
\qt{C_T}{GN}&\approx1.01\qt{C_T}{free fermion}\,,\quad \text{for }d=3\,,\quad N=4,8,12,16\,,\\
\qt{C_T}{GN}&\approx1.00\qt{C_T}{free fermion}\,,\quad \text{for }d=3\,,\quad N=20,24,\ldots\,,
\end{align}
These values turn out to be in a good agreement with the large $N$ result.

On the other hand, the corrections in the large $N$ expansion to the free energy of the GN model are found in, \eg~\cite{Klebanov:2011gs,Klebanov:2011td,Giombi:2014xxa,Giombi:2015haa}. As pointed out in~\cite{Tarnopolsky:2016vvd,Fraser-Taliente:2025udk}, we can express the one-loop correction using $\Delta=d-1$ in~\eqref{eq:deltaF}, obtaining
\begin{equation}\label{eq:F0GNd}
    \qt{F_0}{GN}=N \qt{F_0}{free fermion}+\text{tr}\,\mathbf 1\,\delta F_0(\Delta=d-1)+\mathcal O(1/N)\,,
\end{equation}
where we multiplied by the dimension of the gamma matrices to account for the degrees of freedom of the Dirac field. From this formula we observe
\begin{alignat}{3}
\qt{F_0}{GN}&=\qt{F_0}{free fermion} \left(N+\frac{2\zeta(3)}{\pi^2\log 4+3\zeta(3)}\right)+\mathcal{O}(1/N)\, ,&& \text{for }d=3\,,\\
\qt{F_0}{GN}&=\qt{F_0}{free fermion}\left(N-\frac{24 \pi ^2 \zeta (3)-24 \zeta (5)}{\pi ^4 \log 64+10 \pi ^2 \zeta (3)+15 \zeta (5)}\right)+\mathcal O(1/N)\,,\ \ &&\text{for }d=5\,.
\end{alignat}

Combining the results above we find the quotients
\begin{alignat}{3}
\qt{\frac{C_T}{F_0}}{GN}&\approx\qt{\frac{C_T}{F_0}}{free fermion} \left(1- \frac{0.0490}{N} \right)+\mathcal{O}(1/N^2)\, ,\quad &&\text{for }d=3\,.\\
\qt{\frac{C_T}{F_0}}{GN}&\approx\qt{\frac{C_T}{F_0}}{free fermion} \left(1+ \frac{0.970}{N} \right)+\mathcal{O}(1/N^2)\, ,\quad &&\text{for }d=5\,.\label{eq:CTF0GN5}
\end{alignat}

There are also results for the free energy for finite values of $N$ in three dimensions. In particular, from the Gross--Neveu--Yukawa model in $d=4-\epsilon$ expansion, one finds that for the GN model---obtained when $\epsilon=1$---the simple approximation~\cite{Giombi:2014xxa}
\begin{equation}
 \qt{F_0}{GN} \approx 0.103\,N + 0.0517 - \frac{\pi N}{96(N+6)}\,,\quad \text{for }d=3\,,
\end{equation}
applies. Using this expression to probe small $N$, we observe
\begin{equation}
    \qt{\frac{C_T}{F_0}}{GN}\approx0.0854\,,\quad \text{for }N=4\,,\quad  \qt{\frac{C_T}{F_0}}{GN}\approx0.0895\,,\quad \text{for }N=8\,,\quad \text{for }d=3\,.
\end{equation}
Combining these values with those for $\qt{C_T}{GN}$ in~\eqref{eq:CTGNepsilon3} and~\eqref{eq:CTGNepsilon5}, we observe that the values across different $N$ can be written as
\begin{equation}
    0.0844\lesssim\qt{\frac{C_T}{F_0}}{GN}\lesssim 0.0940\,,\quad \text{for } d=3\,,\quad \forall N.
\end{equation}
Observe that as we vary $N$, the value of the ratio is greater than the free fermion result in some cases, and smaller in others. This contrasts with the situation in the $O(N)$ models, where the ratio is smaller than the free scalar value for all $N$.

On the other hand, there are no reliable approximations for $\qt{C_T}{GN}$ in $d=5$. Consequently, using only expression~\eqref{eq:CTF0GN5}, we find
\begin{equation}
0.167 \leq \qt{\frac{C_T}{F_0}}{GN} \lesssim 0.251\,,
\qquad \text{for } d=5\,,
\end{equation}
where the upper bound is obtained by extrapolating the large-$N$ result down to $N=4$.

\subsection{$\mathcal{N}=2$ Wess--Zumino models}
The free energy and two-point stress energy tensor coefficient corresponding to the three-dimensional critical WZ model---also known as supersymmetric Ising model~\cite{Bobev:2015vsa}, with cubic superpotential $\mathcal{W}=X^3$, are given respectively by~\cite{Nishioka:2013gza,Giombi:2014xxa}
\begin{equation}
\left.C_T\right|_{X^3}\approx 0.02766 \, , \quad \left.F_0\right|_{X^3}\approx 0.29079 \, ,\quad \text{so} \quad
\left.\frac{C_T}{F_0}\right|_{X^3}\approx 0.0951\,.
\end{equation}

On the other hand, the critical point of a WZ model with superpotential $\mathcal{W}=X \sum_{i=1}^N Z_i^2$ in the large $N$ limit has~\cite{Nishioka:2013gza}
\begin{alignat}{23}
\left.C_T\right|_{XZ^2(N)} &= \frac{3N}{8\pi^2}-\frac{2}{\pi^4}+ \frac{1}{N} \left(\frac{68}{9\pi^2}-\frac{48}{\pi^4} \right)+\mathcal{O}(1/N^2)\, , \\  \left.F_0\right|_{XZ^2(N)}&=\frac{N}{2}\log 2 + \frac{4}{\pi^2 N}+ \mathcal{O}(1/N^2)\,,
\end{alignat}
so
\begin{equation}\label{eq:XZ2quotient}
\left.\frac{C_T}{F_0}\right|_{XZ^2(N)}\approx 0.110- \frac{0.0592}{N}+\mathcal{O}(1/N^2)\,.
\end{equation}
At the same time, in~\cite{Nishioka:2013gza} numerical values for finite $N$ are available. While none of them exceed the upper bound in~\eqref{eq:XZ2quotient}, they can be as low as $\qti{C_T/F_0}{XZ^2}\approx0.0950$---which corresponds to $N=2$. Therefore, we can write
\begin{equation}
0.0950\lesssim\left.\frac{C_T}{F_0}\right|_{XZ^2(N)}\lesssim0.110\,.
\end{equation}
The case $N=2$ is particularly interesting. Besides providing the lower bound within the $XZ^2$ model, it also corresponds to the $XYZ$ model~\cite{Giombi:2014xxa}---defined by the superpotential $\mathcal{W}=XYZ$—which is in turn mirror symmetric to $\mathcal{N}=2$ SQED~\cite{Aharony:1997bx,deBoer:1997ka}. For the latter two models, we have~\cite{Jafferis:2010un,Bobev:2015jxa,Witczak-Krempa:2015jca,Jian:2016zll}
\begin{alignat}{2}
&\Fb{XYZ}=\Fb{{\mathcal{N}}=2 \text{ SQED}}=-3\ell(1/3)
\,, \\
&\cTb{XYZ}=\cTb{{\mathcal{N}}=2\text{ SQED}}=\frac{2}{27\pi^2} \left[16- \frac{9\sqrt{3}}{\pi} \right]
\,,
\end{alignat}
where $\ell(z)\equiv -z \log \left(1-\expp{2\pi \iu z} \right)+\iu/2 \left[\pi z^2+\text{Li}_2(\expp{2\pi \iu z})/\pi \right]-\iu\pi/12$. From this, we obtain
\begin{equation}
\cTFb{XYZ}= \cTFb{{\mathcal{N}}=2\text{ SQED}}\approx 0.0950\,,
\end{equation}
which agrees well with the result from~\cite{Nishioka:2013gza} for the $X Z(2)$ model presented above.

 We also have results for the fixed point corresponding to a superpotential $\mathcal{W}=\sum_{i=1}^N (Z_iZ_i)^2$~\cite{Nishioka:2013gza}
\begin{equation}
\cTb{Z^4}= \frac{3N}{8\pi^2}\, , \quad \Fb{{ Z^4}}= \frac{N}{2}\log 2 \, ,\quad \text{so} \quad
\cTFb{Z^4}=\frac{3}{4\pi^2 \log 2}\approx 0.110\,.
\end{equation}
The results are actually identical to the ones obtained for $N$ free chiral multiplets and also satisfy the bounds.

\subsection{Aharony--Bergman--Jafferis--Maldacena model}

We can also write down the quotient $C_T/F_0$ for the three-dimensional $U(N)_k\times U(N)_{-k}$ Aharony--Bergman--Jafferis--Maldacena (ABJM) theory~\cite{Aharony:2008ug}. There is a large literature on the large-$N$ limit of this theory---see, e.g,.~\cite{Kapustin:2009kz,Fuji:2011km,Marino:2011eh,Hanada:2012si,Marino:2016new,Chester:2021gdw,Bobev:2022eus}. In particular, the partition function on $\mathbb S^3$ is known to admit an Airy-function representation up to exponentially suppressed corrections at large $N$, \ie~\cite{Kapustin:2009kz,Fuji:2011km,Marino:2011eh}. Using it, it is straightforward to obtain\footnote{We can compare the leading term in the expansion directly with the holographic value~\eqref{eq:F0holo3} using the holographic dictionary entry $\Ls^2/G=(2\sqrt{2k}/3) N^{3/2}$~\cite{Marino:2011eh}.}~\cite{Fuji:2011km,Marino:2011eh}
\begin{equation}
    \qti{F_0}{U(1)_k\times U(1)_{-k}}=\frac{\pi\sqrt{2k}}{3}N^{3/2}-\frac{\pi\sqrt{2k}}{6}\left(\frac{k}{8}+\frac{1}{k}\right)N^{1/2}+\mathcal O\!\left(N^{-1/2}\right)\,.
\end{equation}

On the other hand, $C_T$ is obtained by considering the free energy on a squashed sphere~\cite{Hama:2011ea,Nosaka:2015iiw,Chester:2021gdw,Bobev:2021oku,Bobev:2022eus}. From this approach, one finds~\cite{Agmon:2017xes,Chester:2020jay,Bobev:2022eus,Closset:2012vg,Closset:2012ru}
\begin{equation}
   \qti{C_T}{U(1)_k\times U(1)_{-k}}= \frac{2 \sqrt{2k}}{\pi ^3}N^{3/2}+\frac{\left(16-k^2\right)}{4\pi ^3 \sqrt{2k}}N^{1/2}+\mathcal{O}\left(N^{-1/2}\right)\,.
\end{equation}

Using these results, we observe that at large $N$,
\begin{equation}\label{eq:CTF0ABJMNlarge}
 \qti{\frac{C_T}{F_0}}{U(1)_k\times U(1)_{-k}}=\frac{6}{\pi ^4}+\frac{9}{\pi ^4 k}\frac{1}{N}+\mathcal O\left(N^{-2}\right)\,.
\end{equation}
 which tends to the holographic value in~\eqref{eq:CTFO_holo}. 
 
 There are also results for finite values of $N$. In particular, using the expressions obtained in~\cite{Kapustin:2010xq,Marino:2011eh,Nishioka:2013haa,Chester:2014fya} for $N=1$, namely
\begin{equation}
\qti{C_T}{U(1)_k\times U(1)_k}=\frac{3}{2\pi^2}\,,\quad \qti{F_0}{U(1)_k\times U(1)_k}=\log(4k)\,,
\end{equation}
we find
\begin{equation}\label{eq:CTF0ABJMN1}
\cTFb{U(1)_k\times U(1)_{-k}}=\frac{3}{2\pi^2\log (4k)}\approx\frac{0.152}{1.39+\log k}\,.
\end{equation}
Observe that for large $k$ the quotient tends to zero, saturating the lower bound. On the other hand, it reaches its maximum value at $k=1$. There are also available results for $\qt{C_T}{ABJM}$ and $\qt{F_0}{ABJM}$ for $N=2,3,4$ at finite $k$ and large $k$ expansions~\cite{Chester:2014fya,Binder:2020ckj,Alday:2021ymb}. The maximum value of all of the respective quotients is lower than the one achieved for $k=1$ in~\eqref{eq:CTF0ABJMN1} and, in all cases, as $N$ grows it asymptotes the expansion~\eqref{eq:CTF0ABJMNlarge}. Based on these results, we find that in all checked cases the bound
\begin{equation}
0\leq\cTFb{U(N)_k\times U(N)_{-k}}\lesssim0.110 \, ,\quad \forall N,k\,.
\end{equation}
is satisfied.

\subsection{Atiyah--Drinfeld--Hitchin--Manin model}
Let us now check the quotient $C_T/F_0$ in the three-dimensional Atiyah--Drinfeld--Hitchin--Manin (ADHM) theory~\cite{Atiyah:1978ri}, with gauge group $U(N)$ and with $\Nf$ fundamental hypermultiplets. In this model, the free energy $F_0$ has been computed in the large $N$ limit~\cite{Bobev:2023lkx} 
\begin{equation}
\qt{F_0}{ADHM}  = \frac{\pi\sqrt{2\Nf}}{3} N^{3/2} + \frac{\pi(\Nf^2-4)}{8\sqrt{2 \Nf}} N^{1/2}  + \mathcal O\!\left(N^{-1/2}\right)
\end{equation}
whereas the coefficient $C_T$ has the following large $N$ expansion~\cite{Bobev:2023lkx}
\begin{equation}
\qt{C_T}{ADHM} = \frac{2 \sqrt{2\Nf}}{\pi^3} N^{3/2} +
\frac{7 \Nf^2 +8}{4 \pi^3 \sqrt{2 \Nf} }N^{1/2}+
\mathcal O(N^0) .
\end{equation}
Therefore, we have that
\begin{equation}
\label{eq:CTFO_ADHM}
    \qt{\frac{\cT}{F_0}}{ADHM}= \frac{6}{\pi^4} + \frac{3(\Nf^2+5)}{2 \pi^4 \Nf} \frac{1}{N} + 
    \mathcal O(1/N)\,,
\end{equation}
where the leading order coincides with the value of $\qt{\cT/F_0}{holo}$ in~\eqref{eq:CTFO_holo}.
Plugging some values of $N_f$, we obtain
\begin{alignat}{3}
    \qt{\frac{\cT}{F_0}}{ADHM} &= \qt{\frac{\cT}{F_0}}{holo}\left(1+ \frac{3}{2} \frac{1}{N}\right) + 
    \mathcal O\left(1/N^2\right)\,, \qquad & \text{for }\Nf=1\,, \\
    \qt{\frac{\cT}{F_0}}{ADHM}&= \qt{\frac{\cT}{F_0}}{holo}\left(1 + \frac{9}{8} \frac{1}{N }\right)+ 
    \mathcal O\left(1/N^2\right)\,, \qquad & \text{for }\Nf=2\,,\\
    \qt{\frac{\cT}{F_0}}{ADHM}&= \qt{\frac{\cT}{F_0}}{holo}\left(1+ \frac{7}{6} \frac{1}{N}\right)+ 
    \mathcal O\left(1/N^2\right)\,, \qquad & \text{for }\Nf=3\,.
\end{alignat}
Based on these results, a conservative bound for the quotient is given by
\begin{equation}
    0.0616\leq\qt{\frac{C_T}{F_0}}{ADHM}\lesssim0.0769\,,\qquad \forall N\,.
\end{equation}

\subsection{Rank-one Seiberg exceptional theories and the
Morrison--Seiberg theory}

A special class of five-dimensional superconformal field theories are the theories with exceptional $E_n$ $(n=1,\ldots,8)$ flavor symmetry proposed by Seiberg~\cite{Seiberg:1996bd}, and $\tilde E_1=\text U(1)$ by Morrison and Seiberg~\cite{Morrison:1996xf}. For $n<6$, the flavor groups are $E_5= \text{SO}(10)$, $E_4 = \text{SU}(5)$, $E_3=\text{SU}(3)\times \text{SU}(2)$, $E_2=\text{SU}(2)\times \text{SU}(1)$, and $E_1 = \text{SU}(2)$. On the other hand, $E_6$, $E_7$ and $E_8$ correspond to exceptional Lie algebras.

In~\cite{Chang:2017cdx} the values for $C_T$ and $F_0$ are computed, they read
\begin{alignat}{4}
    \qti{C_T}{\tilde E_1}&\approx0.483\,,\quad\qti{C_T}{E_1}\ &\approx&\ 0.471\,,\quad \qti{C_T}{E_2}\ &\approx&\ 0.606\,,\quad \qti{C_T}{E_3}\ &\approx&\ 0.763\,,\\
    \qti{C_T}{E_4}&\approx0.957\,,\quad \qti{C_T}{E_5}\ &\approx&\ 1.21\,,\quad\ \, \qti{C_T}{E_6}\ &\approx&\ 1.56\,,\quad \ \, \qti{C_T}{E_7}\ &\approx&\ 2.11\,,\\
    \qti{C_T}{E_8}&\approx3.28\,,
\end{alignat}
and
\begin{alignat}{4}
    \qti{F_0}{\tilde E_1}&\approx5.09\,,\quad\qti{F_0}{E_1}\ &\approx&\ 5.10,
    \quad \qti{F_0}{E_2}\ &\approx&\ 6.15\,,\quad \qti{F_0}{E_3}\ &\approx&\ 7.40\,,\\
    \qti{F_0}{E_4}&\approx8.97\,,\quad \qti{F_0}{E_5}\ &\approx&\ 11.0\,,\quad \qti{F_0}{E_6}\ &\approx&\ 13.9\,,\quad \qti{F_0}{E_7}\ &\approx&\ 18.5\,,\\
    \qti{F_0}{E_8}&\approx28.5\,,
\end{alignat}
respectively. Using these values we obtain
\begin{alignat}{4}
    \qti{\frac{C_T}{F_0}}{\tilde E_1}&\approx0.0949\,,\quad\qti{\frac{C_T}{F_0}}{E_1}\ &\approx&\ 0.0923\,,\quad \qti{\frac{C_T}{F_0}}{E_2}\ &\approx&\ 0.0985\,,\quad \qti{\frac{C_T}{F_0}}{E_3}\ &\approx&\ 0.103\,,\\
    \qti{\frac{C_T}{F_0}}{E_4}&\approx0.106\,,\quad \qti{\frac{C_T}{F_0}}{E_5}\ &\approx&\ 0.109\,,\quad\ \, \qti{\frac{C_T}{F_0}}{E_6}\ &\approx&\ 0.112\,,\quad \ \, \qti{\frac{C_T}{F_0}}{E_7}\ &\approx&\ 0.114\,,\\
    \qti{\frac{C_T}{F_0}}{E_8}&\approx0.115\,.
\end{alignat}

\subsection{$T_N$ theories}
Another class of five-dimensional superconformal field theories are the $T_N$ theories. These non-Lagrangian models 
can be engineered as $(p, q)$ 5-brane webs in Type IIB string theory and possess a global symmetry group $SU(N)\times SU(N)\times SU(N)$~\cite{Benini:2009gi,Bao:2013pwa,Mitev:2014isa}. The values of $C_T$ and $F_0$ have been computed numerically using supersymmetric localization techniques in~\cite{Fluder:2018chf}. 
In our conventions, the results read
\begin{alignat}{12}
&\qti{C_T}{T_3}\ &\approx&\ &1.60&\,,\quad
&\qti{C_T}{T_4}\ \ &\approx&\ &\ \ \ \ \ \ 6.67&\,,\quad 
&\qti{C_T}{T_5}\ &\approx&\ &18.4&\,,\\ 
&\qti{C_T}{T_6}\ &\approx&\ &40.7&\,,\quad
&\qti{C_T}{T_7}\ \ &\approx&\ &\ \ \ \ \ \ 78.8&\,,\quad
&\qti{C_T}{T_8}\ &\approx&\ &138&\,,\\ 
&\qti{C_T}{T_9}\ &\approx&\ &225&\,,\quad
&\qti{C_T}{T_{10}}\ &\approx&\ &\ \ \ \ \ \ \ 349&\,,\quad
&\qti{C_T}{T_{11}}\ &\approx&\ &518&\,,\\
&\qti{C_T}{T_{12}}\ &\approx&\ &740&\,,\quad
&\qti{C_T}{T_{13}}\ &\approx&\ & 1.03\cdot 10^3&\,,\quad
&\qti{C_T}{T_{14}}\ &\approx&\ & 1.39\cdot 10^3&\,,\\
&\qti{C_T}{T_{15}}\ &\approx&\ & 1.83\cdot 10^3&\,,\quad
&\qti{C_T}{T_{16}}\ &\approx&\ & 2.38\cdot 10^3&\,,\quad
&\qti{C_T}{T_{17}}\ &\approx&\ & 3.06\cdot 10^3&\,,\\
&\qti{C_T}{T_{18}}\ &\approx&\ & 3.85\cdot 10^3&\,,\quad
&\qti{C_T}{T_{19}}\ &\approx&\ & 4.81\cdot 10^3&\,,\quad
&\qti{C_T}{T_{20}}\ &\approx&\ & 5.92\cdot 10^3&\,,\\
&\qti{C_T}{T_{21}}\ &\approx&\ & 7.21\cdot 10^3&\,,\quad
&\qti{C_T}{T_{22}}\ &\approx&\ & 8.71\cdot 10^3&\,. \quad
\end{alignat}
The values of $\qti{F_0}{T_N}$ in turn read
\begin{alignat}{12}
&\qti{F_0}{T_3}\ &\approx&\ &13.9& \,, \quad
&\qti{F_0}{T_4}\ \ &\approx&\ &\ \ \ \ \ \ 63.3& \,, \quad
&\qti{F_0}{T_5}\ &\approx&\ &182& \,, \\ 
&\qti{F_0}{T_6}\ &\approx&\ &411&\,, \quad 
&\qti{F_0}{T_7}\ \ &\approx&\ &\ \ \ \ \ \ \ 805&\,, \quad 
&\qti{F_0}{T_8}\ &\approx&\ &1.42\cdot 10^3&\,, \\
&\qti{F_0}{T_9}\ &\approx&\ &2.34\cdot 10^3&\,, \quad 
&\qti{F_0}{T_{10}}\ &\approx&\ &3.64\cdot 10^3&\,,\quad 
&\qti{F_0}{T_{11}}\ &\approx&\ &5.41\cdot 10^3&\,,\\
&\qti{F_0}{T_{12}}\ &\approx&\ &7.76\cdot 10^3&\,,\quad 
&\qti{F_0}{T_{13}}\ &\approx&\ &1.08\cdot 10^4&\,,\quad 
&\qti{F_0}{T_{14}}\ &\approx&\ &1.46\cdot 10^4&\,,\\
&\qti{F_0}{T_{15}}\ &\approx&\ &1.94\cdot 10^4&\,,\quad
&\qti{F_0}{T_{16}}\ &\approx&\ &2.53\cdot 10^4&\,,\quad
&\qti{F_0}{T_{17}}\ &\approx&\ &3.24\cdot 10^4&\,,\\
&\qti{F_0}{T_{18}}\ &\approx&\ &4.09\cdot 10^4&\,,\quad 
&\qti{F_0}{T_{19}}\ &\approx&\ &5.09\cdot 10^4&\,,\quad
&\qti{F_0}{T_{20}}\ &\approx&\ &6.27\cdot 10^4&\,,\\
&\qti{F_0}{T_{21}}\ &\approx&\ &7.65\cdot 10^4&\,,\quad 
&\qti{F_0}{T_{22}}\ &\approx&\ &9.24\cdot 10^4&\,.\quad 
\end{alignat}
Using these values, the ratio  $\qti{C_T/F_0}{T_N}$ can be calculated, and one finds\footnote{See Fig.~\ref{fig:TNalmo} for a plot of $\qti{C_T/F_0}{T_N}$ as a function of $N$, which makes evident the tendency towards the holographic result.} 
\begin{alignat}{12}
&\qti{\frac{C_T}{F_0}}{T_3}\ &\approx&\ 0.1153&\,,\quad
&\qti{\frac{C_T}{F_0}}{T_4}\ &\approx&\ 0.1053&\,,\quad 
&\qti{\frac{C_T}{F_0}}{T_5}\ &\approx&\ 0.1013&\,,\\
&\qti{\frac{C_T}{F_0}}{T_6}\ &\approx&\ 0.0992&\,, \quad 
&\qti{\frac{C_T}{F_0}}{T_7}\ &\approx&\ 0.0979&\,,\quad 
&\qti{\frac{C_T}{F_0}}{T_8}\ &\approx&\ 0.0971&\,,\\
&\qti{\frac{C_T}{F_0}}{T_9}\ &\approx&\ 0.0965&\,,\quad  
&\qti{\frac{C_T}{F_0}}{T_{10}}\ &\approx&\ 0.0960&\,,\quad
&\qti{\frac{C_T}{F_0}}{T_{11}}\ &\approx&\ 0.0957&\,,\\
&\qti{\frac{C_T}{F_0}}{T_{12}}\ &\approx&\ 0.0954&\,,\quad 
&\qti{\frac{C_T}{F_0}}{T_{13}}\ &\approx&\ 0.0951&\,,\quad
&\qti{\frac{C_T}{F_0}}{T_{14}}\ &\approx&\ 0.0950&\,,\\
&\qti{\frac{C_T}{F_0}}{T_{15}}\ &\approx&\ 0.0949&\,,\quad
&\qti{\frac{C_T}{F_0}}{T_{16}}\ &\approx&\ 0.0947&\,,\quad 
&\qti{\frac{C_T}{F_0}}{T_{17}}\ &\approx&\ 0.0946&\,,\\
&\qti{\frac{C_T}{F_0}}{T_{18}}\ &\approx&\ 0.0945&\,,\quad 
&\qti{\frac{C_T}{F_0}}{T_{19}}\ &\approx&\ 0.0944&\,,\quad 
&\qti{\frac{C_T}{F_0}}{T_{20}}\ &\approx&\ 0.0944&\,,\\
&\qti{\frac{C_T}{F_0}}{T_{21}}\ &\approx&\ 0.0943&\,,\quad 
&\qti{\frac{C_T}{F_0}}{T_{22}}\ &\approx&\ 0.0943&\,.\quad 
\end{alignat}

\subsection{$\#_{N,M}$ theories}
The $\#_{N,M}$ are another  set of non-Lagrangian $d=5$ SCFTs with a global symmetry group $SU(N)\times SU(N)\times SU(M)\times U(1)$~\cite{Aharony:1997bh}. Their respective $\qti{C_T}{\#_{N,M}}$ and $\qti{F_0}{\#_{N,M}}$ have also been computed using similar techniques to the $T_N$ ones in~\cite{Fluder:2018chf}. In our conventions, the results for different values of $N$ and $M$ read
\begin{alignat}{12}
&\qti{C_T}{\#_{2,2}}\ &\approx&\ & 1.25&\,,\quad
&\qti{C_T}{\#_{2,4}}\ \ &\approx&\ 6.01&\,,\quad
&\qti{C_T}{\#_{2,6}} \ \ &\approx&\ 13.9&\,,\\
&\qti{C_T}{\#_{2,8}}\ &\approx&\ & 25.0&\,,\quad 
&\qti{C_T}{\#_{2,10}}\  &\approx&\ 39.2&\,,\quad
&\qti{C_T}{\#_{2,12}} \  &\approx&\ 56.6&\,,\\
&\qti{C_T}{\#_{2,14}}\ &\approx&\ & 77.1&\,,\quad 
&\qti{C_T}{\#_{4,2}}\ \ &\approx&\ 57.4&\,,\quad
&\qti{C_T}{\#_{4,4}} \ \ &\approx&\ 29.3&\,,\\
&\qti{C_T}{\#_{4,6}}\ &\approx&\ & 68.8&\,,\quad 
&\qti{C_T}{\#_{4,8}}\ \ &\approx&\  124&\,,\quad
&\qti{C_T}{\#_{4,10}} \  &\approx&\ 196&\,,\\
&\qti{C_T}{\#_{4,12}}\ &\approx&\ & 283&\,,\quad  
&\qti{C_T}{\#_{4,14}}\  &\approx&\  386&\,,\quad
&\qti{C_T}{\#_{6,2}}\ \  &\approx&\ 12.9&\,,\\
&\qti{C_T}{\#_{6,4}}\  &\approx&\ & 67.6&\,,\quad 
&\qti{C_T}{\#_{6,6}}\ \ &\approx&\ 160&\,,\quad 
&\qti{C_T}{\#_{6,8}}\ \  &\approx&\ 290&\,, \\
&\qti{C_T}{\#_{6,10}}\ &\approx&\ & 458&\,, \quad 
&\qti{C_T}{\#_{6,12}}\  &\approx&\ 663&\,,\quad 
&\qti{C_T}{\#_{6,14}}\ &\approx&\ 905&\,,\\
&\qti{C_T}{\#_{8,2}}\  &\approx&\ & 22.8&\,,\quad 
&\qti{C_T}{\#_{8,4}}\ \ &\approx&\ 121&\,,\quad 
&\qti{C_T}{\#_{8,6}}\ \ &\approx&\ 287&\,,\\
&\qti{C_T}{\#_{8,8}}\  &\approx&\ &  522&\,,\quad 
&\qti{C_T}{\#_{8,10}} \ &\approx&\ 825&\,,\quad 
&\qti{C_T}{\#_{8,12}}\ &\approx&\ 1.19\cdot 10^{3}&\,,\\
&\qti{C_T}{\#_{8,14}}\ &\approx&\ & 1.63\cdot 10^{3}&\,,\quad &\qti{C_T}{\#_{10,2}}\ &\approx&\ 35.2&\,,\quad
&\qti{C_T}{\#_{10,4}}\ &\approx&\ 188&\,,\\
&\qti{C_T}{\#_{10,6}}\ &\approx&\ & 450&\,,\quad
&\qti{C_T}{\#_{10,8}}\ &\approx&\ 819 &\,,\quad
&\qti{C_T}{\#_{10,10}} &\approx&\ 1.30\cdot 10^{3}&\,,\\
&\qti{C_T}{\#_{10,12}}\ &\approx&\ &1.87\cdot 10^{3}&\,,\quad
&\qti{C_T}{\#_{10,14}}&\approx&\ 2.57\cdot 10^{3}&\,,\quad
&\qti{C_T}{\#_{12,2}}\ &\approx&\ 503&\,,
\end{alignat}
\begin{alignat}{12}
&\qti{C_T}{\#_{12,4}}\ &\approx&\ &269&\,,\quad&\qti{C_T}{\#_{12,6}}\ &\approx&\ 647&\,,\quad&\qti{C_T}{\#_{12,8}}\ &\approx&\ 1.18\cdot 10^{3}&\,,\\
&\qti{C_T}{\#_{12,10}}\ &\approx&\ &1.87\cdot 10^{3}&\,,\quad&\qti{C_T}{\#_{12,12}} &\approx&\ 2.71\cdot 10^{3}&\,,\quad&\qti{C_T}{\#_{12,14}} &\approx&\ 3.71\cdot 10^{3}&\,,\\
&\qti{C_T}{\#_{14,2}}\ &\approx&\ &67.9&\,,\quad&\qti{C_T}{\#_{14,4}}\ &\approx&\ 365&\,,\quad &\qti{C_T}{\#_{14,6}}\ &\approx&\ 879&\,,\\
&\qti{C_T}{\#_{14,8}}\ &\approx&\ &1.61\cdot 10^{3}&\,,\quad&\qti{C_T}{\#_{14,10}} &\approx&\ 2.54\cdot 10^{3}&\,,\quad&\qti{C_T}{\#_{14,12}} &\approx&\ 3.69\cdot 10^{3}&\,,\\
&\qti{C_T}{\#_{14,14}} &\approx&\ &5.05\cdot 10^{3}&\,.& & & & & & &
\end{alignat}
The values of $\qti{F_0}{\#N,M}$ read~\cite{Fluder:2018chf}:
\begin{alignat}{14}
&\qti{F_0}{\#_{2,2}}\ &\approx&\ &11.1&\,,\quad
&\qti{F_0}{\#_{2,4}}\ \ &\approx&\ &\ \ \ \ \ \ 55.5&\,,\quad
&\qti{F_0}{\#_{2,6}}\ &\approx&\ &129&\,,\\
&\qti{F_0}{\#_{2,8}}\ &\approx&\ &231&\,,\quad
&\qti{F_0}{\#_{2,10}}\ &\approx&\ &\ \ \ \ \ \ \ 360&\,,\quad
&\qti{F_0}{\#_{2,12}}\ &\approx&\ &518&\,,\\
&\qti{F_0}{\#_{2,14}}\ &\approx&\ &703&\,,\quad
&\qti{F_0}{\#_{4,2}}\ \ &\approx&\ &\ \ \ \ \ \ \ 560&\,,\quad
&\qti{F_0}{\#_{4,4}}\ &\approx&\ &293&\,,\\
&\qti{F_0}{\#_{4,6}}\ &\approx&\ &691&\,,\quad
&\qti{F_0}{\#_{4,8}}\ \ &\approx&\ &1.24\cdot 10^{3}&\,,\quad
&\qti{F_0}{\#_{4,10}}\ &\approx&\ &1.95\cdot 10^{3}&\,,\\
&\qti{F_0}{\#_{4,12}}\ &\approx&\ &2.81\cdot 10^{3}&\,,\quad
&\qti{F_0}{\#_{4,14}}\ &\approx&\ &3.83\cdot 10^{3}&\,,\quad
&\qti{F_0}{\#_{6,2}}\ &\approx&\ &132&\,,\\
&\qti{F_0}{\#_{6,4}}\ &\approx&\ &697&\,,\quad
&\qti{F_0}{\#_{6,6}}\ \ &\approx&\ &1.65\cdot 10^{3}&\,,\quad
&\qti{F_0}{\#_{6,8}}\ &\approx&\ &2.99\cdot 10^{3}&\,,\\
&\qti{F_0}{\#_{6,10}}\ &\approx&\ &4.70\cdot 10^{3}&\,,\quad
&\qti{F_0}{\#_{6,12}}\ &\approx&\ &6.79\cdot 10^{3}&\,,\quad
&\qti{F_0}{\#_{6,14}}\ &\approx&\ &9.24\cdot 10^{3}&\,,\\
&\qti{F_0}{\#_{8,2}}\ &\approx&\ &240&\,,\quad
&\qti{F_0}{\#_{8,4}}\ \ &\approx&\ &1.27\cdot 10^{3}&\,,\quad
&\qti{F_0}{\#_{8,6}}\ &\approx&\ &3.01\cdot 10^{3}&\,,\\
&\qti{F_0}{\#_{8,8}}\ &\approx&\ &5.45\cdot 10^{3}&\,,\quad
&\qti{F_0}{\#_{8,10}}\ &\approx&\ &8.59\cdot 10^{3}&\,,\quad
&\qti{F_0}{\#_{8,12}}\ &\approx&\ &1.24\cdot 10^{4}&\,,\\
&\qti{F_0}{\#_{8,14}}\ &\approx&\ &1.69\cdot 10^{4}&\,,\quad
&\qti{F_0}{\#_{10,2}}\ &\approx&\ &\ \ \ \ \ \ \ 378&\,,\quad
&\qti{F_0}{\#_{10,4}}\ &\approx&\ &2.00\cdot 10^{3}&\,,\\
&\qti{F_0}{\#_{10,6}}\ &\approx&\ &4.76\cdot 10^{3}&\,,\quad
&\qti{F_0}{\#_{10,8}}\ &\approx&\ &8.63\cdot 10^{3}&\,,\quad
&\qti{F_0}{\#_{10,10}}\ &\approx&\ &1.36\cdot 10^{4}&\,,\\
&\qti{F_0}{\#_{10,12}}\ &\approx&\ &1.97\cdot 10^{4}&\,,\quad
&\qti{F_0}{\#_{10,14}} &\approx&\ &2.69\cdot 10^{4}&\,,\quad
&\qti{F_0}{\#_{12,2}}\ &\approx&\ &548&\,,\\
&\qti{F_0}{\#_{12,4}}\ &\approx&\ &2.90\cdot 10^{3}&\,,\quad
&\qti{F_0}{\#_{12,6}}\ &\approx&\ &6.90\cdot 10^{3}&\,,\quad
&\qti{F_0}{\#_{12,8}}\ &\approx&\ &1.25\cdot 10^{4}&\,,\\
&\qti{F_0}{\#_{12,10}}\ &\approx&\ &1.98\cdot 10^{4}&\,,\quad
&\qti{F_0}{\#_{12,12}} &\approx&\ &2.86\cdot 10^{4}&\,,\quad
&\qti{F_0}{\#_{12,14}}\ &\approx&\ &3.90\cdot 10^{4}&\,,\\
&\qti{F_0}{\#_{14,2}}\ &\approx&\ &749&\,,\quad
&\qti{F_0}{\#_{14,4}}\ &\approx&\ &3.97\cdot 10^{3}&\,,\quad
&\qti{F_0}{\#_{14,6}}\ &\approx&\ &9.44\cdot 10^{3}&\,,\\
&\qti{F_0}{\#_{14,8}}\ &\approx&\ &1.71\cdot 10^{4}&\,,\quad
&\qti{F_0}{\#_{14,10}} &\approx&\ &2.70\cdot 10^{4}&\,,\quad
&\qti{F_0}{\#_{14,12}}\ &\approx&\ &3.91\cdot 10^{4}&\,,\\
&\qti{F_0}{\#_{14,14}}\ &\approx&\ &5.34\cdot 10^{4}&\,.
\end{alignat}
From this, we find the ratios $\qti{C_T/F_0}{\#N, M}$\footnote{In Fig.~\ref{fig:TNalmo} a plot of $\qti{C_T/F_0}{\# N, M }$ as a function of $N$ and fixed values of $M$ is presented.}
\begin{alignat}{12}
&\qti{\frac{C_T}{F_0}}{\#_{2,2}}\ &\approx&\ &0.1133&\,,\quad
&\qti{\frac{C_T}{F_0}}{\#_{2,4}}\ &\approx&\ &0.1083&\,,\quad
&\qti{\frac{C_T}{F_0}}{\#_{2,6}}\ &\approx&\ &0.1079&\,,\\
&\qti{\frac{C_T}{F_0}}{\#_{2,8}}\ &\approx&\ &0.1082&\,,\quad
&\qti{\frac{C_T}{F_0}}{\#_{2,10}} &\approx&\ &0.1088&\,,\quad
&\qti{\frac{C_T}{F_0}}{\#_{2,12}}\ &\approx&\ &0.1093&\,,\\
&\qti{\frac{C_T}{F_0}}{\#_{2,14}}\ &\approx&\ &0.1097&\,,\quad
&\qti{\frac{C_T}{F_0}}{\#_{4,2}}\ &\approx&\ &0.1026&\,,\quad
&\qti{\frac{C_T}{F_0}}{\#_{4,4}}\ &\approx&\ &0.1000&\,,
\end{alignat}
\begin{alignat}{12}
&\qti{\frac{C_T}{F_0}}{\#_{4,6}}\ &\approx&\ &0.0997&\,,\quad
&\qti{\frac{C_T}{F_0}}{\#_{4,8}}\ \ &\approx&\ &0.0999&\,,\quad
&\qti{\frac{C_T}{F_0}}{\#_{4,10}}\ &\approx&\ &0.1002&\,,\\
&\qti{\frac{C_T}{F_0}}{\#_{4,12}}\ &\approx&\ &0.1005&\,,\quad
&\qti{\frac{C_T}{F_0}}{\#_{4,14}}\ &\approx&\ &0.1008&\,,\quad
&\qti{\frac{C_T}{F_0}}{\#_{6,2}}\ &\approx&\ &0.0980&\,,\\
&\qti{\frac{C_T}{F_0}}{\#_{6,4}}\ &\approx&\ &0.0970&\,,\quad
&\qti{\frac{C_T}{F_0}}{\#_{6,6}}\ \ &\approx&\ &0.0970&\,,\quad
&\qti{\frac{C_T}{F_0}}{\#_{6,8}}\ &\approx&\ &0.0972&\,,\\
&\qti{\frac{C_T}{F_0}}{\#_{6,10}}\ &\approx&\ &0.0974&\,,\quad
&\qti{\frac{C_T}{F_0}}{\#_{6,12}} \ &\approx&\ &0.0977&\,,\quad
&\qti{\frac{C_T}{F_0}}{\#_{6,14}}\ &\approx&\ &0.0979&\,,\\
&\qti{\frac{C_T}{F_0}}{\#_{8,2}}\ &\approx&\ &0.0951&\,,\quad
&\qti{\frac{C_T}{F_0}}{\#_{8,4}}\ \ &\approx&\ &0.0952&\,,\quad
&\qti{\frac{C_T}{F_0}}{\#_{8,6}}\ &\approx&\ &0.0955&\,,\\
&\qti{\frac{C_T}{F_0}}{\#_{8,8}}\ &\approx&\ &0.0958&\,,\quad
&\qti{\frac{C_T}{F_0}}{\#_{8,10}} \ &\approx&\ &0.0960&\,,\quad
&\qti{\frac{C_T}{F_0}}{\#_{8,12}}\ &\approx&\ &0.0962&\,,\\
&\qti{\frac{C_T}{F_0}}{\#_{8,14}}\ &\approx&\ &0.0964&\,,\quad
&\qti{\frac{C_T}{F_0}}{\#_{10,2}} \ &\approx&\ &0.0932&\,,\quad
&\qti{\frac{C_T}{F_0}}{\#_{10,4}}\ &\approx&\ &0.0939&\,,\\
&\qti{\frac{C_T}{F_0}}{\#_{10,6}}\ &\approx&\ &0.0945&\,,\quad
&\qti{\frac{C_T}{F_0}}{\#_{10,8}} \ &\approx&\ &0.0949&\,,\quad
&\qti{\frac{C_T}{F_0}}{\#_{10,10}}\ &\approx&\ &0.0952&\,,\\
&\qti{\frac{C_T}{F_0}}{\#_{10,12}}\ &\approx&\ &0.0954&\,,\quad
&\qti{\frac{C_T}{F_0}}{\#_{10,14}} &\approx&\ &0.0955&\,,\quad
&\qti{\frac{C_T}{F_0}}{\#_{12,2}}\ &\approx&\ &0.0917&\,,\\
&\qti{\frac{C_T}{F_0}}{\#_{12,4}}\ &\approx&\ &0.0929&\,,\quad
&\qti{\frac{C_T}{F_0}}{\#_{12,6}}\ &\approx&\ &0.0937&\,,\quad
&\qti{\frac{C_T}{F_0}}{\#_{12,8}}\ &\approx&\ &0.0942&\,,\\
&\qti{\frac{C_T}{F_0}}{\#_{12,10}}\ &\approx&\ &0.0945&\,,\quad
&\qti{\frac{C_T}{F_0}}{\#_{12,12}} &\approx&\ &0.0948&\,,\quad
&\qti{\frac{C_T}{F_0}}{\#_{12,14}}\ &\approx&\ &0.0950&\,,\\
&\qti{\frac{C_T}{F_0}}{\#_{14,2}}\ &\approx&\ &0.0906&\,,\quad
&\qti{\frac{C_T}{F_0}}{\#_{14,4}}\ &\approx&\ &0.0921&\,,\quad
&\qti{\frac{C_T}{F_0}}{\#_{14,6}}\ &\approx&\ &0.0931&\,,\\
&\qti{\frac{C_T}{F_0}}{\#_{14,8}}\ &\approx&\ &0.0937&\,,\quad
&\qti{\frac{C_T}{F_0}}{\#_{14,10}} &\approx&\ &0.0941&\,,\quad
&\qti{\frac{C_T}{F_0}}{\#_{14,12}}\ &\approx&\ &0.0943&\,,\\
&\qti{\frac{C_T}{F_0}}{\#_{14,14}}\ &\approx&\ &0.0945&\,.\quad
\end{alignat}

\section{Holographic cone entanglement in $d=5$ }\label{holoCorner}
The EE of a region with a conical entangling surface in a $d=5$ CFT dual to Einstein gravity was computed in~\cite{Myers:2012vs}. The result is given by
\begin{equation}
S_{\delta}=\frac{ \pi L_\star^4}{ G}\left[\frac{H^3\sin^2\Omega}{9\delta^3}-\frac{4H\cos^2\Omega}{9\delta}\right]+a_{(5)}(\Omega)\log\frac{H}{\delta}+\mathcal{O}(\delta^0)  
\end{equation}
where the cone function is given implicitly by the expression
\begin{equation}
 a_{(5)}(\Omega)=\frac{\pi L_\star^4}{G}\left[\frac{4\cos^2\Omega}{9h_0}-\frac{\sin^2\Omega}{3h_0^3}-\int_{0}^{h_0}\diff h\left[\frac{\sin^2\theta}{\dot{h}h^4}\sqrt{1+h^2+\dot{h}^2}+\frac{\sin^2\Omega}{h^4}-\frac{4\cos^2\Omega}{9h^2}\right]  \right] \, .
\end{equation}
Here, $h(0)\equiv h_0$, and $h(\theta)$ parametrizes the non-trivial profile of the corresponding Ryu-Takayanagi surface \cite{Ryu:2006bv}. This is found by  solving the following differential equation,
\begin{eqnarray}
   h(1+h^2) \ddot{h}\sin\theta& \notag &+2h\dot{h}^3\cos\theta+2(2+h^2)\dot{h}^2\sin\theta\\
   & &+2h(1+h^2)\dot{h}\cos\theta+(4+7h^2+3h^4)\sin\theta=0\, ,
   \end{eqnarray}
  for each possible value of the cone opening angle $\Omega$, subject to the conditions
\begin{equation}
    \lim_{\theta\rightarrow\Omega}h(\theta)=0\, , \quad \dot h(0)=0\, .
\end{equation}
The result of the numerical evaluation of $a_{(5)}(\Omega)$ is displayed in Fig.~\ref{fig:hee2}. 

As explained in the main text, the result of the numerical integration passes a highly non-trivial check, namely, as $\Omega\rightarrow\pi/2$, which corresponds to a very open cone, it approaches very precisely the result  $a_{(5)}(\Omega)=  4\pi^4C_T(\Omega-\pi/2)^2/270+\dots$ predicted to hold for general CFT$_5$'s in
\cite{Bueno:2015lza,Faulkner:2015csl}. On the other hand, in the very-sharp cone limit we find a best fit of $a_{(5)}(\Omega)= \nu/\Omega+\dots$ with $\nu\approx1.36\approx {\rm e}^{-2.01} \pi^5/30$, which agrees reasonably well with the result originally reported in \cite{Myers:2012vs}, namely, $\nu\approx {\rm e}^{-2.1} \pi^5/30$.

\bibliographystyle{JHEP-2}
\bibliography{Biblio}

\providecommand{\href}[2]{#2}\begingroup\raggedright\begin{thebibliography}{100}

\bibitem{Haag:1963dh}
R.~Haag and D.~Kastler, {\it {An Algebraic approach to quantum field theory}},
  {\em J. Math. Phys.} {\bf 5} (1964) 848--861.

\bibitem{Casini:2022rlv}
H.~Casini and M.~Huerta, {\it {Lectures on entanglement in quantum field
  theory}},  {\em PoS} {\bf TASI2021} (2023) 002
  [\href{http://arXiv.org/abs/2201.13310}{{\tt 2201.13310}}].

\bibitem{Liu:2012eea}
H.~Liu and M.~Mezei, {\it {A Refinement of entanglement entropy and the number
  of degrees of freedom}},  {\em JHEP} {\bf 04} (2013) 162
  [\href{http://arXiv.org/abs/1202.2070}{{\tt 1202.2070}}].

\bibitem{Calabrese:2004eu}
P.~Calabrese and J.~L. Cardy, {\it {Entanglement entropy and quantum field
  theory}},  {\em J. Stat. Mech.} {\bf 0406} (2004) P06002
  [\href{http://arXiv.org/abs/hep-th/0405152}{{\tt hep-th/0405152}}].

\bibitem{Solodukhin:2008dh}
S.~N. Solodukhin, {\it {Entanglement entropy, conformal invariance and
  extrinsic geometry}},  {\em Phys. Lett.} {\bf B665} (2008) 305--309
  [\href{http://arXiv.org/abs/0802.3117}{{\tt 0802.3117}}].

\bibitem{Safdi:2012sn}
B.~R. Safdi, {\it {Exact and Numerical Results on Entanglement Entropy in
  (5+1)-Dimensional CFT}},  {\em JHEP} {\bf 12} (2012) 005
  [\href{http://arXiv.org/abs/1206.5025}{{\tt 1206.5025}}].

\bibitem{Miao:2015iba}
R.-X. Miao, {\it {Universal Terms of Entanglement Entropy for 6d CFTs}},  {\em
  JHEP} {\bf 10} (2015) 049 [\href{http://arXiv.org/abs/1503.05538}{{\tt
  1503.05538}}].

\bibitem{Dowker:2010bu}
J.~S. Dowker, {\it {Entanglement entropy for even spheres}},
  \href{http://arXiv.org/abs/1009.3854}{{\tt 1009.3854}}.

\bibitem{Casini:2011kv}
H.~Casini, M.~Huerta and R.~C. Myers, {\it {Towards a derivation of holographic
  entanglement entropy}},  {\em JHEP} {\bf 05} (2011) 036
  [\href{http://arXiv.org/abs/1102.0440}{{\tt 1102.0440}}].

\bibitem{Witten:2018zxz}
E.~Witten, {\it {APS Medal for Exceptional Achievement in Research: Invited
  article on entanglement properties of quantum field theory}},  {\em Rev. Mod.
  Phys.} {\bf 90} (2018), no.~4 045003
  [\href{http://arXiv.org/abs/1803.04993}{{\tt 1803.04993}}].

\bibitem{Long:2016vkg}
J.~Long, {\it {On co-dimension two defect operators}},
  \href{http://arXiv.org/abs/1611.02485}{{\tt 1611.02485}}.

\bibitem{Chen:2016mya}
B.~Chen and J.~Long, {\it {R{\'e}nyi mutual information for a free scalar field
  in even dimensions}},  {\em Phys. Rev. D} {\bf 96} (2017), no.~4 045006
  [\href{http://arXiv.org/abs/1612.00114}{{\tt 1612.00114}}].

\bibitem{Chen:2017hbk}
B.~Chen, L.~Chen, P.-x. Hao and J.~Long, {\it {On the Mutual Information in
  Conformal Field Theory}},  {\em JHEP} {\bf 06} (2017) 096
  [\href{http://arXiv.org/abs/1704.03692}{{\tt 1704.03692}}].

\bibitem{Cardy:2013nua}
J.~Cardy, {\it {Some results on the mutual information of disjoint regions in
  higher dimensions}},  {\em J. Phys. A} {\bf 46} (2013) 285402
  [\href{http://arXiv.org/abs/1304.7985}{{\tt 1304.7985}}].

\bibitem{Agon:2015ftl}
C.~Ag{\'o}n and T.~Faulkner, {\it {Quantum Corrections to Holographic Mutual
  Information}},  {\em JHEP} {\bf 08} (2016) 118
  [\href{http://arXiv.org/abs/1511.07462}{{\tt 1511.07462}}].

\bibitem{Agon:2021zvp}
C.~A. Ag\'on, P.~Bueno and H.~Casini, {\it {Is the EMI model a QFT? An inquiry
  on the space of allowed entropy functions}},  {\em JHEP} {\bf 08} (2021) 084
  [\href{http://arXiv.org/abs/2105.11464}{{\tt 2105.11464}}].

\bibitem{Casini:2021raa}
H.~Casini, E.~Test{\'e} and G.~Torroba, {\it {Mutual information
  superadditivity and unitarity bounds}},  {\em JHEP} {\bf 09} (2021) 046
  [\href{http://arXiv.org/abs/2103.15847}{{\tt 2103.15847}}].

\bibitem{Agon:2021lus}
C.~A. Ag{\'o}n, P.~Bueno and H.~Casini, {\it {Tripartite information at long
  distances}},  {\em SciPost Phys.} {\bf 12} (2022), no.~5 153
  [\href{http://arXiv.org/abs/2109.09179}{{\tt 2109.09179}}].

\bibitem{Agon:2022efa}
C.~A. Ag{\'o}n, P.~Bueno, O.~Lasso~Andino and A.~Vilar~L{\'o}pez, {\it {Aspects
  of N-partite information in conformal field theories}},  {\em JHEP} {\bf 03}
  (2023) 246 [\href{http://arXiv.org/abs/2209.14311}{{\tt 2209.14311}}].

\bibitem{Agon:2024zae}
C.~A. Ag\'on, P.~Bueno and G.~van~der Velde, {\it {Long-distance N-partite
  information for fermionic CFTs}},  {\em JHEP} {\bf 12} (2024) 178
  [\href{http://arXiv.org/abs/2409.03821}{{\tt 2409.03821}}].

\bibitem{Agon:2024xvs}
C.~A. Agon, H.~Casini, U.~G{\"u}rsoy and G.~Planella~Planas, {\it {Mutual
  information from modular flow in CFTs}},  {\em JHEP} {\bf 08} (2025) 176
  [\href{http://arXiv.org/abs/2409.01406}{{\tt 2409.01406}}].

\bibitem{Casini:2007dk}
H.~Casini, {\it {Entropy localization and extensivity in the semiclassical
  black hole evaporation}},  {\em Phys. Rev. D} {\bf 79} (2009) 024015
  [\href{http://arXiv.org/abs/0712.0403}{{\tt 0712.0403}}].

\bibitem{Casini:2008wt}
H.~Casini and M.~Huerta, {\it {Remarks on the entanglement entropy for
  disconnected regions}},  {\em JHEP} {\bf 03} (2009) 048
  [\href{http://arXiv.org/abs/0812.1773}{{\tt 0812.1773}}].

\bibitem{Casini:2014yca}
H.~Casini, F.~D. Mazzitelli and E.~Test\'e, {\it {Area terms in entanglement
  entropy}},  {\em Phys. Rev. D} {\bf 91} (2015), no.~10 104035
  [\href{http://arXiv.org/abs/1412.6522}{{\tt 1412.6522}}].

\bibitem{Casini:2015woa}
H.~Casini, M.~Huerta, R.~C. Myers and A.~Yale, {\it {Mutual information and the
  F-theorem}},  {\em JHEP} {\bf 10} (2015) 003
  [\href{http://arXiv.org/abs/1506.06195}{{\tt 1506.06195}}].

\bibitem{Bueno:2021fxb}
P.~Bueno, H.~Casini, O.~L. Andino and J.~Moreno, {\it {Disks globally maximize
  the entanglement entropy in 2 + 1 dimensions}},  {\em JHEP} {\bf 10} (2021)
  179 [\href{http://arXiv.org/abs/2107.12394}{{\tt 2107.12394}}].

\bibitem{Bueno:2023gey}
P.~Bueno, H.~Casini, O.~L. Andino and J.~Moreno, {\it {Conformal Bounds in
  Three Dimensions from Entanglement Entropy}},  {\em Phys. Rev. Lett.} {\bf
  131} (2023), no.~17 171601 [\href{http://arXiv.org/abs/2307.05164}{{\tt
  2307.05164}}].

\bibitem{Mezei:2014zla}
M.~Mezei, {\it {Entanglement entropy across a deformed sphere}},  {\em Phys.
  Rev.} {\bf D91} (2015), no.~4 045038
  [\href{http://arXiv.org/abs/1411.7011}{{\tt 1411.7011}}].

\bibitem{Faulkner:2015csl}
T.~Faulkner, R.~G. Leigh and O.~Parrikar, {\it {Shape Dependence of
  Entanglement Entropy in Conformal Field Theories}},  {\em JHEP} {\bf 04}
  (2016) 088 [\href{http://arXiv.org/abs/1511.05179}{{\tt 1511.05179}}].

\bibitem{Casini:2009sr}
H.~Casini and M.~Huerta, {\it {Entanglement entropy in free quantum field
  theory}},  {\em J. Phys.} {\bf A42} (2009) 504007
  [\href{http://arXiv.org/abs/0905.2562}{{\tt 0905.2562}}].

\bibitem{Ryu:2006bv}
S.~Ryu and T.~Takayanagi, {\it {Holographic derivation of entanglement entropy
  from AdS/CFT}},  {\em Phys. Rev. Lett.} {\bf 96} (2006) 181602
  [\href{http://arXiv.org/abs/hep-th/0603001}{{\tt hep-th/0603001}}].

\bibitem{Ryu:2006ef}
S.~Ryu and T.~Takayanagi, {\it {Aspects of Holographic Entanglement Entropy}},
  {\em JHEP} {\bf 08} (2006) 045
  [\href{http://arXiv.org/abs/hep-th/0605073}{{\tt hep-th/0605073}}].

\bibitem{Nishioka:2009un}
T.~Nishioka, S.~Ryu and T.~Takayanagi, {\it {Holographic Entanglement Entropy:
  An Overview}},  {\em J. Phys. A} {\bf 42} (2009) 504008
  [\href{http://arXiv.org/abs/0905.0932}{{\tt 0905.0932}}].

\bibitem{Hung:2011xb}
L.-Y. Hung, R.~C. Myers and M.~Smolkin, {\it {On Holographic Entanglement
  Entropy and Higher Curvature Gravity}},  {\em JHEP} {\bf 04} (2011) 025
  [\href{http://arXiv.org/abs/1101.5813}{{\tt 1101.5813}}].

\bibitem{Myers:2012vs}
R.~C. Myers and A.~Singh, {\it {Entanglement Entropy for Singular Surfaces}},
  {\em JHEP} {\bf 09} (2012) 013 [\href{http://arXiv.org/abs/1206.5225}{{\tt
  1206.5225}}].

\bibitem{Anastasiou:2024rxe}
G.~Anastasiou, I.~J. Araya, P.~Bueno, J.~Moreno, R.~Olea and A.~Vilar~Lopez,
  {\it {Higher-dimensional Willmore energy as holographic entanglement
  entropy}},  {\em JHEP} {\bf 01} (2025) 081
  [\href{http://arXiv.org/abs/2409.19485}{{\tt 2409.19485}}].

\bibitem{willmore1965note}
T.~J. Willmore, {\it Note on embedded surfaces},  {\em An. Sti. Univ.“Al. I.
  Cuza” Iasi Sect. I a Mat.(NS) B} {\bf 11} (1965), no.~493-496 20.

\bibitem{willmore1996riemannian}
T.~Willmore, {\em Riemannian Geometry}.
\newblock Oxford science publications. Clarendon Press, 1996.

\bibitem{marques2014min}
F.~C. Marques and A.~Neves, {\it Min-max theory and the willmore conjecture},
  {\em Annals of mathematics} (2014) 683--782.

\bibitem{Toda2017Willmore}
M.~Toda, {\em Willmore Energy and Willmore Conjecture}.
\newblock Chapman and Hall/CRC, 2017.

\bibitem{Babich:1992mc}
M.~Babich and A.~Bobenko, {\it {Willmore tori with umbilic lines and minimal
  surfaces in hyperbolic space}}, .

\bibitem{alexakis2008renormalized}
S.~Alexakis and R.~Mazzeo, {\it Renormalized area and properly embedded minimal
  surfaces in hyperbolic 3-manifolds},  2008.

\bibitem{Astaneh:2014uba}
A.~F. Astaneh, G.~Gibbons and S.~N. Solodukhin, {\it {What surface maximizes
  entanglement entropy?}},  {\em Phys. Rev.} {\bf D90} (2014), no.~8 085021
  [\href{http://arXiv.org/abs/1407.4719}{{\tt 1407.4719}}].

\bibitem{guven2005conformally}
J.~Guven, {\it Conformally invariant bending energy for hypersurfaces},  {\em
  Journal of Physics A: Mathematical and General} {\bf 38} (2005), no.~37
  7943--7955.

\bibitem{Graham:2017bew}
C.~R. Graham and N.~Reichert, {\it {Higher-dimensional Willmore energies via
  minimal submanifold asymptotics}},  {\em Asian J. Math.} {\bf 24} (2020),
  no.~4 571--610 [\href{http://arXiv.org/abs/1704.03852}{{\tt 1704.03852}}].

\bibitem{Zhang:2017lcd}
Y.~Zhang, {\it {Graham-Witten's Conformal Invariant for Closed Four Dimensional
  Submanifolds}},  {\em J. Math. Study} {\bf 54} (2021), no.~2 200--226
  [\href{http://arXiv.org/abs/1703.08611}{{\tt 1703.08611}}].

\bibitem{Fonda:2014cca}
P.~Fonda, L.~Giomi, A.~Salvio and E.~Tonni, {\it {On shape dependence of
  holographic mutual information in AdS$_{4}$}},  {\em JHEP} {\bf 02} (2015)
  005 [\href{http://arXiv.org/abs/1411.3608}{{\tt 1411.3608}}].

\bibitem{Fonda:2015nma}
P.~Fonda, D.~Seminara and E.~Tonni, {\it {On shape dependence of holographic
  entanglement entropy in AdS$_{4}$/CFT$_{3}$}},  {\em JHEP} {\bf 12} (2015)
  037 [\href{http://arXiv.org/abs/1510.03664}{{\tt 1510.03664}}].

\bibitem{Seminara:2018pmr}
D.~Seminara, J.~Sisti and E.~Tonni, {\it {Holographic entanglement entropy in
  AdS$_{4}$/BCFT$_{3}$ and the Willmore functional}},  {\em JHEP} {\bf 08}
  (2018) 164 [\href{http://arXiv.org/abs/1805.11551}{{\tt 1805.11551}}].

\bibitem{Anastasiou:2020smm}
G.~Anastasiou, J.~Moreno, R.~Olea and D.~Rivera-Betancour, {\it {Shape
  dependence of renormalized holographic entanglement entropy}},  {\em JHEP}
  {\bf 09} (2020) 173 [\href{http://arXiv.org/abs/2002.06111}{{\tt
  2002.06111}}].

\bibitem{Anastasiou:2021swo}
G.~Anastasiou, I.~J. Araya, J.~Moreno, R.~Olea and D.~Rivera-Betancour, {\it
  {Renormalized holographic entanglement entropy for quadratic curvature
  gravity}},  {\em Phys. Rev. D} {\bf 104} (2021), no.~8 086003
  [\href{http://arXiv.org/abs/2102.11242}{{\tt 2102.11242}}].

\bibitem{Martino:2023oog}
D.~Martino, {\it {A duality theorem for a four dimensional Willmore energy}},
  \href{http://arXiv.org/abs/2308.11433}{{\tt 2308.11433}}.

\bibitem{lan2025analysis}
T.~Lan, D.~Martino and T.~Rivi{\`e}re, {\it The analysis of willmore surfaces
  and its generalizations in higher dimensions},  {\em arXiv preprint
  arXiv:2511.01777} (2025).

\bibitem{Huerta:2022cqw}
M.~Huerta and G.~van~der Velde, {\it {Instability of universal terms in the
  entanglement entropy}},  {\em Phys. Rev. D} {\bf 105} (2022), no.~12 125021
  [\href{http://arXiv.org/abs/2204.09464}{{\tt 2204.09464}}].

\bibitem{Casini:2019kex}
H.~Casini, M.~Huerta, J.~M. Mag{\'a}n and D.~Pontello, {\it {Entanglement
  entropy and superselection sectors. Part I. Global symmetries}},  {\em JHEP}
  {\bf 02} (2020) 014 [\href{http://arXiv.org/abs/1905.10487}{{\tt
  1905.10487}}].

\bibitem{Casini:2019nmu}
H.~Casini, M.~Huerta, J.~M. Mag{\'a}n and D.~Pontello, {\it {Logarithmic
  coefficient of the entanglement entropy of a Maxwell field}},  {\em Phys.
  Rev. D} {\bf 101} (2020), no.~6 065020
  [\href{http://arXiv.org/abs/1911.00529}{{\tt 1911.00529}}].

\bibitem{Bueno:2019mex}
P.~Bueno, H.~Casini and W.~Witczak-Krempa, {\it {Generalizing the entanglement
  entropy of singular regions in conformal field theories}},  {\em JHEP} {\bf
  08} (2019) 069 [\href{http://arXiv.org/abs/1904.11495}{{\tt 1904.11495}}].

\bibitem{Casini:2005zv}
H.~Casini and M.~Huerta, {\it {Entanglement and alpha entropies for a massive
  scalar field in two dimensions}},  {\em J. Stat. Mech.} {\bf 0512} (2005)
  P12012 [\href{http://arXiv.org/abs/cond-mat/0511014}{{\tt
  cond-mat/0511014}}].

\bibitem{Bueno:2015rda}
P.~Bueno, R.~C. Myers and W.~Witczak-Krempa, {\it {Universality of corner
  entanglement in conformal field theories}},  {\em Phys. Rev. Lett.} {\bf 115}
  (2015) 021602 [\href{http://arXiv.org/abs/1505.04804}{{\tt 1505.04804}}].

\bibitem{Witczak-Krempa:2016jhc}
W.~Witczak-Krempa, L.~E. Hayward~Sierens and R.~G. Melko, {\it {Cornering
  gapless quantum states via their torus entanglement}},  {\em Phys. Rev.
  Lett.} {\bf 118} (2017), no.~7 077202
  [\href{http://arXiv.org/abs/1603.02684}{{\tt 1603.02684}}].

\bibitem{Berthiere:2019lks}
C.~Berthiere and W.~Witczak-Krempa, {\it {Relating bulk to boundary
  entanglement}},  {\em Phys. Rev. B} {\bf 100} (2019), no.~23 235112
  [\href{http://arXiv.org/abs/1907.11249}{{\tt 1907.11249}}].

\bibitem{Estienne:2021hbx}
B.~Estienne, J.-M. St\'ephan and W.~Witczak-Krempa, {\it {Cornering the
  universal shape of fluctuations}},  {\em Nature Commun.} {\bf 13} (2022),
  no.~1 287 [\href{http://arXiv.org/abs/2102.06223}{{\tt 2102.06223}}].

\bibitem{casini2005entanglement}
H.~Casini, C.~Fosco and M.~Huerta, {\it Entanglement and alpha entropies for a
  massive dirac field in two dimensions},  {\em Journal of Statistical
  Mechanics: Theory and Experiment} {\bf 2005} (2005), no.~07 P07007.

\bibitem{Bueno:2015lza}
P.~Bueno and R.~C. Myers, {\it {Universal entanglement for higher dimensional
  cones}},  {\em JHEP} {\bf 12} (2015) 168
  [\href{http://arXiv.org/abs/1508.00587}{{\tt 1508.00587}}].

\bibitem{Fradkin:2006mb}
E.~Fradkin and J.~E. Moore, {\it {Entanglement entropy of 2D conformal quantum
  critical points: hearing the shape of a quantum drum}},  {\em Phys. Rev.
  Lett.} {\bf 97} (2006) 050404
  [\href{http://arXiv.org/abs/cond-mat/0605683}{{\tt cond-mat/0605683}}].

\bibitem{Casini:2006hu}
H.~Casini and M.~Huerta, {\it {Universal terms for the entanglement entropy in
  2+1 dimensions}},  {\em Nucl. Phys. B} {\bf 764} (2007) 183--201
  [\href{http://arXiv.org/abs/hep-th/0606256}{{\tt hep-th/0606256}}].

\bibitem{Hirata:2006jx}
T.~Hirata and T.~Takayanagi, {\it {AdS/CFT and strong subadditivity of
  entanglement entropy}},  {\em JHEP} {\bf 02} (2007) 042
  [\href{http://arXiv.org/abs/hep-th/0608213}{{\tt hep-th/0608213}}].

\bibitem{Casini:2008as}
H.~Casini, M.~Huerta and L.~Leitao, {\it {Entanglement entropy for a Dirac
  fermion in three dimensions: Vertex contribution}},  {\em Nucl. Phys. B} {\bf
  814} (2009) 594--609 [\href{http://arXiv.org/abs/0811.1968}{{\tt
  0811.1968}}].

\bibitem{Bueno:2015xda}
P.~Bueno and R.~C. Myers, {\it {Corner contributions to holographic
  entanglement entropy}},  {\em JHEP} {\bf 08} (2015) 068
  [\href{http://arXiv.org/abs/1505.07842}{{\tt 1505.07842}}].

\bibitem{Hofman:2008ar}
D.~M. Hofman and J.~Maldacena, {\it {Conformal collider physics: Energy and
  charge correlations}},  {\em JHEP} {\bf 05} (2008) 012
  [\href{http://arXiv.org/abs/0803.1467}{{\tt 0803.1467}}].

\bibitem{Hofman:2016awc}
D.~M. Hofman, D.~Li, D.~Meltzer, D.~Poland and F.~Rejon-Barrera, {\it {A Proof
  of the Conformal Collider Bounds}},  {\em JHEP} {\bf 06} (2016) 111
  [\href{http://arXiv.org/abs/1603.03771}{{\tt 1603.03771}}].

\bibitem{Casini:2021zgr}
H.~Casini and J.~M. Magan, {\it {On completeness and generalized symmetries in
  quantum field theory}},  {\em Mod. Phys. Lett. A} {\bf 36} (2021), no.~36
  2130025 [\href{http://arXiv.org/abs/2110.11358}{{\tt 2110.11358}}].

\bibitem{El-Showk:2011xbs}
S.~El-Showk, Y.~Nakayama and S.~Rychkov, {\it {What Maxwell Theory in
  D{\ensuremath{<}}{\ensuremath{>}}4 teaches us about scale and conformal
  invariance}},  {\em Nucl. Phys. B} {\bf 848} (2011) 578--593
  [\href{http://arXiv.org/abs/1101.5385}{{\tt 1101.5385}}].

\bibitem{Casini:2014aia}
H.~Casini and M.~Huerta, {\it {Entanglement entropy for a Maxwell field:
  Numerical calculation on a two dimensional lattice}},  {\em Phys. Rev. D}
  {\bf 90} (2014), no.~10 105013 [\href{http://arXiv.org/abs/1406.2991}{{\tt
  1406.2991}}].

\bibitem{Minwalla:1997ka}
S.~Minwalla, {\it {Restrictions imposed by superconformal invariance on quantum
  field theories}},  {\em Adv. Theor. Math. Phys.} {\bf 2} (1998) 783--851
  [\href{http://arXiv.org/abs/hep-th/9712074}{{\tt hep-th/9712074}}].

\bibitem{Schlieder:1972qr}
S.~Schlieder and E.~Seiler, {\it {Remarks on the null plane development of a
  relativistic quantum field theory}},  {\em Commun. Math. Phys.} {\bf 25}
  (1972) 62--72.

\bibitem{Wall:2011hj}
A.~C. Wall, {\it {A proof of the generalized second law for rapidly changing
  fields and arbitrary horizon slices}},  {\em Phys. Rev. D} {\bf 85} (2012)
  104049 [\href{http://arXiv.org/abs/1105.3445}{{\tt 1105.3445}}]. [Erratum:
  Phys.Rev.D 87, 069904 (2013)].

\bibitem{Benedetti:2022aiw}
V.~Benedetti, H.~Casini and P.~J. Martinez, {\it {Mutual information of
  generalized free fields}},  {\em Phys. Rev. D} {\bf 107} (2023), no.~4 046003
  [\href{http://arXiv.org/abs/2210.00013}{{\tt 2210.00013}}].

\bibitem{Bousso:2014uxa}
R.~Bousso, H.~Casini, Z.~Fisher and J.~Maldacena, {\it {Entropy on a null
  surface for interacting quantum field theories and the Bousso bound}},  {\em
  Phys. Rev. D} {\bf 91} (2015), no.~8 084030
  [\href{http://arXiv.org/abs/1406.4545}{{\tt 1406.4545}}].

\bibitem{Atiyah:1978ri}
M.~F. Atiyah, N.~J. Hitchin, V.~G. Drinfeld and Y.~I. Manin, {\it {Construction
  of Instantons}},  {\em Phys. Lett. A} {\bf 65} (1978) 185--187.

\bibitem{Perlmutter:2015vma}
E.~Perlmutter, M.~Rangamani and M.~Rota, {\it {Central Charges and the Sign of
  Entanglement in 4D Conformal Field Theories}},  {\em Phys. Rev. Lett.} {\bf
  115} (2015), no.~17 171601 [\href{http://arXiv.org/abs/1506.01679}{{\tt
  1506.01679}}].

\bibitem{Osborn:1989td}
H.~Osborn, {\it {Derivation of a Four-dimensional $c$ Theorem}},  {\em Phys.
  Lett. B} {\bf 222} (1989) 97--102.

\bibitem{Bueno:2015ofa}
P.~Bueno and W.~Witczak-Krempa, {\it {Bounds on corner entanglement in quantum
  critical states}},  {\em Phys. Rev. B} {\bf 93} (2016) 045131
  [\href{http://arXiv.org/abs/1511.04077}{{\tt 1511.04077}}].

\bibitem{Anastasiou:2022pzm}
G.~Anastasiou, I.~J. Araya, A.~Argando\~na and R.~Olea, {\it {CFT correlators
  from shape deformations in Cubic Curvature Gravity}},  {\em JHEP} {\bf 11}
  (2022) 031 [\href{http://arXiv.org/abs/2208.00093}{{\tt 2208.00093}}].

\bibitem{Beccaria:2015uta}
M.~Beccaria and A.~A. Tseytlin, {\it {Conformal a-anomaly of some non-unitary
  6d superconformal theories}},  {\em JHEP} {\bf 09} (2015) 017
  [\href{http://arXiv.org/abs/1506.08727}{{\tt 1506.08727}}].

\bibitem{Osborn:2016bev}
H.~Osborn and A.~Stergiou, {\it {C$_{T}$ for non-unitary CFTs in higher
  dimensions}},  {\em JHEP} {\bf 06} (2016) 079
  [\href{http://arXiv.org/abs/1603.07307}{{\tt 1603.07307}}].

\bibitem{Giombi:2024zrt}
S.~Giombi, E.~Himwich, A.~Katsevich, I.~Klebanov and Z.~Sun, {\it {Sphere free
  energy of scalar field theories with cubic interactions}},  {\em JHEP} {\bf
  12} (2025) 133 [\href{http://arXiv.org/abs/2412.14086}{{\tt 2412.14086}}].

\bibitem{Strominger:2001pn}
A.~Strominger, {\it {The dS / CFT correspondence}},  {\em JHEP} {\bf 10} (2001)
  034 [\href{http://arXiv.org/abs/hep-th/0106113}{{\tt hep-th/0106113}}].

\bibitem{Witten:2001kn}
E.~Witten, {\it {Quantum gravity in de Sitter space}},  in {\em {Strings 2001:
  International Conference}}, 6, 2001.
\newblock \href{http://arXiv.org/abs/hep-th/0106109}{{\tt hep-th/0106109}}.

\bibitem{Maldacena:2002vr}
J.~M. Maldacena, {\it {Non-Gaussian features of primordial fluctuations in
  single field inflationary models}},  {\em JHEP} {\bf 05} (2003) 013
  [\href{http://arXiv.org/abs/astro-ph/0210603}{{\tt astro-ph/0210603}}].

\bibitem{Nakata:2020luh}
Y.~Nakata, T.~Takayanagi, Y.~Taki, K.~Tamaoka and Z.~Wei, {\it {New holographic
  generalization of entanglement entropy}},  {\em Phys. Rev. D} {\bf 103}
  (2021), no.~2 026005 [\href{http://arXiv.org/abs/2005.13801}{{\tt
  2005.13801}}].

\bibitem{Mollabashi:2020yie}
A.~Mollabashi, N.~Shiba, T.~Takayanagi, K.~Tamaoka and Z.~Wei, {\it {Pseudo
  Entropy in Free Quantum Field Theories}},  {\em Phys. Rev. Lett.} {\bf 126}
  (2021), no.~8 081601 [\href{http://arXiv.org/abs/2011.09648}{{\tt
  2011.09648}}].

\bibitem{Mollabashi:2021xsd}
A.~Mollabashi, N.~Shiba, T.~Takayanagi, K.~Tamaoka and Z.~Wei, {\it {Aspects of
  pseudoentropy in field theories}},  {\em Phys. Rev. Res.} {\bf 3} (2021),
  no.~3 033254 [\href{http://arXiv.org/abs/2106.03118}{{\tt 2106.03118}}].

\bibitem{Doi:2022iyj}
K.~Doi, J.~Harper, A.~Mollabashi, T.~Takayanagi and Y.~Taki, {\it
  {Pseudoentropy in dS/CFT and Timelike Entanglement Entropy}},  {\em Phys.
  Rev. Lett.} {\bf 130} (2023), no.~3 031601
  [\href{http://arXiv.org/abs/2210.09457}{{\tt 2210.09457}}].

\bibitem{Narayan:2022afv}
K.~Narayan, {\it {de Sitter space, extremal surfaces, and time entanglement}},
  {\em Phys. Rev. D} {\bf 107} (2023), no.~12 126004
  [\href{http://arXiv.org/abs/2210.12963}{{\tt 2210.12963}}].

\bibitem{Doi:2023zaf}
K.~Doi, J.~Harper, A.~Mollabashi, T.~Takayanagi and Y.~Taki, {\it {Timelike
  entanglement entropy}},  {\em JHEP} {\bf 05} (2023) 052
  [\href{http://arXiv.org/abs/2302.11695}{{\tt 2302.11695}}].

\bibitem{Caputa:2024gve}
P.~Caputa, B.~Chen, T.~Takayanagi and T.~Tsuda, {\it {Thermal pseudo-entropy}},
   {\em JHEP} {\bf 01} (2025) 003 [\href{http://arXiv.org/abs/2411.08948}{{\tt
  2411.08948}}].

\bibitem{Fujiki:2025rtx}
K.~Fujiki, M.~Kohara, K.~Shinmyo, Y.-k. Suzuki and T.~Takayanagi, {\it
  {Entropic Interpretation of Einstein Equation in dS/CFT}},
  \href{http://arXiv.org/abs/2511.07915}{{\tt 2511.07915}}.

\bibitem{Anastasiou:2025rvz}
G.~Anastasiou, I.~J. Araya, A.~Das and J.~Moreno, {\it {Universality of
  pseudoentropy for deformed spheres in dS/CFT}},
  \href{http://arXiv.org/abs/2512.02164}{{\tt 2512.02164}}.

\bibitem{Anastasiou:2026bbf}
G.~Anastasiou, I.~J. Araya, A.~Das and J.~Moreno, {\it {Renormalized
  pseudoentropy in dS/CFT}},  \href{http://arXiv.org/abs/2602.17989}{{\tt
  2602.17989}}.

\bibitem{Osborn:1993cr}
H.~Osborn and A.~C. Petkou, {\it {Implications of conformal invariance in field
  theories for general dimensions}},  {\em Annals Phys.} {\bf 231} (1994)
  311--362 [\href{http://arXiv.org/abs/hep-th/9307010}{{\tt hep-th/9307010}}].

\bibitem{Klebanov:2011gs}
I.~R. Klebanov, S.~S. Pufu and B.~R. Safdi, {\it {F-Theorem without
  Supersymmetry}},  {\em JHEP} {\bf 10} (2011) 038
  [\href{http://arXiv.org/abs/1105.4598}{{\tt 1105.4598}}].

\bibitem{Giombi:2014xxa}
S.~Giombi and I.~R. Klebanov, {\it {Interpolating between $a$ and $F$}},  {\em
  JHEP} {\bf 03} (2015) 117 [\href{http://arXiv.org/abs/1409.1937}{{\tt
  1409.1937}}].

\bibitem{Diab:2016spb}
K.~Diab, L.~Fei, S.~Giombi, I.~R. Klebanov and G.~Tarnopolsky, {\it {On
  ${C}_{J}$ and ${C}_{T}$ in the Gross–Neveu and O(N) models}},  {\em J.
  Phys.} {\bf A49} (2016), no.~40 405402
  [\href{http://arXiv.org/abs/1601.07198}{{\tt 1601.07198}}].

\bibitem{Seiberg:1996bd}
N.~Seiberg, {\it {Five-dimensional SUSY field theories, nontrivial fixed points
  and string dynamics}},  {\em Phys. Lett. B} {\bf 388} (1996) 753--760
  [\href{http://arXiv.org/abs/hep-th/9608111}{{\tt hep-th/9608111}}].

\bibitem{Morrison:1996xf}
D.~R. Morrison and N.~Seiberg, {\it {Extremal transitions and five-dimensional
  supersymmetric field theories}},  {\em Nucl. Phys. B} {\bf 483} (1997)
  229--247 [\href{http://arXiv.org/abs/hep-th/9609070}{{\tt hep-th/9609070}}].

\bibitem{Chang:2017cdx}
C.-M. Chang, M.~Fluder, Y.-H. Lin and Y.~Wang, {\it {Spheres, Charges,
  Instantons, and Bootstrap: A Five-Dimensional Odyssey}},  {\em JHEP} {\bf 03}
  (2018) 123 [\href{http://arXiv.org/abs/1710.08418}{{\tt 1710.08418}}].

\bibitem{Aharony:1997bh}
O.~Aharony, A.~Hanany and B.~Kol, {\it {Webs of (p,q) five-branes,
  five-dimensional field theories and grid diagrams}},  {\em JHEP} {\bf 01}
  (1998) 002 [\href{http://arXiv.org/abs/hep-th/9710116}{{\tt
  hep-th/9710116}}].

\bibitem{Benini:2009gi}
F.~Benini, S.~Benvenuti and Y.~Tachikawa, {\it {Webs of five-branes and N=2
  superconformal field theories}},  {\em JHEP} {\bf 09} (2009) 052
  [\href{http://arXiv.org/abs/0906.0359}{{\tt 0906.0359}}].

\bibitem{Mitev:2014isa}
V.~Mitev and E.~Pomoni, {\it {Toda 3-Point Functions From Topological
  Strings}},  {\em JHEP} {\bf 06} (2015) 049
  [\href{http://arXiv.org/abs/1409.6313}{{\tt 1409.6313}}].

\bibitem{Fluder:2018chf}
M.~Fluder and C.~F. Uhlemann, {\it {Precision Test of AdS$_6$/CFT$_5$ in Type
  IIB String Theory}},  {\em Phys. Rev. Lett.} {\bf 121} (2018), no.~17 171603
  [\href{http://arXiv.org/abs/1806.08374}{{\tt 1806.08374}}].

\bibitem{Parisi:1975im}
G.~Parisi, {\it {The Theory of Nonrenormalizable Interactions. 1. The Large N
  Expansion}},  {\em Nucl. Phys. B} {\bf 100} (1975) 368--388.

\bibitem{Bekaert:2011cu}
X.~Bekaert, E.~Meunier and S.~Moroz, {\it {Towards a gravity dual of the
  unitary Fermi gas}},  {\em Phys. Rev. D} {\bf 85} (2012) 106001
  [\href{http://arXiv.org/abs/1111.1082}{{\tt 1111.1082}}].

\bibitem{Fitzpatrick:2013sya}
A.~L. Fitzpatrick, J.~Kaplan and D.~Poland, {\it {Conformal Blocks in the Large
  $D$ Limit}},  {\em JHEP} {\bf 08} (2013) 107
  [\href{http://arXiv.org/abs/1305.0004}{{\tt 1305.0004}}].

\bibitem{Cordova:2018eba}
C.~C{\'o}rdova, G.~B. De~Luca and A.~Tomasiello, {\it {AdS$_{8}$ solutions in
  type II supergravity}},  {\em JHEP} {\bf 07} (2019) 127
  [\href{http://arXiv.org/abs/1811.06987}{{\tt 1811.06987}}].

\bibitem{Fei:2014yja}
L.~Fei, S.~Giombi and I.~R. Klebanov, {\it {Critical $O(N)$ models in
  $6-\epsilon$ dimensions}},  {\em Phys. Rev.} {\bf D90} (2014), no.~2 025018
  [\href{http://arXiv.org/abs/1404.1094}{{\tt 1404.1094}}].

\bibitem{Zamolodchikov:1986gt}
A.~B. Zamolodchikov, {\it {Irreversibility of the Flux of the Renormalization
  Group in a 2D Field Theory}},  {\em JETP Lett.} {\bf 43} (1986) 730--732.
  [Pisma Zh. Eksp. Teor. Fiz.43,565(1986)].

\bibitem{Buchel:2009sk}
A.~Buchel, J.~Escobedo, R.~C. Myers, M.~F. Paulos, A.~Sinha and M.~Smolkin,
  {\it {Holographic GB gravity in arbitrary dimensions}},  {\em JHEP} {\bf 03}
  (2010) 111 [\href{http://arXiv.org/abs/0911.4257}{{\tt 0911.4257}}].

\bibitem{Casini:2012ei}
H.~Casini and M.~Huerta, {\it {On the RG running of the entanglement entropy of
  a circle}},  {\em Phys. Rev.} {\bf D85} (2012) 125016
  [\href{http://arXiv.org/abs/1202.5650}{{\tt 1202.5650}}].

\bibitem{Hartman:2015lfa}
T.~Hartman, S.~Jain and S.~Kundu, {\it {Causality Constraints in Conformal
  Field Theory}},  {\em JHEP} {\bf 05} (2016) 099
  [\href{http://arXiv.org/abs/1509.00014}{{\tt 1509.00014}}].

\bibitem{El-Showk:2012cjh}
S.~El-Showk, M.~F. Paulos, D.~Poland, S.~Rychkov, D.~Simmons-Duffin and
  A.~Vichi, {\it {Solving the 3D Ising Model with the Conformal Bootstrap}},
  {\em Phys. Rev. D} {\bf 86} (2012) 025022
  [\href{http://arXiv.org/abs/1203.6064}{{\tt 1203.6064}}].

\bibitem{Rattazzi:2008pe}
R.~Rattazzi, V.~S. Rychkov, E.~Tonni and A.~Vichi, {\it {Bounding scalar
  operator dimensions in 4D CFT}},  {\em JHEP} {\bf 12} (2008) 031
  [\href{http://arXiv.org/abs/0807.0004}{{\tt 0807.0004}}].

\bibitem{Poland:2018epd}
D.~Poland, S.~Rychkov and A.~Vichi, {\it {The Conformal Bootstrap: Theory,
  Numerical Techniques, and Applications}},  {\em Rev. Mod. Phys.} {\bf 91}
  (2019) 015002 [\href{http://arXiv.org/abs/1805.04405}{{\tt 1805.04405}}].

\bibitem{Casini:2017vbe}
H.~Casini, E.~Test\'e and G.~Torroba, {\it {Markov Property of the Conformal
  Field Theory Vacuum and the a Theorem}},  {\em Phys. Rev. Lett.} {\bf 118}
  (2017), no.~26 261602 [\href{http://arXiv.org/abs/1704.01870}{{\tt
  1704.01870}}].

\bibitem{Komargodski:2011vj}
Z.~Komargodski and A.~Schwimmer, {\it {On Renormalization Group Flows in Four
  Dimensions}},  {\em JHEP} {\bf 12} (2011) 099
  [\href{http://arXiv.org/abs/1107.3987}{{\tt 1107.3987}}].

\bibitem{Rychkov:2016iqz}
S.~Rychkov, {\em {EPFL Lectures on Conformal Field Theory in D{\ensuremath{>}}=
  3 Dimensions}}.
\newblock SpringerBriefs in Physics. 1, 2016.

\bibitem{Imamura:2011wg}
Y.~Imamura and D.~Yokoyama, {\it {N=2 supersymmetric theories on squashed
  three-sphere}},  {\em Phys. Rev. D} {\bf 85} (2012) 025015
  [\href{http://arXiv.org/abs/1109.4734}{{\tt 1109.4734}}].

\bibitem{Deser:1993yx}
S.~Deser and A.~Schwimmer, {\it {Geometric classification of conformal
  anomalies in arbitrary dimensions}},  {\em Phys. Lett. B} {\bf 309} (1993)
  279--284 [\href{http://arXiv.org/abs/hep-th/9302047}{{\tt hep-th/9302047}}].

\bibitem{Duff:1977ay}
M.~J. Duff, {\it {Observations on Conformal Anomalies}},  {\em Nucl. Phys. B}
  {\bf 125} (1977) 334--348.

\bibitem{Dowker:1976zf}
J.~S. Dowker and R.~Critchley, {\it {The Stress Tensor Conformal Anomaly for
  Scalar and Spinor Fields}},  {\em Phys. Rev. D} {\bf 16} (1977) 3390.

\bibitem{Aros:2026gms}
R.~Aros, F.~Bugini, D.~E. Diaz and C.~N{\'u}{\~n}ez-Barra, {\it {Universal
  relation between $C_{T}$ and the CFT Weyl anomaly}},
  \href{http://arXiv.org/abs/2603.00321}{{\tt 2603.00321}}.

\bibitem{Liu:1998bu}
H.~Liu and A.~A. Tseytlin, {\it {D = 4 superYang-Mills, D = 5 gauged
  supergravity, and D = 4 conformal supergravity}},  {\em Nucl. Phys.} {\bf
  B533} (1998) 88--108 [\href{http://arXiv.org/abs/hep-th/9804083}{{\tt
  hep-th/9804083}}].

\bibitem{Henningson:1998gx}
M.~Henningson and K.~Skenderis, {\it {The Holographic Weyl anomaly}},  {\em
  JHEP} {\bf 07} (1998) 023 [\href{http://arXiv.org/abs/hep-th/9806087}{{\tt
  hep-th/9806087}}].

\bibitem{Wilson:1971dc}
K.~G. Wilson and M.~E. Fisher, {\it {Critical exponents in 3.99 dimensions}},
  {\em Phys. Rev. Lett.} {\bf 28} (1972) 240--243.

\bibitem{tHooft:1973alw}
G.~'t~Hooft, {\it {A Planar Diagram Theory for Strong Interactions}},  {\em
  Nucl. Phys. B} {\bf 72} (1974) 461.

\bibitem{Petkou:1994ad}
A.~Petkou, {\it {Conserved currents, consistency relations and operator product
  expansions in the conformally invariant O(N) vector model}},  {\em Annals
  Phys.} {\bf 249} (1996) 180--221
  [\href{http://arXiv.org/abs/hep-th/9410093}{{\tt hep-th/9410093}}].

\bibitem{Jack:1983sk}
I.~Jack and H.~Osborn, {\it {Background Field Calculations in Curved
  Space-time. 1. General Formalism and Application to Scalar Fields}},  {\em
  Nucl. Phys. B} {\bf 234} (1984) 331--364.

\bibitem{Cappelli:1990yc}
A.~Cappelli, D.~Friedan and J.~I. Latorre, {\it {C theorem and spectral
  representation}},  {\em Nucl. Phys. B} {\bf 352} (1991) 616--670.

\bibitem{Fei:2014xta}
L.~Fei, S.~Giombi, I.~R. Klebanov and G.~Tarnopolsky, {\it {Three loop analysis
  of the critical O(N) models in 6-{\ensuremath{\varepsilon}} dimensions}},
  {\em Phys. Rev. D} {\bf 91} (2015), no.~4 045011
  [\href{http://arXiv.org/abs/1411.1099}{{\tt 1411.1099}}].

\bibitem{Fei:2015kta}
L.~Fei, S.~Giombi, I.~R. Klebanov and G.~Tarnopolsky, {\it {Critical Sp(N )
  models in 6 {\ensuremath{-}} {\ensuremath{\epsilon}} dimensions and higher
  spin dS/CFT}},  {\em JHEP} {\bf 09} (2015) 076
  [\href{http://arXiv.org/abs/1502.07271}{{\tt 1502.07271}}].

\bibitem{Simmons-Duffin:2016gjk}
D.~Simmons-Duffin, {\it {The Conformal Bootstrap}},  in {\em {Theoretical
  Advanced Study Institute in Elementary Particle Physics}: {New Frontiers in
  Fields and Strings}}, pp.~1--74, 2017.
\newblock \href{http://arXiv.org/abs/1602.07982}{{\tt 1602.07982}}.

\bibitem{Poland:2022qrs}
D.~Poland and D.~Simmons-Duffin, {\it {Snowmass White Paper: The Numerical
  Conformal Bootstrap}},  in {\em {Snowmass 2021}}, 3, 2022.
\newblock \href{http://arXiv.org/abs/2203.08117}{{\tt 2203.08117}}.

\bibitem{Hartman:2022zik}
T.~Hartman, D.~Mazac, D.~Simmons-Duffin and A.~Zhiboedov, {\it {Snowmass White
  Paper: The Analytic Conformal Bootstrap}},  in {\em {Snowmass 2021}}, 2,
  2022.
\newblock \href{http://arXiv.org/abs/2202.11012}{{\tt 2202.11012}}.

\bibitem{Kos:2013tga}
F.~Kos, D.~Poland and D.~Simmons-Duffin, {\it {Bootstrapping the $O(N)$ vector
  models}},  {\em JHEP} {\bf 06} (2014) 091
  [\href{http://arXiv.org/abs/1307.6856}{{\tt 1307.6856}}].

\bibitem{El-Showk:2014dwa}
S.~El-Showk, M.~F. Paulos, D.~Poland, S.~Rychkov, D.~Simmons-Duffin and
  A.~Vichi, {\it {Solving the 3d Ising Model with the Conformal Bootstrap II.
  c-Minimization and Precise Critical Exponents}},  {\em J. Stat. Phys.} {\bf
  157} (2014) 869 [\href{http://arXiv.org/abs/1403.4545}{{\tt 1403.4545}}].

\bibitem{Chang:2024whx}
C.-H. Chang, V.~Dommes, R.~S. Erramilli, A.~Homrich, P.~Kravchuk, A.~Liu, M.~S.
  Mitchell, D.~Poland and D.~Simmons-Duffin, {\it {Bootstrapping the 3d Ising
  stress tensor}},  {\em JHEP} {\bf 03} (2025) 136
  [\href{http://arXiv.org/abs/2411.15300}{{\tt 2411.15300}}].

\bibitem{Klebanov:2011td}
I.~R. Klebanov, S.~S. Pufu, S.~Sachdev and B.~R. Safdi, {\it {Entanglement
  Entropy of 3-d Conformal Gauge Theories with Many Flavors}},  {\em JHEP} {\bf
  05} (2012) 036 [\href{http://arXiv.org/abs/1112.5342}{{\tt 1112.5342}}].

\bibitem{Giombi:2015haa}
S.~Giombi, I.~R. Klebanov and G.~Tarnopolsky, {\it {Conformal QED$_d$,
  $F$-Theorem and the $\epsilon$ Expansion}},  {\em J. Phys. A} {\bf 49}
  (2016), no.~13 135403 [\href{http://arXiv.org/abs/1508.06354}{{\tt
  1508.06354}}].

\bibitem{Tarnopolsky:2016vvd}
G.~Tarnopolsky, {\it {Large $N$ expansion of the sphere free energy}},  {\em
  Phys. Rev. D} {\bf 96} (2017), no.~2 025017
  [\href{http://arXiv.org/abs/1609.09113}{{\tt 1609.09113}}].

\bibitem{Fraser-Taliente:2025udk}
L.~Fraser-Taliente, {\it {The sphere free energy of the vector models to order
  $1/N$}},  \href{http://arXiv.org/abs/2507.16896}{{\tt 2507.16896}}.

\bibitem{Bobev:2015vsa}
N.~Bobev, S.~El-Showk, D.~Mazac and M.~F. Paulos, {\it {Bootstrapping the
  Three-Dimensional Supersymmetric Ising Model}},  {\em Phys. Rev. Lett.} {\bf
  115} (2015), no.~5 051601 [\href{http://arXiv.org/abs/1502.04124}{{\tt
  1502.04124}}].

\bibitem{Nishioka:2013gza}
T.~Nishioka and K.~Yonekura, {\it {On RG Flow of $tau_{RR}$ for Supersymmetric
  Field Theories in Three-Dimensions}},  {\em JHEP} {\bf 05} (2013) 165
  [\href{http://arXiv.org/abs/1303.1522}{{\tt 1303.1522}}].

\bibitem{Aharony:1997bx}
O.~Aharony, A.~Hanany, K.~A. Intriligator, N.~Seiberg and M.~J. Strassler, {\it
  {Aspects of N=2 supersymmetric gauge theories in three-dimensions}},  {\em
  Nucl. Phys. B} {\bf 499} (1997) 67--99
  [\href{http://arXiv.org/abs/hep-th/9703110}{{\tt hep-th/9703110}}].

\bibitem{deBoer:1997ka}
J.~de~Boer, K.~Hori, Y.~Oz and Z.~Yin, {\it {Branes and mirror symmetry in N=2
  supersymmetric gauge theories in three-dimensions}},  {\em Nucl. Phys. B}
  {\bf 502} (1997) 107--124 [\href{http://arXiv.org/abs/hep-th/9702154}{{\tt
  hep-th/9702154}}].

\bibitem{Jafferis:2010un}
D.~L. Jafferis, {\it {The Exact Superconformal R-Symmetry Extremizes Z}},  {\em
  JHEP} {\bf 05} (2012) 159 [\href{http://arXiv.org/abs/1012.3210}{{\tt
  1012.3210}}].

\bibitem{Bobev:2015jxa}
N.~Bobev, S.~El-Showk, D.~Mazac and M.~F. Paulos, {\it {Bootstrapping SCFTs
  with Four Supercharges}},  {\em JHEP} {\bf 08} (2015) 142
  [\href{http://arXiv.org/abs/1503.02081}{{\tt 1503.02081}}].

\bibitem{Witczak-Krempa:2015jca}
W.~Witczak-Krempa and J.~Maciejko, {\it {Optical conductivity of topological
  surface states with emergent supersymmetry}},  {\em Phys. Rev. Lett.} {\bf
  116} (2016), no.~10 100402 [\href{http://arXiv.org/abs/1510.06397}{{\tt
  1510.06397}}]. [Addendum: Phys.Rev.Lett. 117, 149903 (2016)].

\bibitem{Jian:2016zll}
S.-K. Jian, C.-H. Lin, J.~Maciejko and H.~Yao, {\it {Emergence of
  supersymmetric quantum electrodynamics}},  {\em Phys. Rev. Lett.} {\bf 118}
  (2017), no.~16 166802 [\href{http://arXiv.org/abs/1609.02146}{{\tt
  1609.02146}}].

\bibitem{Aharony:2008ug}
O.~Aharony, O.~Bergman, D.~L. Jafferis and J.~Maldacena, {\it {N=6
  superconformal Chern-Simons-matter theories, M2-branes and their gravity
  duals}},  {\em JHEP} {\bf 10} (2008) 091
  [\href{http://arXiv.org/abs/0806.1218}{{\tt 0806.1218}}].

\bibitem{Kapustin:2009kz}
A.~Kapustin, B.~Willett and I.~Yaakov, {\it {Exact Results for Wilson Loops in
  Superconformal Chern-Simons Theories with Matter}},  {\em JHEP} {\bf 03}
  (2010) 089 [\href{http://arXiv.org/abs/0909.4559}{{\tt 0909.4559}}].

\bibitem{Fuji:2011km}
H.~Fuji, S.~Hirano and S.~Moriyama, {\it {Summing Up All Genus Free Energy of
  ABJM Matrix Model}},  {\em JHEP} {\bf 08} (2011) 001
  [\href{http://arXiv.org/abs/1106.4631}{{\tt 1106.4631}}].

\bibitem{Marino:2011eh}
M.~Marino and P.~Putrov, {\it {ABJM theory as a Fermi gas}},  {\em J. Stat.
  Mech.} {\bf 1203} (2012) P03001 [\href{http://arXiv.org/abs/1110.4066}{{\tt
  1110.4066}}].

\bibitem{Hanada:2012si}
M.~Hanada, M.~Honda, Y.~Honma, J.~Nishimura, S.~Shiba and Y.~Yoshida, {\it
  {Numerical studies of the ABJM theory for arbitrary N at arbitrary coupling
  constant}},  {\em JHEP} {\bf 05} (2012) 121
  [\href{http://arXiv.org/abs/1202.5300}{{\tt 1202.5300}}].

\bibitem{Marino:2016new}
M.~Marino, {\it {Localization at large N in Chern{\textendash}Simons-matter
  theories}},  {\em J. Phys. A} {\bf 50} (2017), no.~44 443007
  [\href{http://arXiv.org/abs/1608.02959}{{\tt 1608.02959}}].

\bibitem{Chester:2021gdw}
S.~M. Chester, R.~R. Kalloor and A.~Sharon, {\it {Squashing, Mass, and
  Holography for 3d Sphere Free Energy}},  {\em JHEP} {\bf 04} (2021) 244
  [\href{http://arXiv.org/abs/2102.05643}{{\tt 2102.05643}}].

\bibitem{Bobev:2022eus}
N.~Bobev, J.~Hong and V.~Reys, {\it {Large N partition functions of the ABJM
  theory}},  {\em JHEP} {\bf 02} (2023) 020
  [\href{http://arXiv.org/abs/2210.09318}{{\tt 2210.09318}}].

\bibitem{Hama:2011ea}
N.~Hama, K.~Hosomichi and S.~Lee, {\it {SUSY Gauge Theories on Squashed
  Three-Spheres}},  {\em JHEP} {\bf 05} (2011) 014
  [\href{http://arXiv.org/abs/1102.4716}{{\tt 1102.4716}}].

\bibitem{Nosaka:2015iiw}
T.~Nosaka, {\it {Instanton effects in ABJM theory with general R-charge
  assignments}},  {\em JHEP} {\bf 03} (2016) 059
  [\href{http://arXiv.org/abs/1512.02862}{{\tt 1512.02862}}].

\bibitem{Bobev:2021oku}
N.~Bobev, A.~M. Charles, K.~Hristov and V.~Reys, {\it {Higher-derivative
  supergravity, AdS$_{4}$ holography, and black holes}},  {\em JHEP} {\bf 08}
  (2021) 173 [\href{http://arXiv.org/abs/2106.04581}{{\tt 2106.04581}}].

\bibitem{Agmon:2017xes}
N.~B. Agmon, S.~M. Chester and S.~S. Pufu, {\it {Solving M-theory with the
  Conformal Bootstrap}},  {\em JHEP} {\bf 06} (2018) 159
  [\href{http://arXiv.org/abs/1711.07343}{{\tt 1711.07343}}].

\bibitem{Chester:2020jay}
S.~M. Chester, R.~R. Kalloor and A.~Sharon, {\it {3d $ \mathcal{N} $ = 4 OPE
  coefficients from Fermi gas}},  {\em JHEP} {\bf 07} (2020) 041
  [\href{http://arXiv.org/abs/2004.13603}{{\tt 2004.13603}}].

\bibitem{Closset:2012vg}
C.~Closset, T.~T. Dumitrescu, G.~Festuccia, Z.~Komargodski and N.~Seiberg, {\it
  {Contact Terms, Unitarity, and F-Maximization in Three-Dimensional
  Superconformal Theories}},  {\em JHEP} {\bf 10} (2012) 053
  [\href{http://arXiv.org/abs/1205.4142}{{\tt 1205.4142}}].

\bibitem{Closset:2012ru}
C.~Closset, T.~T. Dumitrescu, G.~Festuccia and Z.~Komargodski, {\it
  {Supersymmetric Field Theories on Three-Manifolds}},  {\em JHEP} {\bf 05}
  (2013) 017 [\href{http://arXiv.org/abs/1212.3388}{{\tt 1212.3388}}].

\bibitem{Kapustin:2010xq}
A.~Kapustin, B.~Willett and I.~Yaakov, {\it {Nonperturbative Tests of
  Three-Dimensional Dualities}},  {\em JHEP} {\bf 10} (2010) 013
  [\href{http://arXiv.org/abs/1003.5694}{{\tt 1003.5694}}].

\bibitem{Nishioka:2013haa}
T.~Nishioka and I.~Yaakov, {\it {Supersymmetric Renyi Entropy}},  {\em JHEP}
  {\bf 10} (2013) 155 [\href{http://arXiv.org/abs/1306.2958}{{\tt 1306.2958}}].

\bibitem{Chester:2014fya}
S.~M. Chester, J.~Lee, S.~S. Pufu and R.~Yacoby, {\it {The $ \mathcal{N}=8 $
  superconformal bootstrap in three dimensions}},  {\em JHEP} {\bf 09} (2014)
  143 [\href{http://arXiv.org/abs/1406.4814}{{\tt 1406.4814}}].

\bibitem{Binder:2020ckj}
D.~J. Binder, S.~M. Chester, M.~Jerdee and S.~S. Pufu, {\it {The 3d $
  \mathcal{N} $ = 6 bootstrap: from higher spins to strings to membranes}},
  {\em JHEP} {\bf 05} (2021) 083 [\href{http://arXiv.org/abs/2011.05728}{{\tt
  2011.05728}}].

\bibitem{Alday:2021ymb}
L.~F. Alday, S.~M. Chester and H.~Raj, {\it {ABJM at strong coupling from
  M-theory, localization, and Lorentzian inversion}},  {\em JHEP} {\bf 02}
  (2022) 005 [\href{http://arXiv.org/abs/2107.10274}{{\tt 2107.10274}}].

\bibitem{Bobev:2023lkx}
N.~Bobev, J.~Hong and V.~Reys, {\it {Large N partition functions of 3d
  holographic SCFTs}},  {\em JHEP} {\bf 08} (2023) 119
  [\href{http://arXiv.org/abs/2304.01734}{{\tt 2304.01734}}].

\bibitem{Bao:2013pwa}
L.~Bao, V.~Mitev, E.~Pomoni, M.~Taki and F.~Yagi, {\it {Non-Lagrangian Theories
  from Brane Junctions}},  {\em JHEP} {\bf 01} (2014) 175
  [\href{http://arXiv.org/abs/1310.3841}{{\tt 1310.3841}}].

\end{thebibliography}\endgroup

\end{document}